\let\savedaddtocontents\addtocontents
\documentclass[aps,prx,reprint,superscriptaddress]{revtex4-2}

\usepackage[english]{babel}
\usepackage{amsmath}
\usepackage{amsthm}
\usepackage{amssymb}
\usepackage{braket}
\usepackage{graphicx}
\usepackage{dsfont}
\usepackage{qcircuit}
\usepackage{xcolor}
\usepackage{booktabs}
\usepackage{verbatim}
\usepackage{tikz}

\usepackage{mdframed}

\let\addtocontents\savedaddtocontents

\usepackage{hyperref}

\hypersetup{
    colorlinks,
    linkcolor={red},
    citecolor={blue},
    urlcolor={blue}
}

\mdfdefinestyle{nolines}{%
  hidealllines=true,%
  nobreak=true, 
  leftmargin=0pt,
  rightmargin=0pt,
  innerleftmargin=0pt,
  innerrightmargin=0pt,
}

\newtheorem{theorem}{Theorem}
\newtheorem{corollary}[theorem]{Corollary}
\newtheorem{lemma}[theorem]{Lemma}
\newtheorem{proposition}[theorem]{Proposition}
\newtheorem{definition}[theorem]{Definition}

\newtheorem{assumption}[theorem]{Assumption}

\DeclareUnicodeCharacter{2217}{\textasteriskcentered}

\makeatletter
\renewcommand{\tableofcontents}{%
  \begin{center}
    {\large \bfseries \contentsname}
  \end{center}
  \vspace{0.3cm}
  \@starttoc{toc}%
}
\makeatother

\begin{document}
\let\oldaddcontentsline\addcontentsline
\renewcommand{\addcontentsline}[3]{}


\title{Quantum algorithm for weakly correlated molecular electronic ground states with sub-quintic scaling in the system size}


\author{Oskar Leimkuhler}
\affiliation{Department of Mathematical Sciences, University of Copenhagen,
Universitetsparken 5, 2100 Copenhagen \O{}, Denmark}

\author{Marvin Kronenberger}
\affiliation{Department of Chemistry and Applied Biosciences, ETH Zurich, Vladimir-Prelog-Weg 2, 8093 Zurich, Switzerland}

\author{Davide Castaldo}
\affiliation{Department of Chemistry and Applied Biosciences, ETH Zurich, Vladimir-Prelog-Weg 2, 8093 Zurich, Switzerland}

\author{Timothy Stroschein}
\affiliation{Department of Chemistry and Applied Biosciences, ETH Zurich, Vladimir-Prelog-Weg 2, 8093 Zurich, Switzerland}

\author{Srushti Patil}
\affiliation{
NNF Quantum Computing Programme, Niels Bohr Institute, University of Copenhagen, Blegdamsvej 17, 2100 Copenhagen \O{}, Denmark}

\author{Rebecca Tomann}
\affiliation{Department of Mathematical Sciences, University of Copenhagen,
Universitetsparken 5, 2100 Copenhagen \O{}, Denmark}

\author{Arta Safari}
\affiliation{Department of Chemistry and Applied Biosciences, ETH Zurich, Vladimir-Prelog-Weg 2, 8093 Zurich, Switzerland}

\author{Marek Miller}
\affiliation{Department of Mathematical Sciences, University of Copenhagen,
Universitetsparken 5, 2100 Copenhagen \O{}, Denmark}

\author{Gemma C. Solomon}
\affiliation{
NNF Quantum Computing Programme, Niels Bohr Institute, University of Copenhagen, Blegdamsvej 17, 2100 Copenhagen \O{}, Denmark}
\affiliation{Nano-Science Center and Department of Chemistry, University of Copenhagen, Universitetsparken 5, 2100 Copenhagen \O{}, Denmark}

\author{Matthias Christandl}
\affiliation{Department of Mathematical Sciences, University of Copenhagen,
Universitetsparken 5, 2100 Copenhagen \O{}, Denmark}

\author{Markus Reiher}
\affiliation{Department of Chemistry and Applied Biosciences, ETH Zurich, Vladimir-Prelog-Weg 2, 8093 Zurich, Switzerland}

\date{October 29, 2026}

\begin{abstract}
Weakly correlated electronic structures are predominant among molecules, and are routinely solved with heuristic classical methods such as the ``gold standard'' CCSD(T) coupled cluster model, incurring a formal time complexity $O(n^7)$ in the number of molecular orbitals and with no guarantee of energy error $\varepsilon$ at or below chemical accuracy. In this work, a quantum algorithm is presented for ground state energy estimation of weakly correlated molecules using Trotterized time evolution, with an end-to-end quantum operations cost that scales as $\tilde{O}(n^{4.8}/\varepsilon^2)$ in the limit of high-order product formulae when run as a low-depth, shot-noise limited algorithm, or $\tilde{O}(n^{4.8} + n^{3.8}/\varepsilon)$ when scaling the simulation time $T\propto 1/\varepsilon$. The algorithm combines adiabatic state preparation using a smoothed Gevrey schedule with eigenstate filtering via quantum prolate diagonalization (QPD). Both steps of the algorithm rely on a size-independent spectral gap assumption, which is validated by numerical studies on alanine chains with up to 64 monomers. We benchmark the performance of our eigenstate filtering algorithm on a set of 28 molecules to compare our approach against prior phase estimation algorithms, finding that QPD achieves state-of-the-art cost tradeoffs for $\varepsilon$-accurate results. Physical quantum resource estimates are obtained for running Trotter step circuits with up to 40 spin-orbitals, providing a clear path for scaling up quantum simulation of chemistry on early fault-tolerant hardware. 
\end{abstract}

\maketitle

\section{Introduction}
\label{sec:introduction}

Quantum algorithms promise to accelerate molecular electronic structure calculations, enabling tractable ground and excited state energy estimates for systems with thousands of orbitals, with applications in enzyme catalysis \cite{reiher2017elucidating}, battery material design \cite{zini_quantum_2023}, and drug discovery \cite{santagati_drug_2024}. To date, the majority of quantum algorithms research for molecular electronic structure has focused on strongly correlated systems, for which classical methods are severely challenged, but which make up only a small fraction of molecules at equilibrium geometries \cite{chan_spiers_2024-1}. A notable exception is Ref. \cite{chen_framework_2026}, which argued a polynomial speedup in weakly correlated systems using quantum heuristic methods based on coupled-cluster and the Bethe-Salpeter equations. In the fault-tolerant regime, the dominant paradigm for end-to-end molecular electronic ground state calculations with rigorous error guarantees is to combine a quantum subroutine for state preparation with subsequent eigenstate filtering. Approaches to the eigenstate filtering step include quantum phase estimation (QPE) \cite{aspuru-guzik_simulated_2005-1,kitaev_quantum_1995,lee_even_2021-1,su_fault-tolerant_2021-1,von_burg_quantum_2021}, as well as single-ancilla variants\,\cite{kitaev1997quantum,parrish_quantum_2019,ding2023simultaneous, ding_even_2023, ding2024quantum,klymko2022real,cortes_quantum_2022,li2023adaptive,yi2024quantum, shen2025estimating,ding2026quantum}. While significant progress has been made to reduce the asymptotic cost of eigenstate filtering, the state preparation step remains largely unresolved in strongly correlated systems. As discussed in Ref. \cite{lee_evaluating_2023-2}, the mean-field Slater determinant can have vanishing overlap with the true ground state, and current post-HF state preparation strategies are based on heuristic ground state approximations such as MPS \cite{berry_rapid_2025}. Furthermore, strongly correlated systems often have small spectral gaps, causing the asymptotic runtime of rigorous methods such as adiabatic state preparation (ASP) to diverge \cite{lidar_adiabatic_2009,aspuru-guzik_simulated_2005-1}.

In this work, we direct our attention to classes of weakly correlated systems which exhibit a robust, size-independent spectral gap, $E_1-E_0 \geq \Delta$, and for which the energy scales extensively in the system size for a fixed basis set in the asymptotic limit (i.e., as $O(n)$, where $n$ is the number of orbitals). We demonstrate that these assumptions enable asymptotic runtime guarantees for ground state energy estimation using Trotterized time evolution, targeting the early fault-tolerant regime where the significant prefactor overhead of asymptotically optimal block-encoding techniques \cite{su_fault-tolerant_2021-1} is not yet achievable at scale. First, we demonstrate fast state preparation with a rigorous ground state overlap guarantee via ASP, with a simulation time $T_\text{ASP}=\tilde{O}(n/\Delta^2)$. Second, we show that the eigenstate filtering step can be performed using quantum prolate diagonalization (QPD) \cite{stroschein_ground_2025} with a simulation time $T_\text{QPD}=O(1/\Delta)$, which is independent of $n$ under the spectral gap assumption. This enables ground state estimation with rigorous error guarantees with a shot count that scales as $\tilde{O}(1/\varepsilon^2)$, where $\varepsilon$ is the target chemical accuracy threshold. Alternatively, in the fully fault-tolerant regime our algorithm can be run with a simulation time $T_\text{QPD}=O(1/\varepsilon)$, requiring a shot count of $\tilde{O}(1)$, matching the theoretical lower bound for Heisenberg-limited phase estimation or alternative eigenstate filtering methods \cite{kimura_fisher_2026}. The algorithmic framework described in this work also allows for the selection of any intermediate scaling regime, to optimally balance the coherence time on quantum hardware against the shot budget.

By leveraging Trotterization schemes based on the double-factorized form of the electronic structure Hamiltonian \cite{motta_low_2021-1}, we bound the asymptotic scaling of the total quantum operations cost for ground state energy estimation, defined as the cumulative total number of logical quantum gates over all circuit runs.
Using a second-order symmetric product formula we obtain an end-to-end quantum operations cost of $\tilde{O}(n^{6.2})$ in the system size, which can be reduced to $\tilde{O}(n^{5.5})$ with a fourth-order product formula, or $\tilde{O}(n^{4.8})$ in the limit of high-order product formulae. In this limit the wall clock time can be made sub-quartic in $n$ by exploiting gate parallelization. We emphasize that these scalings are obtained using a standard norm-bound for the commutator sums in the analysis of the Trotter error, which are well known to be highly conservative \cite{childs_theory_2021,poulin_trotter_2015,motta_low_2021-1,martinez2023assessment, mehendale2025estimating, kronenberger2026trotter}. By comparison, the ``gold standard'' classical heuristic method CCSD(T) has a formal computational scaling of $O(n^7)$. While error bounds have been derived for this method\,\cite{rohwedder2013error,laestadius2018analysis,faulstich2019analysis,hassan2023analysis}, a formal result on how these bounds scale with computational resources is still lacking.
Although classical methods are subject to continual improvement, e.g., using local-basis approximations \cite{riplinger_efficient_2013}, these heuristic improvements incur additional errors \cite{liakos2020comprehensive}. Furthermore, the asymptotic scaling for the quantum algorithm is also subject to future improvement, since the commutator scaling for the Trotter error analysis may also be significantly tightened by exploiting the locality of interactions in weakly correlated molecular electronic ground states. We thereby establish high-accuracy ground state energy estimates for weakly correlated molecular systems as a genuine use-case for early fault-tolerant quantum computers, offering a potential polynomial quantum speedup in $n$ over the best classical competitor.

We round out our analysis with a series of numerical studies. In Section \ref{sec:spectral_gap} the size-independent spectral gap assumption is established using HF calculations for alanine chains with up to 64 monomers, and coupled-cluster calculations along the full adiabatic pathway demonstrate the absence of a gap closure or an avoided crossing. In Section \ref{sec:benchmark_main} we benchmark quantum prolate diagonalization for eigenvalue estimation on a set of 28 molecular systems, attaining state-of-the art runtime estimates compared to alternative single-ancilla methods and textbook QPE. In particular, we demonstrate that both the Hamiltonian simulation time and the total runtime are independent of the system size, which confers a significant advantage over these alternatives. Section \ref{sec:resource_estimates} provides a scaling analysis for the Hamiltonian one-norm in active spaces of the arginine tryptophane amino acid dimer with up to 300 spin-orbitals. Explicit gate counts are provided for compiled circuits to implement a single Trotter step on up to 40 system qubits, and runtime estimates are obtained for the fault-tolerant execution of these circuits using a surface code, demonstrating that they can be implemented within minutes under standard assumptions on the physical error rate and the cycle time.

\section{Theory}

\subsection{Complete set of assumptions}
\label{sec:assumptions}

This work addresses the molecular electronic structure problem within the Born-Oppenheimer (clamped nuclei) and finite basis approximations, with $\eta$ electrons in $n$ spin-orbitals, where the Hamiltonian is written in second quantized form as
\begin{align}\label{eq:el_ham}
H = \sum_{pq}^n h_{pq} \hat{a}^\dag_p\hat{a}_q + \sum_{pqrs}^n h_{pqrs} \hat{a}^\dag_p\hat{a}^\dag_q\hat{a}_r\hat{a}_s,
\end{align}
with coefficients $h_{pq}$ and $h_{pqrs}$ obtained by one- and two-electron integrals over a set of kinetic and Coulomb potential energy operators describing interacting point charges \cite{helgaker_second_2000}. The ground state energy estimation problem is to approximate the lowest eigenvalue $E_0$, that is, to produce an estimate $\tilde{E}_0$ that satisfies
\begin{align}
|\tilde{E}_0-E_0|\leq \varepsilon,
\label{eq:energy_error_target}
\end{align}
where $\varepsilon$ is a constant additive error threshold. Ref. \cite{OAI:QMA-hardness-of-continuum-Coulomb-energy-with-unit-nuclear-charges-September-24-2026} has recently claimed to prove that this problem is $\mathsf{QMA}$-hard with arbitrary nuclear arrangements in the continuum limit (i.e., with no reference to a finite basis approximation). However, different restrictions on the nuclear arrangement produce a rich variety of molecular and extended material systems, for which the ground state energy estimation problem can exhibit varying complexity. 

In this work we establish efficient ground state energy estimation by a quantum algorithm for a specific class of molecular electronic structure Hamiltonians, which satisfy the following three assumptions.

\begin{assumption}
The Hartree-Fock method converges with low-order polynomial time complexity in $n$, providing the mean-field Hamiltonian $H^{(0)}$, and the HF state $\ket{\psi_0^{(0)}}$.

\label{asmp:hf_efficient}
\end{assumption}

\begin{assumption}
Let $s\in[0,1]$ be a real-valued interpolation parameter, and define
\begin{align}
H^{(s)} = (1-s)H^{(0)} + sH^{(1)},
\label{eq:interp_ham}
\end{align}
where $H^{(1)}$ is the full Hamiltonian in Eq. \ref{eq:el_ham}. Then for all values of $s\in[0,1]$, the spectral gap is bounded below by
\begin{align}
\Delta^{(s)} = E_1^{(s)} - E_0^{(s)} \geq \Delta,
\end{align}
where $\Delta$ is a constant independent of $n$.
\label{asmp:spectral_gap}
\end{assumption}

\begin{assumption}
The energy scale of the Hamiltonian is size-extensive, meaning that within a fixed basis set, the spectral norm scales as $\|H^{(s)}\|=\Theta(n)$ in the asymptotic limit as $\eta$ and $n$ are increased proportionally to the system size.
\label{asmp:size_extensive}
\end{assumption}

These assumptions are considered to hold in general for weakly correlated molecular systems (see \ref{app:weak_correlation}). We demonstrate specific systems where Assumption \ref{asmp:spectral_gap} can be validated numerically in Section \ref{sec:spectral_gap}.

\subsection{Bound on the adiabatic simulation time}
\label{sec:adiabatic_sim}

The first step in any eigenstate filtering algorithm is to prepare an initial state, $\ket{\Phi}$, that has non-vanishing support on the ground state. This step is vital, because the overlap between the mean-field state and the true ground state generally decays exponentially in $n$,
\begin{align}
|\braket{\psi_0^{(0)}|\psi_0^{(1)}} | \propto e^{-\xi n},
\end{align}
where $\xi$ is a system-dependent decay parameter, a phenomenon sometimes referred to as the `Van Vleck orthogonality catastrophe' \cite{kohn_nobel_1999}. This exponential decay in ground state support can be seen even for systems decomposed into non-interacting fragments, as the number of fragments increases. While Hartree-Fock initial states have been shown numerically to exhibit robust ground state support for moderately sized systems \cite{tubman_postponing_2018}, the orthogonality catastrophe presents an obstacle to rigorous end-to-end asymptotic analysis in the large $n$ limit. 

For weakly correlated systems that satisfy Assumption \ref{asmp:spectral_gap}, the size-independent lower bound on the spectral gap enables efficient ground state preparation with exact error bounds by adiabatic state preparation (ASP) \cite{lidar_adiabatic_2009,aspuru-guzik_simulated_2005-1,jansen_bounds_2007}, which exploits the quantum adiabatic theorem to prepare a time-evolved state $\ket{\Phi}$ that is arbitrarily close to $\ket{\psi_0^{(1)}}$, quantified by the 2-norm bound condition
\begin{align}
\| P_\perp\ket{\Phi}\| \leq \epsilon_\perp,
\label{eq:asp_perp_projection}
\end{align}
where $P_\perp = \mathds{1} - \ket{\psi_0^{(1)}}\bra{\psi_0^{(1)}}$ is the projector onto the subspace orthogonal to the true ground state, and $\epsilon_\perp$ is the target accuracy in the state preparation.

The state is prepared by evolving under a time-dependent Hamiltonian, 
\begin{align}
\ket{\Phi} = \mathcal{T} \exp\left( -i T_\text{ASP} \int_{0}^1 H^{(s(\tau))} \, d\tau \right)\ket{\psi^{(0)}_0},
\label{eq:asp_evol}
\end{align}
where $\mathcal{T}$ is the time-ordering superoperator (see e.g. Ref. \cite{wiebe_simulating_2011}), $\tau=t/T_\text{ASP}$ is the renormalized time parameter, and $H^{(s)}$ is as defined in Eq. \ref{eq:interp_ham}. The schedule function $s(\tau)\in[0,1]$ is chosen to be monotonically increasing on $\tau\in[0,1]$. Refs. \cite{elgart_note_2012,jansen_bounds_2007} established near-optimal bounds for the runtime of adiabatic state preparation using schedule functions from the Gevrey class, which have compactly supported derivatives that vanish at $\tau=0$ and $\tau=1$ (see \ref{app:adiabatic}). Combining this result with Assumptions \ref{asmp:hf_efficient}--\ref{asmp:size_extensive} yields a bound on the adiabatic simulation time.

\begin{proposition}
Let $\ket{\Phi}$ be the time-evolved state in Eq. \ref{eq:asp_evol}, where $H^{(s)}$ is defined in Eq. \ref{eq:interp_ham} and $s(\tau)$ is a smooth Gevrey schedule, with $\tau=t/T_\emph{ASP}$ the renormalized time parameter. Given Assumptions \ref{asmp:hf_efficient}--\ref{asmp:size_extensive}, preparing a time-evolved state $\ket{\Phi}$ that satisfies Eq. \ref{eq:asp_perp_projection} to any constant precision $\epsilon_\perp=O(1)$, requires an adiabatic simulation time
\begin{align}
T_\emph{ASP} = \tilde{O}\left(\frac{n}{\Delta^2}\right),
\end{align}
where the dependence on the target precision is $\emph{log}(1/\epsilon_\perp)$.
\label{prop:adiabatic_sim_time}
\end{proposition}

The proof is contained in \ref{app:adiabatic}.

In reality, we will prepare a Trotterized state $\ket{\tilde{\Phi}}$, such that the target precision $\epsilon_\perp$ now denotes the acceptable upper bound on the sum of two error components,
\begin{align}
 \|P_\perp \ket{\tilde{\Phi}}\| \leq \epsilon_{\perp}^\text{ad} + \epsilon_{\perp}^\text{trot} \leq \epsilon_\perp,
\end{align}
where $\|P_\perp\ket{\Phi}\|\leq \epsilon_\perp^\text{ad}$, and $\|\ket{\tilde{\Phi}}-\ket{\Phi}\|\leq \epsilon_{\perp}^\text{trot}$. We analyze the Trotter error contribution in Section \ref{sec:trotter_bounds}.

\subsection{Eigenstate filtering by quantum prolate diagonalization}

Having prepared an input state $\ket{\Phi}$, single-ancilla time-series methods use a Hadamard test to estimate the value of the autocorrelation function,
\begin{align}
C(t) = \braket{\Phi|e^{iHt}|\Phi},
\label{eq:autoc_function}
\end{align}
at a series of discrete time points. Various classical post-processing procedures can then be applied to estimate the eigenphase $E_0$. The first example has been put forward by Kitaev\,\cite{kitaev1997quantum}, and more recently several different algorithms based on the same quantum data have been proposed\,\cite{ding2023simultaneous, ding_even_2023, ding2024quantum,klymko2022real,cortes_quantum_2022,li2023adaptive,yi2024quantum, shen2025estimating,ding2026quantum}. We refer the reader to \ref{app:methods_phase_estimation}, where we provide a summary of the methods we have implemented to assess their performance in estimating molecular electronic ground state energies (see Sec.\,\ref{sec:benchmark_main}) and show how they can all be cast under the same cost model along with standard multi-ancilla QPE\,\cite{nielsen_quantum_2010}.

Here we adapt the framework of quantum prolate diagonalization (QPD) \cite{stroschein_ground_2025} to the case where the input state has arbitrarily high support on the ground state, which is separated from the excited states by a large spectral gap $\Delta$, by projecting onto a spectral filter band $[-W,W]$, of width $W\leq \Delta/2$. In this case we can decompose the autocorrelation function into in-band and out-of-band components,
\begin{align}
C(t) = C_W(t) + C_\perp(t).
\end{align}
The spectral window can be arbitrarily positioned using a level-shift, which amounts to transforming the autocorrelation function as
\begin{align}
C(t) \mapsto e^{i\omega_0 t} C(t),
\end{align}
which incurs no additional error since $e^{i\omega_0 t}$ is simply a unitary phase transformation, and the correct ground state energy is then obtained by subtracting $\omega_0$ from the computed estimate within the window. Note that locating the ground state by `sliding' the position of window across the Hamiltonian spectrum incurs a classical post-processing cost of $O(n/W)$ under Assumption \ref{asmp:size_extensive}. Given that the ground state has been isolated within the window, the in-band component is
\begin{align}
C_W(t) = |\phi_0|^2 e^{i E_0 t},
\end{align}
where $E_0\in[-W,W]$, and the out-of-band component obeys the pointwise bound $|C_\perp(t)| \leq \epsilon^2_\perp$.

We will demonstrate that the QPD framework enables exact computation of $E_0$ to arbitrary accuracy from a small set of autocorrelation samples over a time window $[0,2T]$, where $T=O(1/W)$, with no classical numerical optimization required.
Prolate spheroidal wave functions have optimal properties for 
the task of approximating an arbitrary band-limited signal $C_W(t)$ from a discrete set of measurements within a finite time window:
they are the sequence of band-limited functions $\{f_p(t)\}$
that is optimally localized on $[-T,T]$ \cite{slepian_prolate_1961,landau_prolate_1961,landau_prolate_1962}. The insight of Stroschein \emph{et al.} \cite{stroschein_ground_2025} was to use prolate spheroidal filter functions to construct a basis of filtered quantum states $\{\ket{v_0},\ldots,\ket{v_{M-1}}\}$, 
\begin{align}
\ket{v_p} = \gamma_p^{\frac{1}{2}}\Big(\int_{-\infty}^{\infty} f_p(t) e^{-iHt} dt\Big) \ket{\Phi},
\end{align}
in which the Hamiltonian can be diagonalized. The constant $\gamma_p \in[0,1]$ is the $p$-th Slepian eigenvalue, which quantifies the concentration of the $p$-th prolate function on the domain $[-T,T]$ (see \ref{app:prolate__theory}). Because the prolate filter functions are strictly band-limited, they project the input state $\ket{\Phi}$ exactly onto the filter band $[-W,W]$.
This then allows for multiple in-band frequencies to be disambiguated by computing the Hamiltonian and overlap matrix elements,
\begin{align}
h_{pq} = \braket{v_p|H|v_q}, \qquad s_{pq} = \braket{v_p|v_q},
\label{eq:subs_matels}
\end{align}
and solving a generalized eigenvalue problem on a classical computer to extract the energy estimates within the subspace. Because the prolates form a basis for all band-limited functions, the generalized eigenvalue problem then yields arbitrarily accurate energies, using a time window parameter $T$ that scales inversely with the minimum separation between in-band energy levels, and with an eigenvalue error that decays exponentially as $M$ is increased past the Nyquist-Shannon number,
\begin{align}
M_0=\frac{2WT}{\pi},
\end{align}
provided that the time-bandwidth product satisfies the inequality $WT\geq \pi$.

If $C_W(t)$ contains only a single in-band frequency $E_0$, then it suffices to choose $T=\pi/W$, since there is no need to distinguish between multiple in-band frequencies. The filtered states $\ket{v_p}$ are then all parallel vectors, and the exact energy can be obtained using only a single basis state, $\ket{v_0}$, such that the eigenphase is given by $E_0 = h_{00}/s_{00}$.
However, the denominator $s_{00}=\braket{v_0|v_0}$ in this expression decreases sharply as $E_0$ approaches the edges of the band at $-W$ or $W$, resulting in an ill-conditioned estimator when $h_{00}=\braket{v_0|H|v_0}$ is obtained up to an additive error term. The numerical instability in this estimator can be resolved by retaining $M$ prolate basis functions, and computing
\begin{align}
E_0 = \frac{\vec{u}^\dag \mathbf{H}\vec{u}}{\|\mathbf{S}\|},
\end{align}
where the subspace matrix pair $(\mathbf{H},\mathbf{S})$, a.k.a. the `matrix pencil', has elements given by Eqs. \ref{eq:subs_matels}, and $\vec{u}$ is the principal eigenvector of $\mathbf{S}$. 

This follows because $\mathbf{S}$ is a rank-1 matrix, and we prove in \ref{app:qpd_outline} that choosing to retain $M>M_0$ provides a guaranteed lower bound on the largest singular value of the overlap matrix,
\begin{align}
\|\mathbf{S}\| = 2T|\phi_0|^2 - O^*(\epsilon_\text{trunc}),
\end{align}
where the asymptotic notation $O^*(\cdot)$ suppresses polynomial factors in $M$, $T$, and $W$. $\epsilon_\text{trunc}$ represents the error in the reconstructed in-band autocorrelation function $C_W$ on $[-T,T]$ due to truncating the expansion to $M$ prolates, which we prove is asymptotically bounded by 
\begin{align}
\epsilon_\text{trunc} = O^*\left(e^{-\frac{\alpha}{2}(M-M_0)}\right), \qquad \alpha = \frac{\pi^2}{\ln(WT)}
\end{align}
(see Lemma \ref{lem:S_norm_trunc_bound} in \ref{app:qpd_outline}). We note that while setting $M>M_0$ provides an energy estimator that is guaranteed to be well-conditioned for any $E_0\in[-W,W]$, it can often be more advantageous in practice to use $M<M_0$ prolates in more general settings where the out-of-band error $\epsilon_\perp$ cannot be arbitrarily suppressed (see Section \ref{sec:discussion} and \ref{app:out_of_band}).

\subsection{Subspace matrices from numerical quadrature}
\label{sec:subspace_quad}

By the double-orthogonality property of the prolate basis, they form an orthogonal basis for all band-limited functions on both time domains $(-\infty,\infty)$ and $[-T,T]$ simultaneously, which means that the matrix elements in Eqs. \ref{eq:subs_matels} can be written exactly as integrals in terms of the in-band component of the autocorrelation function, $C_W(t)$, over the domain $[-T,T]$:
\begin{align}
h_{pq} &= i(\gamma_p\gamma_q)^{-\frac{1}{2}}\iint_{-T}^T f_p(t) \frac{\partial}{\partial t}C_W(t-t') f_q(t') \,dt\,dt', \label{eq:ham_matel_int}\\
s_{pq} &= (\gamma_p\gamma_q)^{-\frac{1}{2}}\iint_{-T}^T f_p(t) C_W(t-t') f_q(t') \,dt\,dt'. 
\label{eq:ovlp_matel_int}
\end{align}
We now demonstrate how to approximate the continuous integrals in Eqs. \ref{eq:ham_matel_int} and \ref{eq:ovlp_matel_int} to arbitrary accuracy by numerical quadrature. Here we assume that we can perfectly measure the band-limited autocorrelation function $C_W(t-t')$, although in reality we can only measure $C(t-t')$, which also includes the small out-of-band component, of magnitude bounded by $\epsilon^2_\perp$, which is treated rigorously in Section \ref{sec:theory_asymptotic_bound}. The time derivative of $C_W$ in Eq. \ref{eq:ham_matel_int} is handled by integrating by parts (see \ref{app:numerical_quad}).
The 2d integrals are then approximated by discrete summations over a 2d quadrature grid, with a set of time points and associated quadrature weights for each of the time coordinates $t$ and $t'$ given by
\begin{align}
\{t_k,w_k\}_{k=1}^K, \qquad t_k\in[-T,T].
\label{eq:quad_grid}
\end{align}
This requires $(K^2-K)/2$ total measurements of $C_W$ on the time domain $[0,2T]$, by Hermiticity of the autocorrelation function, and since the maximum difference $|t_k-t_l|\leq 2T$, and $C_W(0)=1-O(\epsilon^2_\perp)$.

Using the asymptotic notation $O^*(\cdot)$ to suppress polynomial factors in $M,T,K,$ and $W$, the discrete summation yields the matrix expressions
\begin{align}
\mathbf{H} &= \frac{i}{2}\left(\mathbf{Y}\mathbf{A}\mathbf{X}^\dag + \mathbf{X}\mathbf{A}\mathbf{Y}^\dag \right) + O^*(\epsilon_\text{quad}), 
\label{eq:H_mat}\\ 
\mathbf{S} &= \mathbf{X}\mathbf{A}\mathbf{X}^\dag + O^*(\epsilon_\text{quad}),
\label{eq:S_mat}
\end{align}
where $\mathbf{A}$ is the $K\times K$ autocorrelation matrix with elements $a_{kl} = C_W(t_k-t_l)$, $\mathbf{X}$ is an $M\times K$ discretized prolate projection matrix, and $\mathbf{Y}$ is an $M\times K$ discretized prolate-derivative projection matrix, which absorbs the boundary corrections due to integrating by parts. $\mathbf{X}$ and $\mathbf{Y}$ are numerically pre-computable from point evaluations of the prolate functions and their derivatives, and have well-bounded operator norms (further details of the construction and explicit norm bounds are provided in \ref{app:numerical_quad}).

The band-limited functions which appear in Eqs. \ref{eq:ham_matel_int} and \ref{eq:ovlp_matel_int} belong to a Paley-Wiener space, which allows for tight bounds on the quadrature error using standard results in approximation theory via the Paley-Wiener theorem \cite{wiener_fourier_1934} (see \ref{app:prolate_quadrature}). For example, when the quadrature grid in Eq. \ref{eq:quad_grid} is selected according to a Gauss-Legendre quadrature rule \cite{doi:10.1137/1.9781611975949.ch19}, the quadrature error is super-exponentially suppressed in the number of grid-points,
\begin{align}
\epsilon_\text{quad} = O^*(e^{-2K \ln(\beta K)}), \qquad \beta = \frac{2}{eWT},
\end{align}
which is proven rigorously in \ref{app:prolate_quadrature}. Therefore, assuming $WT=\pi$, the quadrature error can be suppressed below $\sim 10^{-6}$ using $K=10$ grid-points, or below $\sim 10^{-10}$ using only $K=12$ grid-points. Since suppressing $\epsilon_\text{quad}$ to any constant additive error threshold requires $K$ to grow only sub-logarithmically in the target precision, the sub-dominant terms in $\epsilon_\text{quad}$ can be safely dropped from the end-to-end error analysis.

\subsection{Asymptotic error bound for the eigenvalue estimate}
\label{sec:theory_asymptotic_bound}

In reality, we cannot perfectly measure the in-band autocorrelation signal $C_W$, instead we obtain the noisy autocorrelation matrix
\begin{align}
\tilde{\mathbf{A}} = \mathbf{A} + \delta\mathbf{A}_\text{stoch} + \delta\mathbf{A}_\text{coh} + \delta \mathbf{A}_\perp,
\end{align}
where $\delta \mathbf{A}_\text{stoch}$ represents shot noise and stochastic gate errors, $\delta \mathbf{A}_\text{coh}$ includes Trotter error and coherent gate errors, and $\delta \mathbf{A}_\perp$ denotes the out-of-band error due to the small support of $\ket{\Phi}$ on the excited states. Substituting $\tilde{\mathbf{A}}$ into Eqs. \ref{eq:H_mat} and \ref{eq:S_mat} then yields the noisy matrix pencil $(\tilde{\mathbf{H}},\tilde{\mathbf{S}})$, from which we obtain the eigenphase estimate
\begin{align}
\tilde{E}_0 = \frac{\vec{u}^\dag \tilde{\mathbf{H}} \vec{u}}{\|\tilde{\mathbf{S}}\|},
\label{eq:noisy_energy_estimator}
\end{align}
where the $\vec{u}$ is now the principal eigenvector of $\tilde{\mathbf{S}}$.
In the following we assume that the element-wise stochastic error terms are independent normally distributed random variables, with mean zero and variance $\epsilon_\text{stoch}^2$, so that $\delta \mathbf{A}_\text{stoch}$ is a random Wigner matrix. We also assume the elements of the coherent error matrix are of magnitude at most $\epsilon_\text{coh}$. Then in the limit where all the error terms are decreased to achieve some target precision, a simple matrix perturbation analysis on Eq. \ref{eq:noisy_energy_estimator} yields the following asymptotic error bound theorem, which is proven in \ref{app:prolate_error_prop}.

\begin{theorem}[Asymptotic error bound for QPD with a single in-band phase]
Let $C(t)$ be the autocorrelation function obtained from the input state $\ket{\Phi}$, with an in-band component $C_W(t)=|\phi_0|^2e^{iE_0 t}$, where $E_0\in[-W,W]$, and an out-of-band component bounded by $|C_\perp(t)|\leq\epsilon^2_\perp$. Let $\tilde{E}_0$ be the energy estimate obtained by Eq. \ref{eq:noisy_energy_estimator} with a suitable choice of $K,M=\tilde{O}(1)$, such that $\epsilon_\emph{trunc}$ and $\epsilon_\emph{quad}$ are negligible compared to other sources of error. Assume that each autocorrelation measurement is obtained with a stochastic error term of variance $\epsilon_\emph{stoch}^2$, and a coherent error term of magnitude at most $\epsilon_\emph{coh}$. Then the eigenvalue estimate $\tilde{E}_0$ satisfies the asymptotic bound
\begin{align}
|\tilde{E}_0-E_0| = O\left( W(1+\mathcal{C}) \left( \frac{\epsilon_\emph{stoch}}{\sqrt{K}} + \epsilon_\emph{coh} + \epsilon_\perp^2 \right) \right),
\end{align}
where $\mathcal{C} = \sqrt{MK}(1+WT) = \tilde{O}(1)$ by choosing $TW=O(1)$.
\label{thm:qpd_error_single_eigenphase}
\end{theorem}
As a corollary, we have that,
assuming ideal (i.e., noiseless) quantum hardware, the stochastic error term is entirely due to shot noise, and can be suppressed as
\begin{align}
\epsilon_{\text{stoch}} = O(N_\text{shots}^{-1/2}),
\end{align}
where $N_\text{shots}$ is the number of shots per autocorrelation evaluation.

\begin{corollary}
Assume that the stochastic error is entirely due to shot noise, and is suppressed by taking a number of measurements $N_\emph{shot}$ per grid-point evaluation of $C(t)$. Choose a spectral bandwidth of $W=O(1/T)$, and allow $T$ as a free parameter. Then holding $\epsilon_\emph{coh}$ and $\epsilon_\perp$ constant, the energy error is asymptotically bounded by
\begin{align}
|\tilde{E}_0-E_0| = O\left(\frac{1}{T}\left(1+\frac{1}{\sqrt{K N_\emph{shot}}}\right)\right).
\end{align}
\label{cor:qpd_error_T_scaling}
\end{corollary}

This reveals a tradeoff between the circuit depth and the shot count: by choosing $W=O(\Delta)$ and $T=O(1/\Delta)$, to suppress $|\tilde{E}_0-E_0|\leq \varepsilon$ requires a total number of measurements $\propto K^2N_\text{shot}=\Omega(1/\varepsilon^2)$, resulting in a total runtime of $\Omega(1/\varepsilon^2)$. However, by choosing $T=\Theta(1/\varepsilon)$, and shrinking the width of the spectral band as $W=O(1/T)$, one can resolve the eigenphase to any error $\varepsilon$ by requiring a total number of measurements $\propto K^2 N_\text{shots}=\tilde{O}(1)$. In other words, the total number of circuit runs is then at worst $\text{polylog}(1/\varepsilon)$, and the total quantum runtime is $\tilde{O}(1/\varepsilon)$, thus achieving Heisenberg-limited scaling. Note that this incurs a classical post-processing cost of $O(n/\varepsilon)$ to locate the ground state energy, by `sliding' a narrow frequency band using the level shift across the range of the Hamiltonian spectrum (which grows as $O(n)$ under the size-extensivity assumption), until the ground state is detected inside the band.

\subsection{Bounding the Trotter error}
\label{sec:trotter_bounds}

Trotter error describes the error from a Hamiltonian simulation,
where a Trotter product formula is employed to discretize the time
evolution operator $e^{-iHt}$ \cite{lloyd_universal_1996,childs_theory_2021}, assuming a decomposition 
\begin{align}
H = \sum_{k=1}^L H_k,
\end{align}
such that $e^{-iH_k t}$ has an exact quantum circuit description for all $k$.
In our analysis we assume the Lie-Trotter-Suzuki construction, \cite{suzuki_general_1991} although other Trotterization schemes could be substituted, since the asymptotic error bounds depend only on the order of the non-canceling commutator terms in the BCH expansion.
For generality, we refer to the Trotterized product $S_p$ as a $p$-th order product formula, which is given by a Lie-Trotter-Suzuki formula of order $p-1$. The formulae for $p=1$ and $p=2$ are given by
\begin{align}
S_1(t) &= \prod_{k=1}^{L} e^{-i H_k t}, \label{eq:first_order_pf}\\
S_2(t) &= \bar{S}_1(t/2) S_1(t/2), \label{eq:sec_order_pf}
\end{align}
where $\bar{S}_1$ reverses the order of application for the unitary operators in Eq. \ref{eq:first_order_pf},
and higher orders can be defined through a recursive formula \cite{suzuki_general_1991}. The unitary time evolution operator $e^{-iHt}$ can be then approximated to arbitrary accuracy by a product of $N$ $p$-th order Trotter steps,
\begin{align}
e^{-iHt} \approx \left(S_p(\delta t)\right)^{N}, \qquad \delta t = \frac{t}{N}.
\end{align}
Trotter error can be quantified by various measures, ranging from
differences in the time-evolved wave function, to differences in the spectrum of the effective Hamiltonian $H_{\mathrm{eff}}$ that is simulated by Trotterized simulation \cite{kronenberger2026trotter}. For our asymptotic analysis, we use a standard operator norm characterization, which contributes to the coherent error term $\epsilon_\text{coh}$ in Theorem \ref{thm:qpd_error_single_eigenphase}, and which obeys the asymptotic $p$-th order commutator bound \cite{childs_theory_2021},
\begin{align}
\|e^{-iHt} - \left(S_p(\delta t)\right)^{N} \| = O\left( \frac{t^{p+1} \aleph_p}{N^p} \right),
\label{eq:trotter_error_asymptotic}
\end{align}
where $\aleph_p$ is the sum of all $p$-th order nested commutators between $H_1,\ldots,H_L$. Through repeated application of standard matrix norm relations, we can bound $\aleph_p = O(\lambda^{p+1})$, where $\lambda =\sum_k\|H_k\|$ is the induced 1-norm of the Hamiltonian. Then rearranging for $N$ yields an asymptotic bound on the number of Trotter steps to implement $e^{-iHt}$ up to any constant additive error $\epsilon_\text{trot}$ (quantified through the norm difference expression in Eq. \ref{eq:trotter_error_asymptotic}):
\begin{align}
N = O\left( \frac{(\lambda t)^{1+1/p}}{\epsilon_\text{trot}^{1/p}} \right).
\label{eq:trotter_stepcount}
\end{align}
We stress that the one-norm bound for $\aleph_p$ neglects all sign cancellations in the nested commutator sum, and has been observed to provide a highly conservative upper bound on the required Trotter step complexity in practice \cite{childs_theory_2021,poulin_trotter_2015,motta_low_2021-1}. For the gate count analysis in Section \ref{sec:resource_estimates}, we will mainly consider Trotter product formulae with $p\leq 4$, due to the availability of analytical bounds with small prefactors \cite{childs_theory_2021}, and extensions to higher-order formulae will only be considered in terms of asymptotic complexity.

In addition to the time-independent Hamiltonian evolution used to measure the autocorrelation function in Eq. \ref{eq:autoc_function}, we also require to Trotterize the time-dependent Hamiltonian evolution in Eq. \ref{eq:asp_evol} in order to prepare the input state, $\ket{\Phi}$. A widely adopted framework to discretize this time-dependent Hamiltonian simulation operator is that of standard time-dependent product formulae, first introduced by Suzuki \cite{suzuki_general_1993} and further developed in Ref. \cite{wiebe_higher_2010}. We refer the reader to \ref{app:adiabatic_product_formulae} for the details of the construction. In this time-dependent case, the Trotter error is bounded by a sum over nested commutators (the `spatial commutator terms', which have the same mathematical form as $\aleph_p$) as well their time derivatives \cite{mizuta_theory_2026}. We prove in \ref{app:adiabatic_commutator_bound} that, for the smooth Gevrey schedule described in Section \ref{sec:adiabatic_sim}, the terms involving time derivatives grow sub-dominantly to the spatial commutator terms, provided that we choose $T_\text{ASP}=\Omega(L/\lambda)$, which leads to the following simplification in bounding the Trotter step complexity.

\begin{proposition}
Let the approximate evolved state $\ket{\tilde{\Phi}}$ be obtained by substituting a $p$-th order standard time-dependent product formula for the time-dependent adiabatic evolution operator in Eq. \ref{eq:asp_evol}, defined by interpolating between $H^{(0)}$ and $H^{(1)}$ using a smoothed $G^2$ Gevrey schedule, with $T_\emph{ASP}=\Omega(L/\lambda_\emph{max})$, where $\lambda_\text{max} = \max_s \lambda^{(s)}$. Then a total number of Trotter steps
\begin{align}
N = O\left( \frac{(\lambda_\emph{max}T_\emph{ASP} )^{1+1/p}}{(\epsilon_\perp^{\emph{trot}})^{1/p}} \right)
\end{align}
is sufficient to bound the adiabatic state preparation error by $\| \ket{\tilde{\Phi}} - \ket{\Phi} \| \leq \epsilon_\perp^{\emph{trot}}$.
\label{prop:adiabatic_trotter_stepcount}
\end{proposition}

The scaling of the induced one-norm $\lambda$ as a function of $n$ depends heavily on the spatial geometry of the system \cite{koridon_orbital_2021}, the choice of orbital basis \cite{koridon_orbital_2021, ollitrault2025improving}, and the Hamiltonian decomposition strategy \cite{von_burg_quantum_2021, lee_even_2021-1, rocca2024reducing}. In this work, we employ the double-factorized Hamiltonian representation introduced by Motta \emph{et al.} \cite{motta_low_2021-1}, which enables asymptotic reductions in the gate complexity per Trotter step as discussed in Section \ref{sec:end-to-end-main}. By double-factorizing $\bar{H}=\sum_k \bar{H}_k$, we can write the interpolated Hamiltonian in Eq. \ref{eq:interp_ham} as
\begin{align}
H^{(s)} = H^{(0)} + s\bar{H}_1 + s \sum_{k=2}^L G_k A_k^2 G_k^\dag,
\label{eq:el_ham_df}
\end{align}
where $H^{(0)}$ and $\bar{H}_1$ are one-body Hamiltonians, the $G_k$ are orbital rotation operators, and the $A_k$ are diagonal one-body Hamiltonians,
\begin{align}
A_k = \sum_{p=1}^{\rho_k} h^{(k)}_p \hat{n}_p.
\label{eq:df_one_body_ops}
\end{align}
While a dense factorization of the electron repulsion integral tensor would incur a classical time complexity of $O(n^6)$, for $L<n^2$ this classical cost can be reduced to $O(Ln^4)$ using an iterative solver (see \ref{sec:circuit_lowrank}). The induced one-norm is then obtained as 
\begin{align}
\lambda^{(s)} = \|H^{(0)}\| + s\|\bar{H}_1\|+ s\sum_{k=2}^L \|A_k\|^2,
\label{eq:lambda_bound}
\end{align}
implying $\lambda_\text{max} = \lambda^{(1)}$. Because Trotter error arises from non-canceling nested commutators between Hamiltonian terms which all conserve particle number, one could obtain a tighter bound on the Trotter error by replacing the spectral norms in Eq. \ref{eq:lambda_bound} with the fermionic seminorm \cite{su_nearly_2021} (see \ref{app:fermionic_seminorm}).

\subsection{End-to-end quantum operations cost}
\label{sec:end-to-end-main}

\begin{figure}[h]
    \centering
    \includegraphics[width=0.77\linewidth]{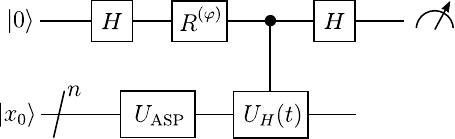}
    \caption{Hadamard test circuit to estimate the autocorrelation function $C(t)$ in Eq. \ref{eq:autoc_function} up to an additive error $\epsilon_\text{stoch} + \epsilon_\text{coh}$, denoting stochastic and coherent error contributions, using $O(1/\epsilon_\text{stoch}^2)$ shots. The system register is initialized in a computational basis state $\ket{x_0}$ of Hamming weight $\eta$, and the Trotterized adiabatic time evolution circuit $U_\text{ASP}$ then prepares the input state $\ket{\tilde{\Phi}}$. The Trotterized time evolution circuit $U_{H}(t)$ is then run for a simulation time $t\in[0,2T]$ with control from the ancilla qubit, which is measured in the $Z$-basis at the end of the computation to extract $\text{Re}[C(t)]=P(1)-P(0)$ when the ancilla phase is set to $\varphi=0$, or $\text{Im}[C(t)]=P(1)-P(0)$ by setting $\varphi=\pi/2$.}
    \label{fig:autoc_had_test}
\end{figure}

We now combine Proposition \ref{prop:adiabatic_sim_time}, Theorem \ref{thm:qpd_error_single_eigenphase}, and Corollary \ref{cor:qpd_error_T_scaling}, together with Equation \ref{eq:trotter_stepcount} and Proposition \ref{prop:adiabatic_trotter_stepcount}, to yield a complete asymptotic bound for the end-to-end quantum operations cost
required to produce an energy estimate $\tilde{E}_0$ satisfying Eq. \ref{eq:energy_error_target}. We define the total quantum operations cost $N_\text{quop}$ as the cumulative total number of quantum gates summed over all circuit runs. The Hadamard test quantum circuit used to estimate the autocorrelation function is shown in Figure \ref{fig:autoc_had_test}, which produces an estimate of the in-band autocorrelation function $C_W(t)$ up to an additive error $O(\epsilon_\text{stoch} + \epsilon_\text{coh} + \epsilon_\perp^2)$, using a number of shots per circuit of $N_\text{shot}=O(1/\epsilon_\text{stoch}^2)$. To estimate $\tilde{E}_0$ according to Eq. \ref{eq:noisy_energy_estimator} using the numerical quadrature scheme outlined in Section \ref{sec:subspace_quad}, the total number of circuit runs is then $O(K^2 N_\text{shot})$. 

Using $N_\text{gate}^{\text{ASP}}$ to denote the number of gates in the Trotterized adiabatic time evolution circuit $U_\text{ASP}$ to prepare the input state $\ket{\tilde{\Phi}}$, and $N_\text{gate}^{\text{QPD}}$ to denote the number of gates in the deepest Trotterized time evolution circuit $U_H(2T)$ in Figure \ref{fig:autoc_had_test}, we can express the total quantum operations cost as
\begin{align}
N_\text{quop} = O\left( K^2 N_\text{shot} \left( N_\text{gate}^{\text{ASP}} + N_\text{gate}^{\text{QPD}} \right) \right).
\end{align}

We highlight two possible scaling regimes that follow from Corollary \ref{cor:qpd_error_T_scaling}, which we call \emph{wide-band} ($W\!\propto\!\Delta$, $T\!\propto\!\Delta^{-1}$) and \emph{narrow-band} ($W\!\propto\!\varepsilon$, $T\!\propto\!\varepsilon^{-1}$) respectively. In the wide-band, short-time regime, we set the bandwidth parameter $W=O(\Delta)$, and the maximum QPD evolution time as $T=O(1/\Delta)$. In this case, the number of shots required to resolve $\epsilon_\text{stoch} = O(\varepsilon)$ is $O(1/\varepsilon^2)$, where $\varepsilon$ is the target accuracy of the energy estimate. In the narrow-band, long-time regime, we instead shrink the filter bandwidth as $W=O(\varepsilon)$, and set $T=\Theta(1/\varepsilon)$. In this case, the number of required shots per circuit is reduced to $O(1)$, by Corollary \ref{cor:qpd_error_T_scaling}. Assuming that we choose the time-bandwidth product to satisfy $WT=O(1)$, then we can safely choose $K=\tilde{O}(1)$ to suppress the quadrature error to be negligible compared to other sources of error. Then in the wide-band and narrow-band scaling regimes, we have
\begin{align}
T\!\propto\!\Delta^{-1}\!: \,\, N_\text{quop} &= \tilde{O}\left(\frac{1}{\varepsilon^2}  \left( N_\text{gate}^{\text{ASP}} + N_\text{gate}^{\text{QPD}} \right)\right), \\
T\!\propto\!\varepsilon^{-1}\!: \,\, N_\text{quop} &= \tilde{O}\left( N_\text{gate}^{\text{ASP}} + N_\text{gate}^{\text{QPD}} \right).
\end{align}

To analyze the gate count for the Trotterized time evolution circuits, we use the double-factorized form of the Hamiltonian in Eq. \ref{eq:el_ham_df}, where the maximum of the induced one-norm occurs at $s=1$, which we denote as $\lambda$ for compactness. We also define $\rho = \max_k \rho_k$, where $\rho_k$ is the number of terms retained in the second factorization step (see Eq. \ref{eq:df_one_body_ops}). Time evolution under the one-body Hamiltonian $H^{(0)}$ can be implemented with $O(n^2)$ gates per Trotter step \cite{kivlichan_quantum_2018}, while the remaining two-body terms can be implemented with a combined cost of $O(n\rho L)$ gates per Trotter step \cite{motta_low_2021-1} (see \ref{app:df_analytic_gc}). The total number of gates is then
\begin{align}
N_\text{gate}^{\text{ASP}} &= O\left(\frac{(n^2+n\rho L)(\lambda T_\text{ASP})^{1+1/p}}{\epsilon^{1/p}_\perp}\right), \\
N_\text{gate}^{\text{QPD}} &= O\left(\frac{(n^2+n\rho L)(\lambda T)^{1+1/p}}{\epsilon^{1/p}_\text{coh}}\right),
\end{align}
where $\epsilon_\perp$ and $\epsilon_\text{coh}$ provide upper bounds on the required Trotter error for the ASP and QPD steps respectively. In the wide-band regime, we set $T=O(1/\Delta)$, and we require to suppress $\epsilon^2_\perp, \epsilon_\text{coh} =O( \varepsilon)$, while in the narrow-band regime, we set $T=O(1/\varepsilon)$, which suppresses all the error terms by shrinking the bandwidth parameter $W$, thus requiring $\epsilon^2_\perp, \epsilon_\text{coh} = \tilde{O}(1)$ (see Theorem \ref{thm:qpd_error_single_eigenphase} and Corollary \ref{cor:qpd_error_T_scaling}). Substituting these bounds, as well as the bound on the adiabatic simulation time $T_\text{ASP}$ from Proposition \ref{prop:adiabatic_sim_time} under Assumptions \ref{asmp:hf_efficient}--\ref{asmp:size_extensive}, we arrive at the final asymptotic expressions for the total quantum operations cost under both scaling regimes,
\begin{widetext}
\begin{align}
T\!\propto\!\Delta^{-1}\!: \,\, N_\text{quop} &= \tilde{O}\left((n^2+n\rho L)\left(\frac{(\lambda n)^{1+1/p}}{\Delta^{2+2/p}\varepsilon^{2+1/(2p)}} + \frac{\lambda^{1+1/p}}{\Delta^{1+1/p}\varepsilon^{2+1/p}} \right)\right), \label{eq:wideband_e2e_scaling}\\
T\!\propto\!\varepsilon^{-1}\!: \,\, N_\text{quop} &= \tilde{O}\left((n^2+n\rho L)\left(\frac{(\lambda n)^{1+1/p}}{\Delta^{2+2/p}} + \frac{\lambda^{1+1/p}}{\varepsilon^{1+1/p}} \right)\right). \label{eq:narrowband_e2e_scaling}
\end{align}
\end{widetext}
In the limit of high order product formulae ($p\rightarrow \infty$), these reduce to the simpler expressions,
\begin{align}
T\!\propto\!\Delta^{-1}\!: \,\, N_\text{quop} &= \tilde{O}\left((n\!+\!\rho L)\left(\frac{\lambda n^2}{\Delta^2\varepsilon^2} + \frac{\lambda n}{\Delta\varepsilon^{2}} \right)\right), \\
T\!\propto\!\varepsilon^{-1}\!:\,\,  N_\text{quop} &= \tilde{O}\left((n\!+\!\rho L)\left(\frac{\lambda n^2}{\Delta^2} + \frac{\lambda n}{\varepsilon} \right)\right).
\end{align}

Equations \ref{eq:wideband_e2e_scaling} and \ref{eq:narrowband_e2e_scaling} follow directly from Assumptions \ref{asmp:hf_efficient}--\ref{asmp:size_extensive}, without approximation. In realistic systems, approximate asymptotic scalings in $n$ and $\varepsilon$ can be obtained by estimating the asymptotic behavior of $L$, $\rho$, and $\lambda$ in the large $n$ limit. Substituting the empirical scalings $L=O(n)$ and $\rho=O(1)$ from Ref. \cite{motta_low_2021-1}, and our numerical results showing a best-case scaling of $\lambda=O(n^{1.8})$ for the systems studied in Sections \ref{sec:spectral_gap} and \ref{sec:resource_estimates}, and using $\Delta=\Omega(1)$ from Assumption \ref{asmp:spectral_gap}, we obtain the following empirical asymptotic scalings for high order product formulae in terms of $n$ and $\varepsilon$:
\begin{align}
\label{eq:final_empirical_scaling_main_wide}
T\!\propto\!\Delta^{-1}\!: \,\, N_\text{quop} &= \tilde{O}\left(\frac{n^{4.8}}{\varepsilon^2} \right), \\
T\!\propto\!\varepsilon^{-1}\!: \,\, N_\text{quop} &= \tilde{O}\left(n^{4.8} + \frac{n^{3.8}}{\varepsilon} \right).
\label{eq:final_empirical_scaling_main_narrow}
\end{align}

In addition to the total quantum operations cost, the other main consideration is the circuit layer depth, which we denote by $D$. The the circuit runs in QPD are independent and can thus be parallelized, in which case $D$ may be the determining factor for the end-to-end runtime, subject to constraints on hardware availability. Reducing the circuit depth is also vital for running quantum circuits on early fault-tolerant quantum hardware, which have a limited maximum coherence time. Using the low-rank factorization, each Trotter step can be performed in depth $O(n + \rho L)$, which simplifies to $O(n)$ assuming the empirical scalings $\rho=O(1)$ and $L=O(n)$. Then in the limit of high order product formulae, with $\lambda=O(n^{1.8})$, we get the following empirical circuit depth scalings,
\begin{align}
T\!\propto\!\Delta^{-1}\!: \,\, D = \tilde{O}\left( \frac{n^{3.8}}{\Delta^2} + \frac{n^{2.8}}{\Delta} \right), \\
T\!\propto\!\varepsilon^{-1}\!: \,\, D = \tilde{O}\left( \frac{n^{3.8}}{\Delta^2} + \frac{n^{2.8}}{\varepsilon} \right).
\end{align}
This explains why the wide-band, short-time regime may be preferable for early fault-tolerant hardware if one requires high accuracy estimates where $\varepsilon \ll \Delta$. This comes with a tradeoff in the number of shots, which must then scale as $O(1/\varepsilon^2)$, compared to the $\tilde{O}(1)$ scaling of the total number of circuit runs in the narrow-band regime. The algorithmic framework presented in this work also allows for the selection of any intermediate scaling regime, for example scaling the window as $W\propto \varepsilon^a$, where $a\in[0,1]$, thus requiring a simulation time $T=\Theta(\varepsilon^{-a})$ and a number of shots per circuit $N_\text{shot}=O(\varepsilon^{2(1-a)})$, to achieve an optimal balance between the hardware coherence time and the shot budget.

\section{Size-independent spectral gaps in weakly correlated molecules}
\label{sec:spectral_gap}

To investigate the spectral-gap assumption (Assumption \ref{asmp:spectral_gap}), we examine capped poly-L-alanine chains in an ideal $\alpha$-helical conformation as a representative model family of weakly correlated molecular systems. We consider three complementary quantities: Hartree-Fock HOMO-LUMO orbital gaps as a function of chain length, spectral gaps along an adiabatic pathway, and correlated excitation energies for the smallest systems. Together, these calculations provide numerical evidence for the plausibility of favorable gap behavior in weakly correlated molecules. Computational details are provided in Section \ref{sec:si_gap_methods}.

\begin{figure}[h]
    \centering
    \includegraphics[width=0.85\linewidth]{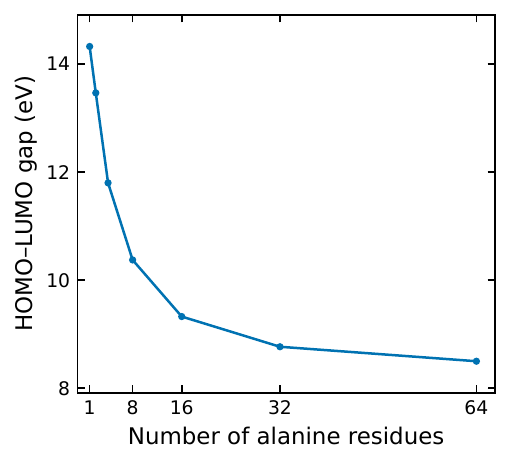}
    \caption{Hartree–Fock HOMO–LUMO gap as a function of the number of alanine residues in capped poly-L-alanine chains with ideal alpha-helical conformation.}
    \label{fig:hf_gap_ala}
\end{figure}
We first examine the dependence of the HF HOMO-LUMO gap on the number of alanine residues, ranging from 1 to 64. As shown in Fig. \ref{fig:hf_gap_ala}, the HOMO-LUMO gap decreases from 14.32~eV for a single alanine residue to 8.50~eV for 64 residues. However, the decrease becomes progressively smaller with increasing chain length, with gaps of 9.32, 8.76, and 8.50~eV for 16, 32, and 64 residues, respectively. The orbital gap therefore remains substantial over the investigated range, with an apparent flattening of its size dependence for longer chains. Additional calculations using the CPCM implicit solvation model for water show an even weaker dependence on chain length, with HOMO-LUMO gaps ranging from 14.67 to 14.45~eV for chains containing 4 to 64 residues. Although the HF HOMO-LUMO gap does not directly determine the excitation gap, it's weak size dependence provides an encouraging indication that increasing the molecular size does not necessarily lead to a closing of the spectral gap for these types of systems.

However, a favorable gap at the beginning of the adiabatic evolution does not guarantee that the gap remains open throughout the adiabatic interpolation. Previous numerical results \cite{kremenetski2021simulation, sugisaki2022adiabatic} have shown that the behavior of the spectral gap can depend strongly on the electronic structure. In particular, Sugisaki \textit{et al.} found that for the stretched N$_2$ molecule, the minimum spectral gap can occur at intermediate points along the interpolation rather than the final Hamiltonian. Furthermore, stretching the N-N bond showed to reduce the gap. While these observations arise in systems with substantial strong-correlation effects, they highlight the importance of examining the entire adiabatic pathway rather than relying on the spectral properties of its endpoints.

We therefore investigate the spectral gap along the adiabatic interpolation as defined in Eq. \ref{eq:interp_ham} and its sensitivity to active-space size. For a  single capped alanine residue, we consider five active spaces ranging from CAS(4,4) to CAS(12,12).

\begin{figure}[h]
    \centering
    \includegraphics[width=0.9\linewidth]{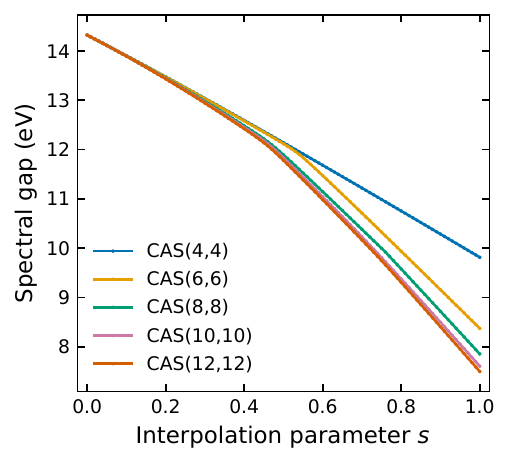}
    \caption{Spectral gap along the linear adiabatic interpolation between the HF Fock operator ($s=0$) and the interaction electronic Hamiltonian ($s=1$). }
    \label{fig:asp_active_space}
\end{figure}

The resulting spectral gaps are shown in Fig. \ref{fig:asp_active_space}. All five scans show a decreasing gap, with the smallest gap occurring at the end point $(s=1)$. The endpoint gap decreases from 9.81~eV for CAS(4,4) to 7.50~eV for CAS(12,12). Although the absolute magnitude of the gap depends on the active-space size, the qualitative behavior along the interpolation remains unchanged, with no intermediate minimum or gap closure observed.

To complement these calculations, we also estimate the lowest correlated excitation energies for the one- and two-residue systems using DLPNO-STEOM-CCSD/def2-SVP. The lowest triplet excitation energies are approximately 5.20 and 5.02~eV, respectively, while the lowest singlet excitation energies are 5.50 and 5.38~eV. This provides further indication that the inclusion of a more extensive treatment of electron correlation does not lead to a vanishing spectral gap. 
Taken together, these results provide numerical support for favorable spectral gap behavior in weakly correlated molecular systems.  

\section{Benchmarking QPD for eigenstate filtering}
\label{sec:benchmark_main}

We now move on studying the performance of QPD as compared to textbook QPE \cite{nielsen_quantum_2010} and some recently developed single-ancilla phase estimation algorithms, namely MM-QCELS \cite{ding2023simultaneous}, QMEGS \cite{ding2024quantum}, ODMD \cite{shen2025estimating} and UVQPE \cite{klymko2022real}. For the sake of completeness, we have discussed the algorithmic details of all these methods in \ref{app:methods_phase_estimation}. In all our numerical experiments, we neglected the effects of 
Trotter and gate errors
in the time evolution (i.e., we set $\epsilon_{\text{coh}} = 0$) and accounted for stochastic errors of variance $\epsilon_{\text{stoch}}^2$ arising purely from shot noise. Regarding state preparation, we did not implement an ASP circuit and instead assumed 
$\ket{\Phi} = \ket{\psi_0^{(0)}}$
(i.e., the initial state is the Hartree-Fock determinant). The results presented here correspond to 19 different molecules for which 
$|\phi_0| = |\braket{\psi_0^{(1)}|\Phi}|\geq 0.9$. As we will see, this is sufficient to simulate the regime considered in our end-to-end analysis at moderate system sizes. In ~\ref{app:additional_benchmark}, we report the same analysis for a smaller subset of nine molecular systems with $|\phi_0| < 0.9$, in order to assess the performance of QPD relative to other prominent phase-estimation algorithms when the assumptions made in this paper do not hold. We refer to \ref{app:comp-detail-benchmark} for further information
regarding the implementation of the methods and computational details regarding the results presented in the following.

\begin{figure*}
    \centering
    \includegraphics[width=0.95\textwidth]{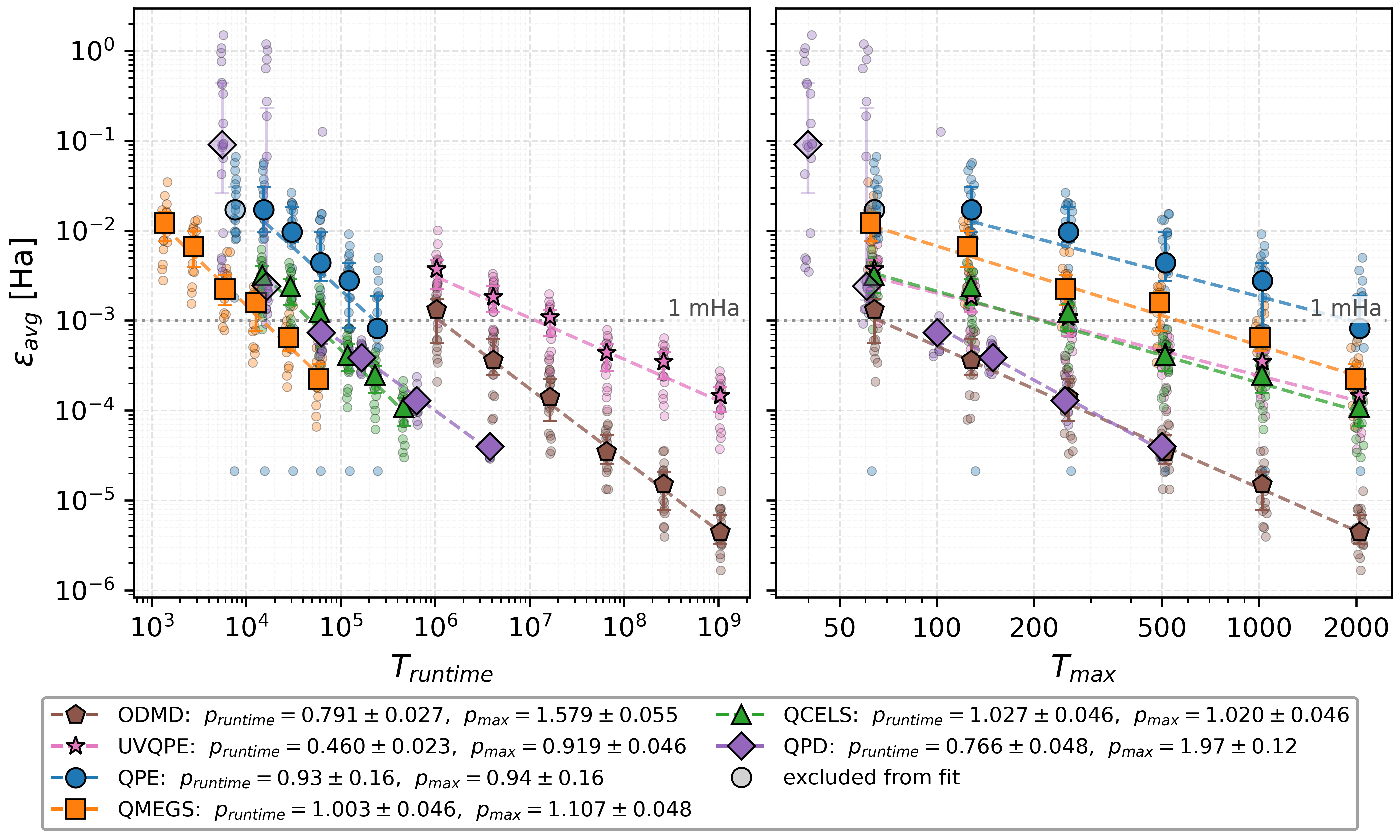}
    \caption{Benchmarking the computational cost of molecular ground-state phase estimation. The legend identifies the different phase estimation algorithms compared in this study (see \ref{app:methods_phase_estimation}). (left) Average estimated ground-state energy as a function of $T_{\mathrm{runtime}}$ (see Eq.\,\ref{eq:Trun_main}). (right) Average estimated ground-state energy as a function of $T_{\mathrm{max}}$. Faint dots represent the average estimates obtained from 10 independent realizations of each algorithm for a given molecule in our benchmark. Larger markers indicate the median of the averaged errors, and error bars represent the interquartile range. Dashed lines show power-law fits of the form $\epsilon_{\mathrm{avg}}(T)=AT^{-p}$. Empty markers correspond to data point not considered for the fit. Further details on the computational procedure are provided in \ref{app:comp-detail-benchmark}.}
    \label{fig:scaling_vs_tmax_truntime}
\end{figure*}

In Fig.\,\ref{fig:scaling_vs_tmax_truntime}, we study the average error in the ground-state energy estimate as a function of the maximum evolution time $T_{\max}$ and the runtime $T_{\text{runtime}}$, defined as

\begin{equation}
\label{eq:Trun_main}
    T_{\text{runtime}} = \sum_{k=1}^{N_t} N_{\text{shots}}^{(k)}\,|t_k|,
\end{equation}
where $N_t$ is the number of independent samples of $C(t)$ measured, and $N_{\text{shots}}^{(k)}$ is the number of shots acquired for the sample $C(t_k)$ (see also ~\ref{app:methods_phase_estimation}). We note that this quantity can be directly related to $N_{\text{quop}}$ (see Sec.\,\ref{sec:end-to-end-main}) once a state-preparation routine and a quantum simulation technique have been specified for a given circuit.

We start by considering the right panel, where we report $\varepsilon_{\text{avg}}$ as a function of $T_{\text{runtime}}$. As we can see, QMEGS (orange squares) is able to achieve chemical accuracy ($\varepsilon \leq 1\,\text{mHa}$) using the fewest quantum resources, with a runtime approximately six times shorter than those of QPD (purple diamonds) and QCELS (green triangles). The performances of the latter two methods are largely comparable in terms of $T_{\text{runtime}}$, and both are almost one order of magnitude faster than standard QPE (blue dots). Moreover, ODMD (brown pentagons) and UVQPE (pink stars) require runtimes that are two to three orders of magnitude higher. Looking more specifically at QPD  we can notice that the first two points exhibit a sharp increase in accuracy as we increase $T_\text{runtime}$. The reason is that those points correspond to $T_\text{max}$ values too short for the prolate filtering condition for suppressing of out-of-band noise to hold (see \ref{app:out_of_band}). While the second point is already out of this exponentially (in $T_\text{max}$) sharp phase transition region, we excluded it from all our power-law fits (dashed lines) in this and all the following analysis.

Regarding the power-law fits, we find that the exponent obtained for our scaling, $(p_{\mathrm{runtime}} = 0.766 \pm 0.048$), lies between the exponents in Eqs.\,\ref{eq:final_empirical_scaling_main_wide} and \ref{eq:final_empirical_scaling_main_narrow}, corresponding to a dense-limit scaling in the nomenclature of Ref.\,\cite{dutkiewicz2022heisenberg}.  In our experiments, we set the filter window width to a constant value, $W=0.35\,\,\text{Ha}$. 
Nevertheless, even under these conditions, we observe asymptotic scalings faster than that predicted by Eq.\,\ref{eq:final_empirical_scaling_main_wide}. It is possible, as also done in Ref.\,\cite{Castaldo2026ChemicallyAccurate} for QPD, to derive lower-bounds for the attainable scaling that only depend on the sampling design (Cramer-Rao lower bounds\,\cite{dutkiewicz2022heisenberg, kimura_fisher_2026}). These kind of bounds for QPD using the shot schedule of this work would predict $p=0.7$, i.e. they would slightly underestimate the performance of the algorithm. We attribute this phenomenon to the fact that the prior knowledge needed to run QPD correctly is completely disregarded by an error analysis that only considers the information content of the acquired data on which the estimator is built.

Similarly, when considering the scaling derived in this work, we should also note that, on the one hand, our error bounds account for discretization errors and shot noise affecting the matrix elements, while out-of-band contributions are bounded by the ASP. On the other hand, although our bounds capture the stabilization of the GEP arising from the fact that the QPD $\varepsilon$-accuracy scales inversely with the minimum eigenvalue $\sigma_m$ of the projected overlap matrix, they do not fully capture the additional stabilization of the projection step as $T_{\max}$ increases. In particular, current analytical bound read $\sigma_m^{-1}\approx (|\phi_0|^2 T_{\max})^{-1}$, with the numerical results yielding $\sigma_m^{-1} \propto T_{\mathrm{runtime}}^{-0.406 \pm 0.034}$ and $\sigma_m^{-1} \propto T_{\max}^{-1.047 \pm 0.088}$ (see \ref{app:out_of_band}--\ref{app:numerical_quad} and \ref{app:additional_benchmark}).

We notice that these considerations regarding the conditioning of the overlap matrix apply also to UVQPE and ODMD, since Toeplitz and Hankel matrices admit Vandermonde decompositions \cite{boley1997general} for which we can apply the results of Ref.\,\cite{moitra2015super}. 
To conclude this discussion on the performance in terms of runtime we want to highlight that, even though also in the case of ODMD and UVQPE the GEPs stabilize increasing $T_\text{max}$, these latter methods are less stable w.r.t. shot noise (with ODMD more stable than UVQPE, as found in Ref.\,\cite{shen2025estimating}) as demonstrated by the fact that to obtain these results we needed to set $N_\text{shots} = 500$ for these methods, as compared to the maximum number of shots used in our QPD run that was $N_\text{shots} = 45$. Again we refer to \ref{app:comp-detail-benchmark} for more computational details. 

We now turn to the right panel, where we study $\varepsilon_\text{avg}$ as a function of $T_\text{max}$. As shown in the figure, the concentration properties of prolates yield one of the best performances in our benchmark set, on par with ODMD, which also achieves chemically accurate estimates for values of $T_\text{max} < 10^2\,a.u.$. This can be contrasted with the value of $T_\text{max}\approx 500\,a.u.$ required by QMEGS, whose advantage over QPD in terms of $T_\text{runtime}$ is offset by its deeper circuits. Finally, we note that all single-ancilla methods achieve chemical accuracy at values of $T_\text{max}$ that are one order of magnitude lower than those required by standard QPE, consistent with the findings reported by the respective authors in their original papers.

\begin{figure*}
    \centering
    \includegraphics[width=0.95\textwidth]{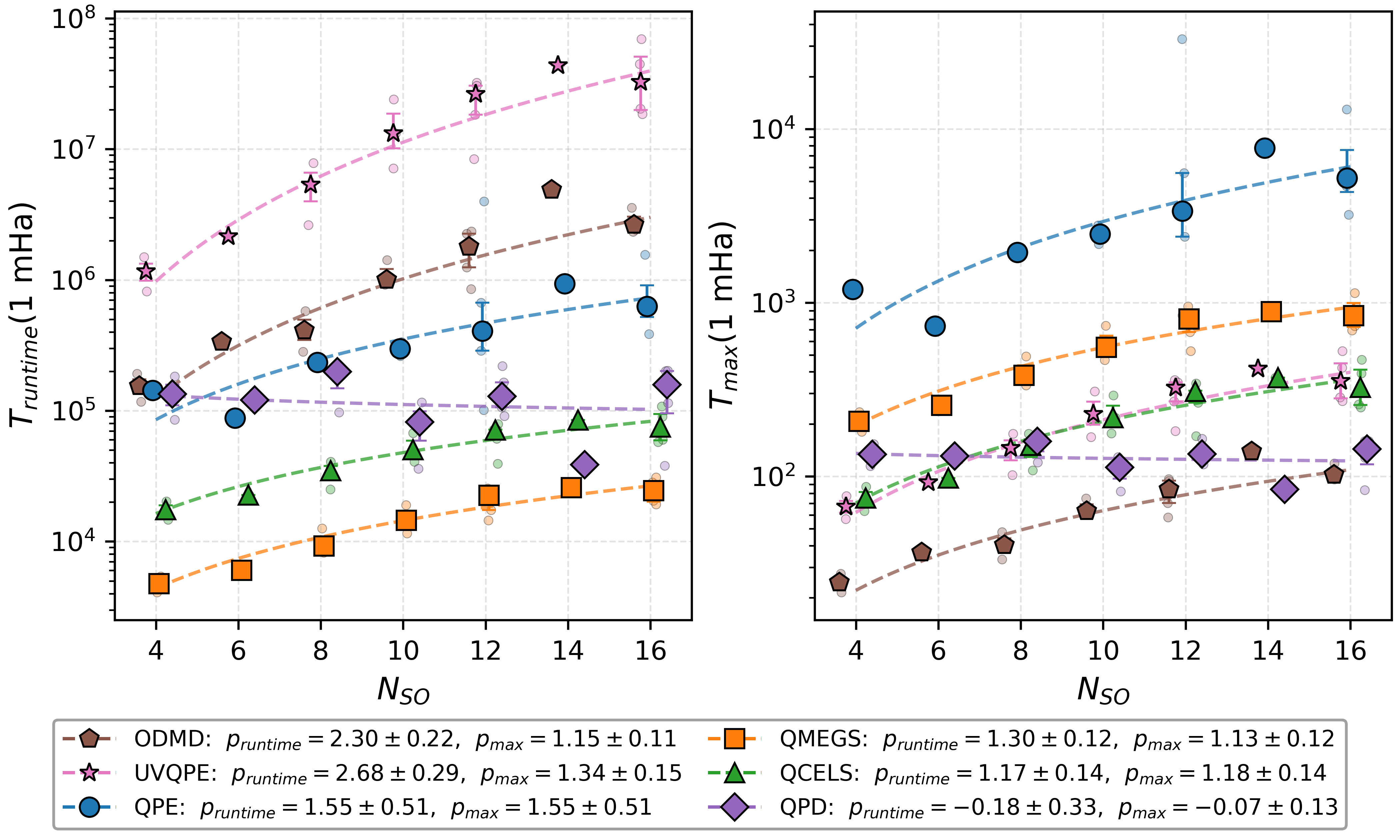}
    \caption{Computational cost scaling with system size. The legend identifies the different  phase estimation algorithms compared in this study (see \ref{app:methods_phase_estimation}). (left) $T_\text{runtime}$ needed to achieve an error estimate of $\epsilon = 1\,\text{mHa}$ of error as a function of the number of spin-orbitals $N_{SO}$. (right) $T_\text{max}$ needed to achieve an error estimate of $\epsilon = 1\text\,{mHa}$ of error as a function of the number of spin-orbitals $N_{SO}$. Faint dots represent the average estimates obtained from 10 independent realizations of each algorithm for a given molecule in our benchmark. Larger markers indicate the median of the averaged errors, and error bars represent the interquartile range. Dashed lines show power-law fits of the form $T(N_{SO})=AN_{SO}^{p}$. Further details on the computational procedure are provided in \ref{app:comp-detail-benchmark}.}
    \label{fig:scaling_vs_active_space_size}
\end{figure*}

A second important aspect of the scaling analysis presented in this paper concerns the dependence of the algorithms on system size. In particular, to achieve the scalings described by Eqs.\,\ref{eq:final_empirical_scaling_main_wide}–\ref{eq:final_empirical_scaling_main_narrow}, we assume that the minimum energy gap $\Delta$ in $C_W(t)$ is independent of system size and/or that this condition can always be satisfied by means of the ASP. Since $T_\text{max}$ in QPD must be chosen such that $T_\text{max} \geq \pi\Delta^{-1}$, under this assumption, $T_\text{max}$ is constant with respect to system size, i.e., $T_\text{max} = O(1)$. In Fig.\,\ref{fig:scaling_vs_active_space_size}, we verify this statement numerically and compare the results with those obtained using the methods discussed so far. In particular, both looking at the minimal $T_\text{runtime}$ and $T_\text{max}$ required to achieve $\varepsilon_\text{avg} = 1\,\text{mHa}$, we can see that both $T_\text{max}=O(1)$ and $T_\text{runtime}=O(1)$ are verified. If we instead look at the other algorithms we can see that there is an explicit dependency of this quantities in terms of system size (here as in the number of spin-orbitals included in the active space $N_{SO}$). The origin of this behavior is different for the different methods. All methods except QPD require to operate on the rescaled Hamiltonian $\widetilde H=(H-e_0)/\lambda$, so they return an estimate of $\widetilde E_0$ with an error $\tilde\varepsilon$ in rescaled units. Mapping back to the original energy scale multiplies this error by the scaling factor, $\varepsilon=\lambda\,\tilde\varepsilon$. At a fixed rescaled time budget, $\tilde\varepsilon$ for QPE, QMEGS and QCELS is not dependent on the system size. For these methods the growth of the physical error $\varepsilon$ with system size reflects the growth of $\lambda$ with the number of orbitals rather than a degradation of the algorithms themselves (see \ref{app:additional_benchmark}). Equivalently, reaching a fixed target accuracy $\varepsilon^\ast$ requires a rescaled accuracy $\tilde\varepsilon^\ast=\varepsilon^\ast/\lambda$, and hence rescaled evolution times that grow with system size as $\lambda$ does. This analysis is supported by the fact that the scaling exponents of these methods are very similar to the scaling exponent of the scaling factor $\lambda$ used in our benchmark set, i.e. $p_{\lambda} = 1.32\,\pm\,0.16$ (see Fig.\,\ref{fig:normalization_factor_scaling}). Interestingly, the behavior of ODMD and UVQPE differs depending on whether one considers the scaling of \(T_{\max}\) or \(T_{\mathrm{runtime}}\) required to achieve chemical accuracy as a function of \(N_{SO}\). In particular, for the former, we observe that the exponents for both methods are comparable to \(\lambda_p\), which can therefore be explained using the same argument presented above. Turning to the runtime exponents, \(p_{\mathrm{runtime}}^{\mathrm{ODMD}} = 2.30 \pm 0.22\) and \(p_{\mathrm{runtime}}^{\mathrm{UVQPE}} = 2.68 \pm 0.29\), we observe an approximately linear scaling overhead arising. Understanding the origin of this additional dependency is beyond the scope of the current work, focusing instead on the characterization of an end-to-end scaling of QPD, and we leave it for future research.

\begin{figure*}
    \centering
    \includegraphics[width=0.95\textwidth]{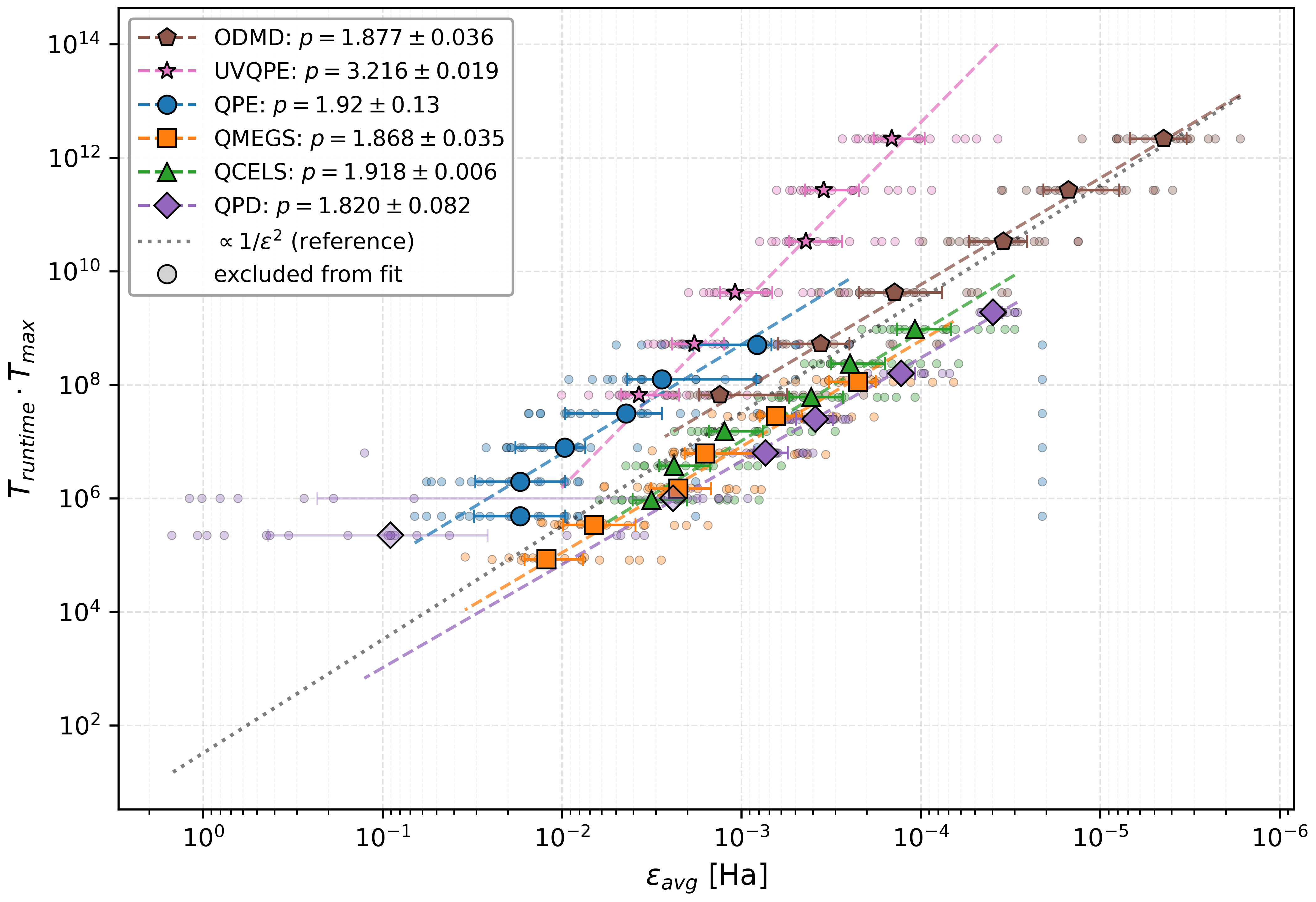}
    \caption{Combined cost $T_\text{runtime}\,T_\text{max}$ needed to reach an average ground-state energy error $\varepsilon_\text{avg}$. Faint dots represent the individual molecules at each computational budget with error estimates averaged from 10 independent realizations of each algorithm. Larger markers indicate the median of the averaged errors, and error bars represent the interquartile range. Dashed lines represent power-law fits $T_\text{runtime}\,T_\text{max}(\varepsilon_\text{avg}) = C\,\varepsilon_\text{avg}^{-p}$. Empty markers correspond to data points not considered for the fit. Each fitting is performed independently for each molecule and the legend gives the mean $\pm$ its standard error. Dotted line represents the reference scaling corresponding to the bound derived in Ref.\,\cite{kimura_fisher_2026}.}
    \label{fig:cost_depth_comparison}
\end{figure*}

We conclude this section by discussing the results shown in Fig.\,\ref{fig:cost_depth_comparison}. Here, we examine the cost product $T_\text{runtime}T_\text{max}$ as a function of the estimation error $\varepsilon$. This quantity allows us to compare the overall performance of different algorithms, particularly in light of the trade-off between circuit depth (i.e., $T_\text{max}$) and total runtime, which, as discussed above, is unavoidable.

To make this statement more rigorous, we also draw on a recent result by Kimura and Mitarai \cite{kimura_fisher_2026}, who derived the following lower bound on this quantity:

\begin{equation}
\label{km_bound}
    T_\text{max}\cdot T_\text{runtime} \geq \frac{c}{\varepsilon^2}
\end{equation}
where $c$ is an algorithm-dependent constant.

In Fig.\,\ref{fig:cost_depth_comparison} we have reported as a dotted line a reference for Eq.\,\ref{km_bound} while all the others methods have been depicted similarly to the previous figures (see captions). As we can see all the methods, except for UVQPE, show scaling exponents very close to the theoretical limit of $p=2$. Moreover, we notice that, when looking at this aggregated metric, QPD performance is the best within our benchmark set closely followed by QMEGS and QCELS.

To conclude this section, we note that the analysis presented here assesses the performance of different phase estimation algorithms using metrics that are agnostic to the quantum simulation routine employed, the logical and physical compilation strategies, the QPU architecture, and the error correction scheme adopted. When these factors are taken into account, the differences in performance between the various methods may become even more pronounced. This is because, for a circuit of depth $D$ to be executed correctly with a constant probability, the logical error rate must satisfy $\epsilon_l \leq D^{-1}$\cite{aharonov2025importance}. Therefore, as the circuit depth increases, the code distance must also increase, leading to greater computational overhead\cite{sivak2023real, wang2026demonstration}.

In the next section, we address this gap in the analysis by providing explicit resource estimates for the QPD algorithm.

\section{Quantum resource estimates}
\label{sec:resource_estimates}

\subsection{Scaling of the double factorized Hamiltonian 1-norm}\label{sec:numerics_df_onorm}

Loose estimation of Trotter error, which allows the subsequent estimation of the number of required Trotter steps (see Eq. \ref{eq:trotter_stepcount}),
is possible by determining the induced 1-norm of the Hamiltonian,
which is defined in Eq. \ref{eq:df_ham_onorm}.
For our end-to-end runtime estimation as a function
of system size, measured by the number of active
space orbitals $n$, we hence need to determine
the polynomial scaling $O(n^a)$ of the function
$\lambda_{\mathrm{DF}}(n)$.

\begin{figure}[h]
    \centering
    \includegraphics[width=0.95\linewidth]{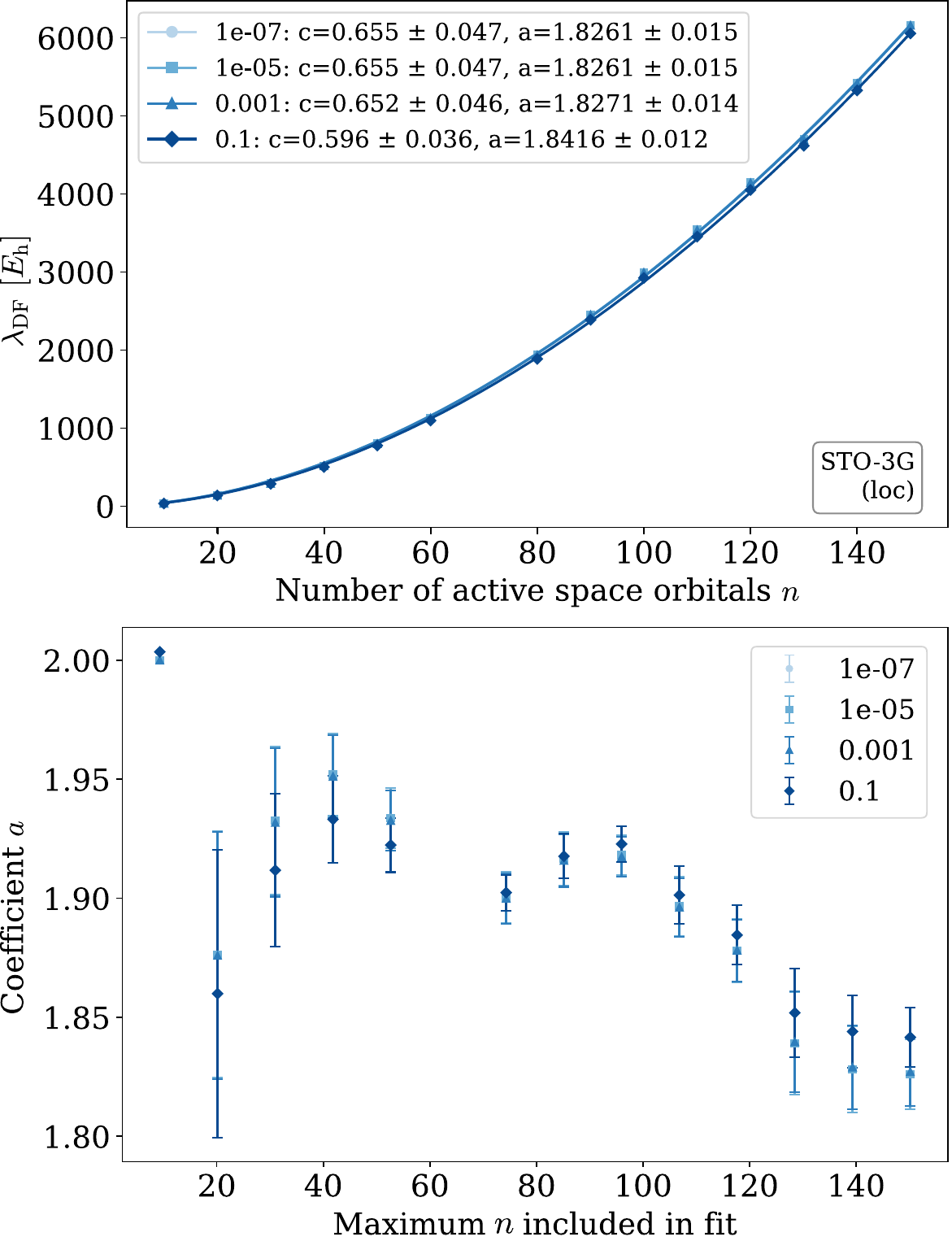}
    \caption{(Top): plots of the induced Hamiltonian 1-norm $\lambda_{\mathrm{DF}}$ as a function of the number of spatial orbitals in the active space.
    The number of active space electrons equals the number of
    active space spatial orbitals.
    A minimal STO-3G basis set was employed,
    with a MP2 natural orbital basis.
    The active orbital space was subsequently localized by the Foster-Boys
    localization scheme \cite{foster1960canonical}.
    Raw data points are shown by various dot symbols.
    Fits of the function $\lambda_{\mathrm{DF}}(n) = cn^a$ for the different
    double factorization thresholds, and
    computed by the scipy library \cite{virtanen2020scipy},
    are plotted as solid lines.
    The parameters for the prefactor $c$ and the polynomial
    coefficient $a$ obtained by the fitting are provided
    in the figure legend.
    (Bottom): Convergence of the polynomial fit coefficient $c$,
    as only a subset of the data points (up to $n$ active space orbitals)
    are considered.}
    \label{fig:df_onorm_scaling_main}
\end{figure}

To estimate the scaling of the induced 1-norm as a function
of the number of spatial orbitals, we computed induced 1-norms
for up to 150 spatial orbitals, using various basis sets,
as well as MP2 natural versus localized active space orbitals.
As a system, we consider the arginine tryptophane amino
acid dimer structure, embedded in a water solvent
that was also employed in Ref. \cite{Castaldo2026ChemicallyAccurate}.
Additional details on the calculation of the 1-norm
for the employed DF Hamiltonian can be found in
\ref{app:numerics_df_onorm_scaling}.
A subset of our results is shown in
Fig. \ref{fig:df_onorm_scaling_main},
with additional results presented
in figs. \ref{fig:df_onorm_scaling}
and \ref{fig:df_onorm_fit_coeff_convergence}.

We can observe in Fig. \ref{fig:df_onorm_scaling_main}
that as $n$ increases, the value of the polynomial coefficient
of $\lambda_{\mathrm{DF}}(n) = O(n^a)$ in a
localized basis appears to continue to decrease.
In the asymptotic limit, one might consequently assume
even smaller values of $a$ compared to those we reported.
We hence make the tentative, realistic assumption of a scaling
of $\lambda_{\mathrm{DF}}(n) = O(n^{1.8})$.

\subsection{Quantum gate counts per Trotter step}
\label{subsec:trotter_steps}

\begin{figure}
    \centering
    \includegraphics[width=\linewidth]{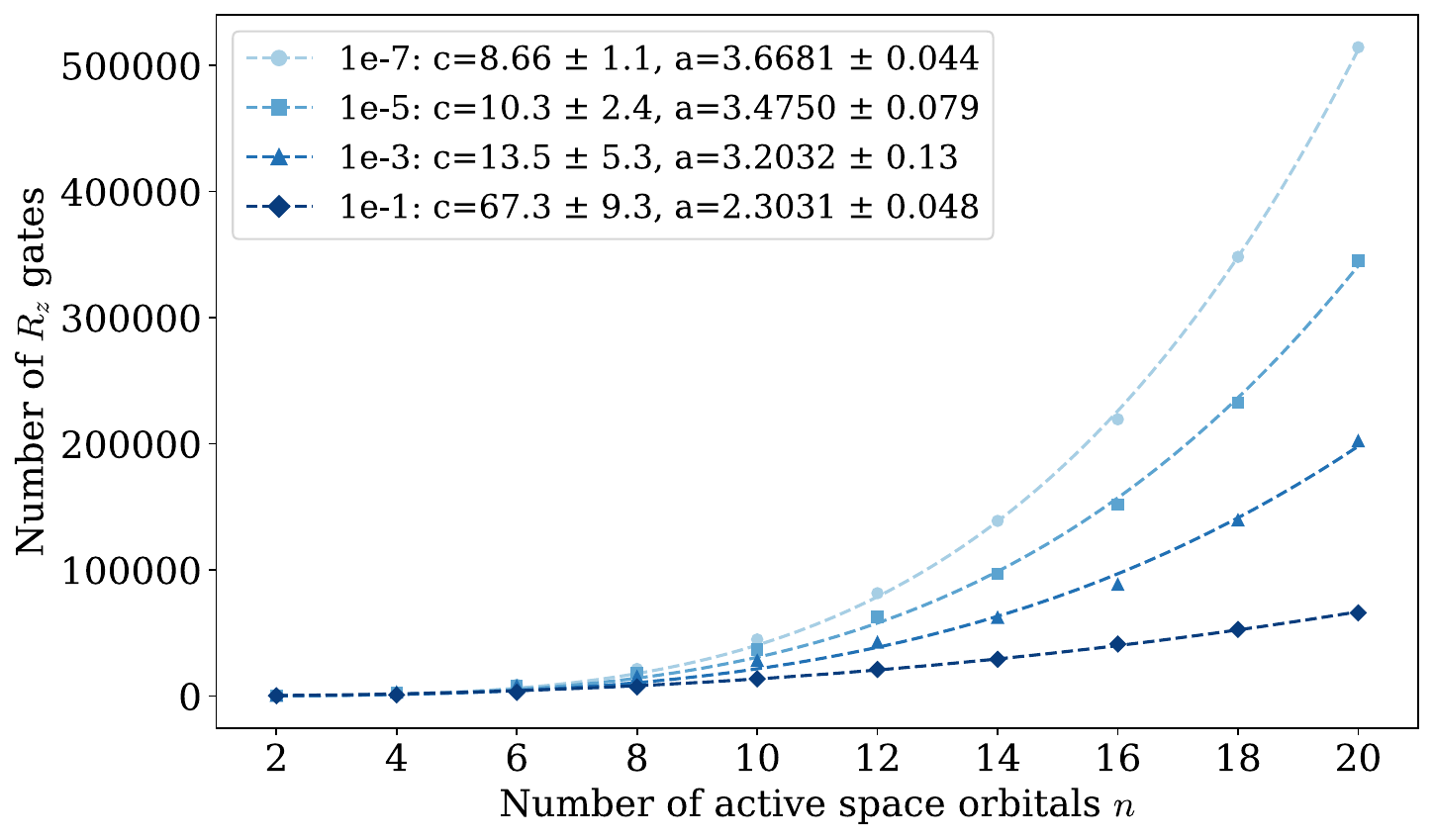}
    \caption{Total number of $R_z$ gates for quantum circuits of a single Trotter step with a
    double factorized Hamiltonian with different double factorization truncation thresholds.
    Raw data points are shown by various dot symbols.
    Fits of the function $R_z(n) = cn^a$ for the different
    double factorization thresholds, and
    computed by the scipy library \cite{virtanen2020scipy},
    are plotted as solid lines.
    The parameters for the prefactor $c$ and the polynomial
    coefficient $a$ obtained by the fitting are provided
    in the figure legend.}
    \label{fig:rz_growth_with_n}
\end{figure}

We have compiled a software stack for the explicit
compilation of qiskit circuits for the controlled evolution
of a single first order Trotter step (as defined in Eq. \ref{eq:first_order_pf}) of the double factorized Hamiltonian,
which employs the PySCF package \cite{sun2018pyscf}, and ffsim libraries \cite{sung2026ffsim}.
Circuits are compiled to a gate set of $\{ R_z(\theta), X, SX, CNOT \}$,
from which we can straightforwardly estimate the number of
T gates, under the assumption of a target rotation
gate synthesis error (see \ref{sec:physical_resource_estimate}).
Additional details regarding the software stack
are provided in \ref{app:compilation_details}. For the arginine-tryptophane structures, which we also
considered for our numerical study in Fig. \ref{fig:df_onorm_scaling_main},
we compiled explicit quantum circuits for active orbital
spaces of various sizes. The number of $R_z$ gates in these circuits
are reported in Fig. \ref{fig:rz_growth_with_n}.
The observed scalings in terms of the parameter
$a$ are discussed in
detail in section \ref{app:comp_circuit_depths}.

\subsection{Physical resource estimates}
\label{sec:physical_resource_estimate}

In this section we estimate the fault-tolerant resource requirements of the Trotter circuits considered in this work.
We consider four surface-code-based architectures that are representative
of the two connectivity regimes of current hardware: the baseline,
compact and fast data blocks \cite{litinski2019}, which rely on
nearest-neighbour lattice surgery, and the active-volume architecture
\cite{litinski2022av}, which assumes long-range connections between
logical qubits.
The details of each architecture are given in~\ref{app:resources}.

We report the resources for a single Trotter step as a function of the
system size.
The Trotter circuits consist mostly of arbitrary-angle $R_z$ rotations,
which we decompose into sequences of Clifford and $T$ gates with the
GridSynth algorithm \cite{ross2016}; a rotation synthesised to precision
$\epsilon_{\rm gs}$ costs $3\log_2(1/\epsilon_{\rm gs}) + O(\log\log(1/\epsilon_{\rm gs}))$
$T$ gates.
We choose the precision such that the accumulated synthesis error of a
Trotter step, the number of $R_z$ rotations $N_{\rm rot}$ times
$\epsilon_{\rm gs}$, stays below $10^{-4}$, i.e.\
$\epsilon_{\rm gs} = 10^{-4}/N_{\rm rot}$.
Since $N_{\rm rot}$ grows from 192 to $1.4\times10^{5}$ across the series,
$\epsilon_{\rm gs}$ tightens from $5\times10^{-7}$ to $7\times10^{-10}$
and the mean cost per rotation from 64 to 95 $T$ gates, giving a $T$ count
per Trotter step of $1.2\times10^{4}$ for 4 spin orbitals up to
$1.3\times10^{7}$ for 40 spin orbitals at a coefficient cutoff of
$10^{-3}$, and up to $2.1\times10^{7}$ at the tightest cutoff.
Using this decomposition we first recompile each circuit into the
Clifford$+T$ gate set, and then rewrite it into the instruction set of the
given architecture \cite{patil2026}: a sequence of Pauli product rotations executed by
lattice surgery for the three data blocks, and a schedule of logical
blocks for the active-volume architecture.

\begin{figure*}[t]
  \centering
  \includegraphics[width=\textwidth]{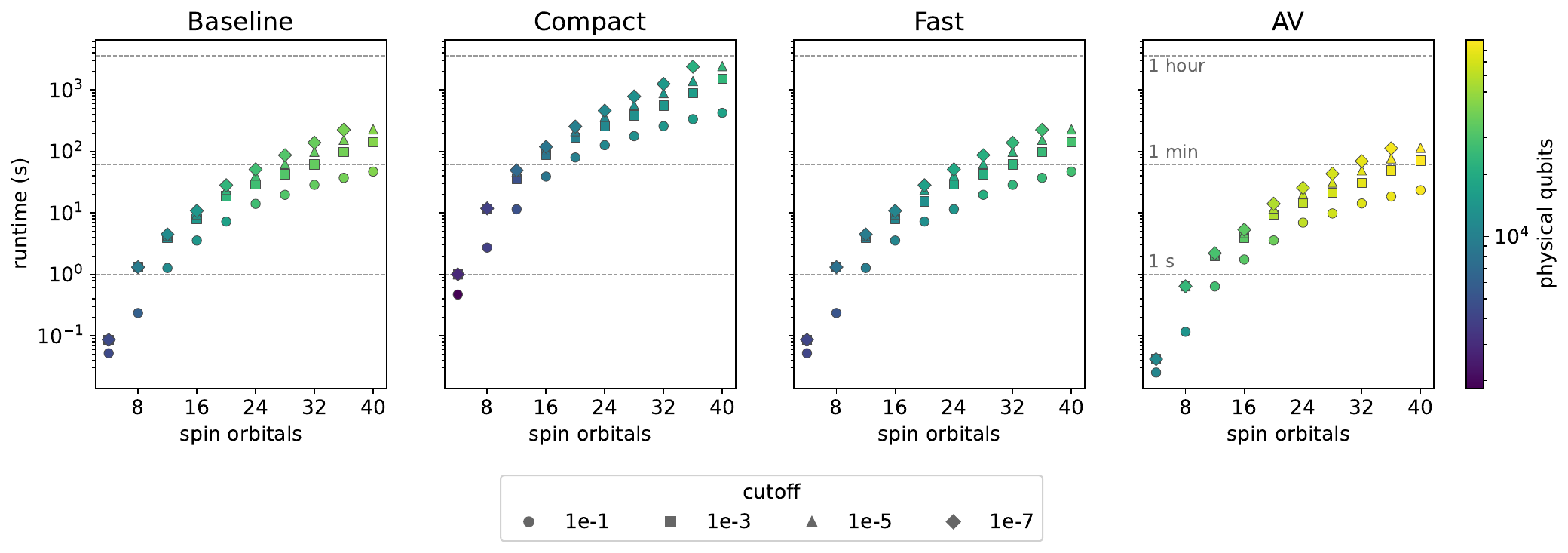}
  \caption{Time-to-solution of one first order product formula Trotter step versus the number of spin orbitals for the CAS($n$,$n$) series on the baseline, compact, fast and active-volume architectures at a physical error rate $p=10^{-4}$.
  Marker shapes distinguish the Hamiltonian coefficient cutoffs and the colour scale gives the total number of physical qubits, data block and factories included.  Each architecture uses the smallest number of 15-to-1 distillation factories at which magic state production is not the bottleneck.  
  Dashed lines mark one second, one minute and one hour.}
  \label{fig:runtime-vs-orbitals-floor}
\end{figure*}

Figure~\ref{fig:runtime-vs-orbitals-floor} shows the resulting
time-to-solution of one Trotter step versus the number of spin orbitals
for the four architectures, with the total physical qubit count, data
block and factories included, indicated by the colour scale.
Each architecture is given the smallest number of 15-to-1 factories at
which the computation is no longer limited by magic state production:
one for the compact block, and two or three for the others depending on
the code distance (see~\ref{app:factory}).
The runtime is then set by the rate at which the data block consumes
$T$ states, so that each panel is close to a straight line in the $T$
count: the baseline and fast blocks execute one $T$ gate per logical
cycle, the compact block one every nine, and the active-volume
architecture two.
Across the series the runtime grows by three orders of magnitude, from
$0.1$\,s for 4 spin orbitals to $142$\,s for 40 spin orbitals on the
baseline block at a coefficient cutoff of $10^{-3}$, following the $T$
count, while the coefficient cutoff shifts the runtime by only a factor
of 2--3 at fixed system size.
The code distance increases from $d=7$ to $d=11$ along the series
($d=13$ for the compact block at 40 spin orbitals), which adds a mild
extra growth through the logical cycle time $d\cdot1\,\mu$s and through
the $2d^{2}$ physical qubits per tile.
The four architectures trade qubits against time.
For the 40-spin-orbital system the baseline block needs $4.4\times10^{4}$
physical qubits for $142$\,s; the fast block reaches the same runtime
with $2.9\times10^{4}$ qubits owing to its smaller footprint, and the
compact block with $2.4\times10^{4}$ qubits is an order of magnitude
slower ($1509$\,s), since it can consume only one $T$ state every nine
logical cycles.
The active-volume architecture halves the runtime to $71$\,s by executing
two $T$ gates per logical cycle, but requires $8.9\times10^{4}$ qubits,
because the Pauli product rotations of these circuits act on nearly all
qubits once the Clifford operations are commuted through, so that the
workspace has to hold two such rotations side by side
(see~\ref{app:av}).
Overall, a single Trotter step of the largest system completes within
minutes on $10^{4}$ to $10^{5}$ physical qubits at $p=10^{-4}$, and the
time-to-solution for a full energy estimate follows by multiplying with
the Trotter step count and the number of sequential circuit runs.

\section{Discussion and Conclusion}
\label{sec:discussion}

In this work we have shown how a quantum algorithm can exploit the size-independent spectral gap of suitable weakly correlated systems to reduce the gate complexity of ground state preparation and energy estimation. Our key result is an end-to-end quantum algorithm for ground state energy estimation based on Trotterized time evolution, with an asymptotic quantum operations complexity that scales as $\tilde{O}(n^{4.8}/\varepsilon^2)$ in the limit of high order product formulae when run as a low-depth, shot-noise limited algorithm, or as $\tilde{O}(n^{4.8} + n^{3.8}/\varepsilon)$ when run as a Heisenberg-limited method, scaling the maximum simulation time as $T\propto 1/\varepsilon$. This algorithm enables ground state energy estimation to within any target additive precision $\varepsilon$, with identical scaling in $n$ for any target precision. 

Our algorithm combines both state preparation and eigenvalue estimation within a unified Hamiltonian simulation framework by exploiting the quantum adiabatic theorem, enabled by the large spectral gap in weakly correlated systems. In Section \ref{sec:spectral_gap}, we validated the size-independent gap assumption for alanine chains with up to 64 monomers, and this finding is in line with the existing literature on protein systems \cite{lever_electrostatic_2013} (see \ref{app:weak_correlation} for further discussion of weak correlation and how this relates to the spectral gap). Alternative approaches to state preparation have been proposed, such as quasi-adiabatic continuation \cite{wan_fast_2022} and dissipative state preparation \cite{lin_dissipative_2025,zhan2026rapid}. In the latter case, Ref. \cite{zhan2026rapid} demonstrated that the gate complexity can be reduced to $\text{polylog}(n)$ under an assumption on the dissipative mixing time. A direct comparison of state preparation methods in weakly correlated molecular electronic structure Hamiltonians is left to future work.

For the eigenvalue estimation step, we adapted quantum prolate diagonalization (QPD) to a signal that is dominated by a single frequency, as obtained for gapped systems after adiabatic state preparation. QPD itself is more generally applicable: it was introduced as a subspace method that estimates multiple energies while suppressing out-of-band noise \cite{stroschein_ground_2025} (see \ref{app:out_of_band}) and, as shown by our comparison in Sec.\,\ref{sec:benchmark_main} offers, among the algorithms considered, a very competitive tradeoff between circuit depth and runtime for the application considered in this work. We further introduced a discretization of the subspace matrix integrals in QPD based on Gauss-Lendre quadrature. Ref.\,\cite{stroschein_ground_2025} used an alternative discretization scheme based on the prolate interpolation formula, and Ref.\,\cite{Castaldo2026ChemicallyAccurate} combines the latter with sparse sampling to reduce the number of required signal samples. A systematic comparison of these strategies will also be left for future work.

The algorithm makes use of Trotterized Hamiltonian simulation with a doubly factorized Hamiltonian, and is consequently well-suited to early fault-tolerant quantum hardware, avoiding the substantial quantum resource cost overhead required by algorithms that rely on block-encoding \cite{lee_even_2021-1,su_fault-tolerant_2021-1}. Qubitization-based approaches to phase estimation must also scale the simulation time with the relative precision, i.e., as $T=O(\lambda/\varepsilon)$ \cite{low_hamiltonian_2019-2, babbush2018encoding, poulin2018quantum}, where $\lambda$ is the induced one-norm of the Hamiltonain, while our Trotterization-based approach only requires the simulation time to be scaled with the absolute precision in the Heisenberg-limited regime, as $T=O(1/\varepsilon)$.

Trotterized time evolution also provides scope to further improve both the asymptotic complexity and the prefactors in front of the gate count expressions we have derived in this work. To simplify the analysis we upper-bounded the nested commutator sums that arise in the Trotter error expressions for deterministic product formulae in terms of $\lambda$, which is well known to be highly conservative in practice \cite{childs_theory_2021,poulin_trotter_2015,motta_low_2021-1}. Throughout the algorithm, the state prepared on the system register remains in the low-energy subspace of either the interpolated Hamiltonian $H^{(s)}$ or the full Hamiltonian $H^{(1)}$, and recent work has demonstrated further reductions in the Trotter error for time evolution under this constraint \cite{sahinoglu_hamiltonian_2021}. 

A complementary perspective is that, in weakly correlated systems, evaluating the nested commutators in a localized orbital basis is likely to result in significant cancellations due to the spatial decay of interactions in the Hamiltonian. The dominant long-range interactions in weakly correlated protein systems in the ground state are often van der Waals interactions, which decay as the sixth power in the distance between localized orbital centroids. Orbital localization and multipole cancellations have also recently been employed in Ref. \cite{gunther_sublinear-depth_2026} to reduce the gate count for a single Trotter step to $\text{polylog}(n)$, and the incorporation of these or similar techniques is left as an important topic for future work.

In contrast to our rigorous $\text{poly}(n,\varepsilon^{-1})$ asymptotic bound on the end-to-end quantum operations cost for ground state energy estimation, a formal analysis characterizing the computational complexity of the classical coupled-cluster hierarchy and establishing bounds of the form $O(n^{\alpha}\varepsilon^{-\beta})$ is still lacking. 
Although rigorous error bounds for the coupled-cluster equations (and their variants) have been derived in recent years\,\cite{rohwedder2013error,laestadius2018analysis,faulstich2019analysis,hassan2023analysis}, how these errors scale with system size remains an open question, precluding a direct comparison with our results. Nonetheless, the steep increase in computational cost with the level of the coupled-cluster hierarchy positions our results to challenge traditional coupled-cluster methods. While approximate solutions can often be obtained with low-order time complexity in the number of orbitals, $n$ (as low as $O(n)$ in some cases \cite{riplinger_efficient_2013}), often high-accuracy solutions (e.g., as required in thermochemistry benchmarks\,\cite{karton2006w4} or to describe large non-covalent $\pi$ interactions\,\cite{schafer2025understanding} or in near metallic systems\,\cite{neufeld2023highly}) may require higher-order polynomial scalings. For example, CCSDT scales as $O(n^8)$, while CCSDTQ scales as $O(n^{10})$. In future work, we will seek to address this gap in the runtime analysis of classical coupled-cluster methods to provide a systematic framework for comparing quantum and classical methods on an equal footing.

In conclusion, we have presented an end-to-end quantum algorithm for the efficient estimation of ground-state energies of weakly correlated molecular systems based on Trotterized time evolution. The algorithm provides error guarantees and achieves an end-to-end quantum operations complexity that scales below $O(n^5)$ in the limit of high-order product formulas. We emphasize that this scaling can be further improved by incorporating recent advances in Hamiltonian simulation for molecular systems\,\cite{gunther_sublinear-depth_2026, casares2026theory}. Furthermore, QPD can be made more robust to errors in the initial state as recently demonstrated in Ref.\,\cite{stroschein_ground_2025}, thus reducing the cost of state preparation, and there is also scope to further reduce the sampling complexity\,\cite{Castaldo2026ChemicallyAccurate} and the quantum error correction overhead\,\cite{low2026denser, chen2026efficient}. This work thereby establishes high-accuracy ground state energy estimation for weakly correlated molecules as a genuine use-case for quantum computers.

\section{Author contributions}

O.L. devised the end-to-end quantum algorithm, and proved the propositions and theorems in the main text. M.M. and T.S. checked the theorems and propositions and suggested improvements. O.L., D.C. and T.S. collaborated on the analysis of the asymptotic scaling properties of QPD. O.L. and M.K. performed the analysis for the double factorized Hamiltonian. D.C. performed numerical benchmarks on the comparison of QPD and other phase estimation algorithms. M.K. performed numerical experiments on the double factorized Hamiltonians, and performed the quantum circuit compilation at the level of logical qubits. R.T., A.S. and D.C. performed numerical experiments on the constant spectral gap assumption. S.P. performed numerical experiments for the mapping from quantum circuits at the logical qubit level to quantum circuits at the physical qubit level. O.L., M.M., R.T., D.C., M.K., T.S., and S.P. wrote the first version of the manuscript. M.C., G.C.S., and M.R. planned and supervised the project and acquired the financial funding and compute. All authors reviewed the manuscript.

\section*{Acknowledgments}
The authors gratefully acknowledge financial support from XPRIZE through a Phase-2 finalist award.

M.R. gratefully acknowledges generous financial support by the Swiss National Science Foundation (grant no. 200021\_219616). 
M.C., M.M., M.R., O.L., R.T., D.C., M.K. acknowledge support from the Novo Nordisk Foundation (grant no. NNF20OC0059939 ‘Quantum for Life’). 
O.L. and R.T. acknowledge support from the Novo Nordisk Foundation, (Grant No. NNF25OC0105181 `Molecular Recognition from Quantum Computing').
S.P. and G.C.S. acknowledge support from the Novo Nordisk Foundation, Grant number NNF22SA0081175, NNF Quantum Computing Programme and Innovation Fund Denmark (IFD) under File no. 4356-00009B.
D.C. acknowledges support by Grant No. 233975 (SPEED-Q Swiss National Postdoctoral Fellowship), %
M.R. gratefully acknowledges financial support from the Swiss National Science Foundation through Grant No. NF 200021\_219616.

\section*{Statement on LLM use}
LLMs were not used to design the study,
to formulate its research questions,
or to write any part of the manuscript. The authors used LLMs in conversational mode
to survey the relevant literature
and to help devise the propositions, theorems and lemmas,
in particular Lemma~\ref{lem:1d_quadrature_bound} in ~\ref{app:prolate_quadrature}.
All references were checked against their original sources.
The authors declare that the final statements and proofs are their own original work.

\bibliography{refs}

@article{boixo_necessary_2010,
  title = {Necessary Condition for the Quantum Adiabatic Approximation},
  author = {Boixo, S. and Somma, R. D.},
  year = 2010,
  month = mar,
  journal = {Physical Review A},
  volume = {81},
  number = {3},
  pages = {032308},
  publisher = {American Physical Society},
  doi = {10.1103/PhysRevA.81.032308},
  urldate = {2026-08-19}
}

@article{cruz2020optimizing,
  title={Optimizing quantum phase estimation for the simulation of Hamiltonian eigenstates},
  author={Cruz, Pedro MQ and Catarina, Gon{\c{c}}alo and Gautier, Ronan and Fern{\'a}ndez-Rossier, Joaqu{\i}n},
  journal={Quantum Science \& Technology},
  volume={5},
  number={4},
  pages={044005},
  year={2020},
  publisher={IOP Publishing}
}

@article{castaldo2024differentiable,
  title={A differentiable quantum phase estimation algorithm},
  author={Castaldo, Davide and Jahangiri, Soran and Migliore, Agostino and Miguel Arrazola, Juan and Corni, Stefano},
  journal={Quantum Science and Technology},
  volume={9},
  number={4},
  pages={045026},
  year={2024},
  publisher={IOP Publishing}
}

@inproceedings{boley1997general,
  title={A general Vandermonde factorization of a Hankel matrix},
  author={Boley, DL and Luk, FT and Vandevoorde, David},
  booktitle={Int’l Lin. Alg. Soc.(ILAS) Symp. on Fast Algorithms for Control, Signals and Image Processing},
  year={1997}
}

@article{kitaev1997quantum,
  title={Quantum computations: algorithms and error correction},
  author={Kitaev, A Yu},
  journal={Russian Mathematical Surveys},
  volume={52},
  number={6},
  pages={1191--1249},
  year={1997}
}

@article{ding2026quantum,
  title={Quantum Filtering and Analysis of Multiplicities in Eigenvalue Spectra},
  author={Ding, Zhiyan and Lin, Lin and Yang, Yilun and Zhang, Ruizhe},
  journal={PRX Quantum},
  volume={7},
  number={2},
  pages={020318},
  year={2026},
  publisher={APS}
}

@article{aharonov2025importance,
  title={On the importance of error mitigation for quantum computation},
  author={Aharonov, Dorit and Alberton, Ori and Arad, Itai and Atia, Yosi and Bairey, Eyal and Brakerski, Zvika and Cohen, Itsik and Golan, Omri and Gurwich, Ilya and Kenneth, Oded and others},
  journal={arXiv preprint arXiv:2503.17243},
  year={2025}
}

@article{sivak2023real,
  title={Real-time quantum error correction beyond break-even},
  author={Sivak, Volodymyr V and Eickbusch, Alec and Royer, Baptiste and Singh, Shraddha and Tsioutsios, Ioannis and Ganjam, Suhas and Miano, Alessandro and Brock, Benjamin L and Ding, Andy Z and Frunzio, Luigi and others},
  journal={Nature},
  volume={616},
  number={7955},
  pages={50--55},
  year={2023},
  publisher={Nature Publishing Group UK London}
}

@article{wang2026demonstration,
  title={Demonstration of low-overhead quantum error correction codes},
  author={Wang, Ke and Lu, Zhide and Zhang, Chuanyu and Liu, Gongyu and Chen, Jiachen and Wang, Yanzhe and Wu, Yaozu and Xu, Shibo and Zhu, Xuhao and Jin, Feitong and others},
  journal={Nature Physics},
  volume={22},
  number={2},
  pages={308--314},
  year={2026},
  publisher={Nature Publishing Group UK London}
}

@article{hassan2023analysis,
  title={Analysis of the single reference coupled cluster method for electronic structure calculations: the full-coupled cluster equations},
  author={Hassan, Muhammad and Maday, Yvon and Wang, Yipeng},
  journal={Numerische Mathematik},
  volume={155},
  number={1},
  pages={121--173},
  year={2023},
  publisher={Springer}
}

@article{laestadius2018analysis,
  title={Analysis of the extended coupled-cluster method in quantum chemistry},
  author={Laestadius, Andre and Kvaal, Simen},
  journal={SIAM Journal on Numerical Analysis},
  volume={56},
  number={2},
  pages={660--683},
  year={2018},
  publisher={SIAM}
}

@unpublished{Castaldo2026ChemicallyAccurate,
  title  = {Chemically accurate molecular energies from sparse
            quantum prolate diagonalization on a trapped-ion
            quantum computer},
  author = {Castaldo, Davide and
            Kronenberger, Marvin and
            Miller, Marek and
            Forget, S{\'e}l{\`e}ne and
            Solomon, Gemma C. and
            Lindorff-Larsen, Kresten and
            Christandl, Matthias and
            Reiher, Markus},
  year   = {2026},
  note   = {Manuscript in preparation}
}

@article{karton2006w4,
  title={W4 theory for computational thermochemistry: In pursuit of confident sub-kJ/mol predictions},
  author={Karton, Amir and Rabinovich, Elena and Martin, Jan ML and Ruscic, Branko},
  journal={The Journal of chemical physics},
  volume={125},
  number={14},
  year={2006},
  publisher={AIP Publishing}
}

@article{schafer2025understanding,
  title={Understanding discrepancies in noncovalent interaction energies from wavefunction theories for large molecules},
  author={Sch{\"a}fer, Tobias and Irmler, Andreas and Gallo, Alejandro and Gr{\"u}neis, Andreas},
  journal={Nature Communications},
  volume={16},
  number={1},
  pages={9108},
  year={2025},
  publisher={Nature Publishing Group UK, London}
}

@article{neufeld2023highly,
  title={Highly accurate electronic structure of metallic solids from coupled-cluster theory with nonperturbative triple excitations},
  author={Neufeld, Verena A and Berkelbach, Timothy C},
  journal={Physical Review Letters},
  volume={131},
  number={18},
  pages={186402},
  year={2023},
  publisher={APS}
}

@article{faulstich2019analysis,
  title={Analysis of the tailored coupled-cluster method in quantum chemistry},
  author={Faulstich, Fabian M and Laestadius, Andre and Legeza, Ors and Schneider, Reinhold and Kvaal, Simen},
  journal={SIAM Journal on Numerical Analysis},
  volume={57},
  number={6},
  pages={2579--2607},
  year={2019},
  publisher={SIAM}
}

@article{rohwedder2013error,
  title={Error estimates for the Coupled Cluster method∗},
  author={Rohwedder, Thorsten and Schneider, Reinhold},
  journal={ESAIM: Mathematical Modelling and Numerical Analysis},
  volume={47},
  number={6},
  pages={1553--1582},
  year={2013},
  publisher={EDP Sciences}
}

@article{yi2024quantum,
  title={Quantum phase estimation by compressed sensing},
  author={Yi, Changhao and Zhou, Cunlu and Takahashi, Jun},
  journal={Quantum},
  volume={8},
  pages={1579},
  year={2024},
  publisher={Verein zur F{\"o}rderung des Open Access Publizierens in den Quantenwissenschaften}
}

@article{li2023adaptive,
  title={Adaptive low-depth quantum algorithms for robust multiple-phase estimation},
  author={Li, Haoya and Ni, Hongkang and Ying, Lexing},
  journal={Physical Review A},
  volume={108},
  number={6},
  pages={062408},
  year={2023},
  publisher={APS}
}

@inproceedings{moitra2015super,
  title={Super-resolution, extremal functions and the condition number of Vandermonde matrices},
  author={Moitra, Ankur},
  booktitle={Proceedings of the forty-seventh annual ACM symposium on Theory of computing},
  pages={821--830},
  year={2015}
}

@article{casares2026theory,
  title={Theory and practice of Trotter product formulas for quantum chemistry},
  author={Casares, Pablo AM and Maxwell, William and Motlagh, Danial and Choubisa, Hitarth and Niu, Zy and Loaiza, Ignacio and Mueller, Jonathan E and Voigt, Arne-Christian and Arrazola, Juan Miguel and Fomichev, Stepan},
  journal={arXiv preprint arXiv:2606.30741},
  year={2026}
}

@article{low2026denser,
  title={A Denser Planar Surface Code},
  author={Low, Guang Hao and Huggins, William J and Berry, Dominic W and Khattar, Tanuj and White, Alec F and Rubin, Nicholas C and Babbush, Ryan},
  journal={arXiv preprint arXiv:2605.30455},
  year={2026}
}

@article{chen2026efficient,
  title={Efficient magic state cultivation on rp 2},
  author={Chen, Zi-Han and Chen, Ming-Cheng and Lu, Chao-Yang and Pan, Jian-Wei},
  journal={PRX Quantum},
  volume={7},
  number={1},
  pages={010315},
  year={2026},
  publisher={APS}
}

@article{dutkiewicz2022heisenberg,
  title={Heisenberg-limited quantum phase estimation of multiple eigenvalues with few control qubits},
  author={Dutkiewicz, Alicja and Terhal, Barbara M and O'Brien, Thomas E},
  journal={Quantum},
  volume={6},
  pages={830},
  year={2022},
  publisher={Verein zur F{\"o}rderung des Open Access Publizierens in den Quantenwissenschaften}
}

@article{childs_theory_2021,
  title = {A {{Theory}} of {{Trotter Error}}},
  author = {Childs, Andrew M. and Su, Yuan and Tran, Minh C. and Wiebe, Nathan and Zhu, Shuchen},
  year = 2021,
  month = feb,
  journal = {Physical Review X},
  volume = {11},
  number = {1},
  primaryclass = {quant-ph},
  pages = {011020},
  issn = {2160-3308},
  doi = {10.1103/PhysRevX.11.011020},
  urldate = {2026-08-07},
  archiveprefix = {arXiv}
}

@article{cortes_quantum_2022,
  title = {Quantum {{Krylov}} Subspace Algorithms for Ground- and Excited-State Energy Estimation},
  author = {Cortes, Cristian L. and Gray, Stephen K.},
  year = 2022,
  month = feb,
  journal = {Physical Review A},
  volume = {105},
  number = {2},
  pages = {022417},
  publisher = {American Physical Society},
  doi = {10.1103/PhysRevA.105.022417},
  urldate = {2026-08-19}
}

@article{zhan2026rapid,
  title={Rapid quantum ground state preparation via dissipative dynamics},
  author={Zhan, Yongtao and Ding, Zhiyan and Huhn, Jakob and Gray, Johnnie and Preskill, John and Chan, Garnet Kin-Lic and Lin, Lin},
  journal={Physical Review X},
  volume={16},
  number={1},
  pages={011004},
  year={2026},
  publisher={APS}
}

@article{hastings_spectral_2006,
  title = {Spectral {{Gap}} and {{Exponential Decay}} of {{Correlations}}},
  author = {Hastings, Matthew B. and Koma, Tohru},
  year = 2006,
  month = aug,
  journal = {Communications in Mathematical Physics},
  volume = {265},
  number = {3},
  pages = {781--804},
  issn = {1432-0916},
  doi = {10.1007/s00220-006-0030-4},
  urldate = {2026-08-19},
  langid = {english}
}

@article{jansen_bounds_2007,
  title = {Bounds for the Adiabatic Approximation with Applications to Quantum Computation},
  author = {Jansen, Sabine and Ruskai, Mary-Beth and Seiler, Ruedi},
  year = 2007,
  month = oct,
  journal = {Journal of Mathematical Physics},
  volume = {48},
  number = {10},
  pages = {102111},
  issn = {0022-2488},
  doi = {10.1063/1.2798382},
  urldate = {2026-08-19}
}

@article{patel2026optimal,
  title={Optimal coherent quantum phase estimation via tapering},
  author={Patel, Dhrumil and Tan, Shi Jie Samuel and Suba{\c{s}}{\i}, Yi{\u{g}}it and Sornborger, Andrew T},
  journal={PRX Quantum},
  volume={7},
  number={2},
  pages={020302},
  year={2026},
  publisher={APS}
}

@article{loaiza2025nonlinear,
  title={Nonlinear spectroscopy via generalized quantum phase estimation},
  author={Loaiza, Ignacio and Motlagh, Danial and Hejazi, Kasra and Zini, Modjtaba Shokrian and Delgado, Alain and Arrazola, Juan Miguel},
  journal={Quantum},
  volume={9},
  pages={1822},
  year={2025},
  publisher={Verein zur F{\"o}rderung des Open Access Publizierens in den Quantenwissenschaften}
}

@article{rendon2022effects,
  title={Effects of cosine tapering window on quantum phase estimation},
  author={Rendon, Gumaro and Izubuchi, Taku and Kikuchi, Yuta},
  journal={Physical Review D},
  volume={106},
  number={3},
  pages={034503},
  year={2022},
  publisher={APS}
}

@article{ding2023simultaneous,
  title={Simultaneous estimation of multiple eigenvalues with short-depth quantum circuit on early fault-tolerant quantum computers},
  author={Ding, Zhiyan and Lin, Lin},
  journal={Quantum},
  volume={7},
  pages={1136},
  year={2023},
  publisher={Verein zur F{\"o}rderung des Open Access Publizierens in den Quantenwissenschaften}
}

@article{mezic2004comparison,
  title={Comparison of systems with complex behavior},
  author={Mezi{\'c}, Igor and Banaszuk, Andrzej},
  journal={Physica D: Nonlinear Phenomena},
  volume={197},
  number={1-2},
  pages={101--133},
  year={2004},
  publisher={Elsevier}
}

@article{schmid2010dynamic,
  title={Dynamic mode decomposition of numerical and experimental data},
  author={Schmid, Peter J},
  journal={Journal of fluid mechanics},
  volume={656},
  pages={5--28},
  year={2010},
  publisher={Cambridge University Press}
}

@article{shen2025estimating,
  title={Estimating eigenenergies from quantum dynamics: A unified noise-resilient measurement-driven approach},
  author={Shen, Yizhi and Camps, Daan and Szasz, Aaron and Darbha, Siva and Klymko, Katherine and Williams, David B and Tubman, Norm M and Van Beeumen, Roel and others},
  journal={Quantum},
  volume={9},
  pages={1836},
  year={2025},
  publisher={Verein zur F{\"o}rderung des Open Access Publizierens in den Quantenwissenschaften}
}

@article{klymko2022real,
  title={Real-time evolution for ultracompact Hamiltonian eigenstates on quantum hardware},
  author={Klymko, Katherine and Mejuto-Zaera, Carlos and Cotton, Stephen J and Wudarski, Filip and Urbanek, Miroslav and Hait, Diptarka and Head-Gordon, Martin and Whaley, K Birgitta and Moussa, Jonathan and Wiebe, Nathan and others},
  journal={PRX Quantum},
  volume={3},
  number={2},
  pages={020323},
  year={2022},
  publisher={APS}
}

@article{ding2024quantum,
  title={Quantum Multiple Eigenvalue Gaussian filtered Search: an efficient and versatile quantum phase estimation method},
  author={Ding, Zhiyan and Li, Haoya and Lin, Lin and Ni, HongKang and Ying, Lexing and Zhang, Ruizhe},
  journal={Quantum},
  volume={8},
  pages={1487},
  year={2024},
  publisher={Verein zur F{\"o}rderung des Open Access Publizierens in den Quantenwissenschaften}
}

@misc{kimura_fisher_2026,
  title = {Fisher Information Based Lower Bounds on the Cost of Quantum Phase Estimation},
  author = {Kimura, Ryosuke and Mitarai, Kosuke},
  year = 2026,
  month = mar,
  number = {arXiv:2603.12706},
  primaryclass = {quant-ph},
  publisher = {arXiv},
  doi = {10.48550/arXiv.2603.12706},
  urldate = {2026-08-21},
  archiveprefix = {arXiv}
}

@article{kivlichan_quantum_2018,
  title = {Quantum {{Simulation}} of {{Electronic Structure}} with {{Linear Depth}} and {{Connectivity}}},
  author = {Kivlichan, Ian D. and McClean, Jarrod and Wiebe, Nathan and Gidney, Craig and {Aspuru-Guzik}, Al{\'a}n and Chan, Garnet Kin-Lic and Babbush, Ryan},
  year = 2018,
  month = mar,
  journal = {Physical Review Letters},
  volume = {120},
  number = {11},
  pages = {110501},
  publisher = {American Physical Society},
  doi = {10.1103/PhysRevLett.120.110501},
  urldate = {2024-12-06}
}

@article{lidar_adiabatic_2009,
  title = {Adiabatic Approximation with Exponential Accuracy for Many-Body Systems and Quantum Computation},
  author = {Lidar, D. A. and Rezakhani, A. T. and Hamma, A.},
  year = 2009,
  month = oct,
  journal = {Journal of Mathematical Physics},
  volume = {50},
  number = {10},
  primaryclass = {quant-ph},
  pages = {102106},
  issn = {0022-2488, 1089-7658},
  doi = {10.1063/1.3236685},
  urldate = {2026-08-19},
  archiveprefix = {arXiv}
}

@article{lloyd_universal_1996,
  title = {Universal {{Quantum Simulators}}},
  author = {Lloyd, Seth},
  year = 1996,
  month = aug,
  journal = {Science},
  volume = {273},
  number = {5278},
  pages = {1073--1078},
  publisher = {American Association for the Advancement of Science},
  doi = {10.1126/science.273.5278.1073},
  urldate = {2026-08-19}
}

@article{motta_low_2021-1,
  title = {Low Rank Representations for Quantum Simulation of Electronic Structure},
  author = {Motta, Mario and Ye, Erika and McClean, Jarrod R. and Li, Zhendong and Minnich, Austin J. and Babbush, Ryan and Chan, Garnet Kin-Lic},
  year = 2021,
  month = may,
  journal = {npj Quantum Information},
  volume = {7},
  number = {1},
  pages = {83},
  publisher = {Nature Publishing Group},
  issn = {2056-6387},
  doi = {10.1038/s41534-021-00416-z},
  urldate = {2026-08-07},
  copyright = {2021 The Author(s)},
  langid = {english}
}

@misc{nielsen_quantum_2010,
  title = {Quantum {{Computation}} and {{Quantum Information}}: 10th {{Anniversary Edition}}},
  shorttitle = {Quantum {{Computation}} and {{Quantum Information}}},
  author = {Nielsen, Michael A. and Chuang, Isaac L.},
  year = 2010,
  month = dec,
  journal = {Higher Education from Cambridge University Press},
  publisher = {Cambridge University Press},
  doi = {10.1017/CBO9780511976667},
  urldate = {2025-02-27},
  isbn = {9780511976667},
  langid = {english}
}

@misc{stroschein_ground_2025,
  title = {Ground and Excited-State Energies with Analytic Errors and Short Time Evolution on a Quantum Computer},
  author = {Stroschein, Timothy and Castaldo, Davide and Reiher, Markus},
  year = 2025,
  month = jul,
  journal = {arXiv.org},
  doi = {10.1103/zy63-92c4},
  urldate = {2026-08-11},
  langid = {english}
}

@article{su_nearly_2021,
  title = {Nearly Tight {{Trotterization}} of Interacting Electrons},
  author = {Su, Yuan and Huang, Hsin-Yuan and Campbell, Earl T.},
  year = 2021,
  month = jul,
  journal = {Quantum},
  volume = {5},
  primaryclass = {quant-ph},
  pages = {495},
  issn = {2521-327X},
  doi = {10.22331/q-2021-07-05-495},
  urldate = {2026-08-07},
  archiveprefix = {arXiv}
}

@misc{wan_fast_2022,
  title = {Fast Digital Methods for Adiabatic State Preparation},
  author = {Wan, Kianna and Kim, Isaac H.},
  year = 2022,
  month = mar,
  number = {arXiv:2004.04164},
  primaryclass = {quant-ph},
  publisher = {arXiv},
  doi = {10.48550/arXiv.2004.04164},
  urldate = {2026-08-19},
  archiveprefix = {arXiv}
}

@article{slepian_prolate_1961,
  title = {Prolate Spheroidal Wave Functions, Fourier Analysis and Uncertainty — {{I}}},
  author = {Slepian, D. and Pollak, H. O.},
  date = {1961-01},
  journal = {The Bell System Technical Journal},
  volume = {40},
  number = {1},
  pages = {43--63},
  issn = {0005-8580},
  doi = {10.1002/j.1538-7305.1961.tb03976.x},
  url = {https://ieeexplore.ieee.org/document/6773659},
  urldate = {2026-09-07}
}

@article{landau_prolate_1961,
  author  = {Landau, H. J. and Pollak, H. O.},
  journal = {Bell Syst. Tech. J.},
  title = {{P}rolate spheroidal wave functions, fourier analysis and uncertainty — {II}},
  year    = {1961},
  volume  = {40},
  number  = {1},
  pages   = {65-84},
  doi     = {10.1002/j.1538-7305.1961.tb03977.x}
}

@article{landau_prolate_1962,
  author={Landau, H. J. and Pollak, H. O.},
  journal={Bell Syst. Tech. J.}, 
  title = {{P}rolate spheroidal wave functions, fourier analysis and uncertainty — {III}: {T}he dimension of the space of essentially time- and band-limited signals}, 
  year={1962},
  volume={41},
  number={4},
  pages={1295-1336},
  doi={10.1002/j.1538-7305.1962.tb03279.x}}

@article{walter_sampling_2003,
  title = {Sampling {{With Prolate Spheroidal Wave Functions}}},
  author = {Walter, Gilbert G. and Shen, Xiaoping A.},
  date = {2003-01-01},
  journal = {Sampling Theory in Signal and Image Processing},
  shortjournal = {STSIP},
  volume = {2},
  number = {1},
  pages = {25--52},
  issn = {1530-6429},
  doi = {10.1007/BF03549384},
  url = {https://doi.org/10.1007/BF03549384},
  urldate = {2026-09-02},
  langid = {english}
}

@article{lever_electrostatic_2013,
  title = {Electrostatic Considerations Affecting the Calculated {{HOMO}}–{{LUMO}} Gap in Protein Molecules},
  author = {Lever, Greg and Cole, Daniel J and Hine, Nicholas D M and Haynes, Peter D and Payne, Mike C},
  date = {2013-03},
  journal = {Journal of Physics: Condensed Matter},
  shortjournal = {J. Phys.: Condens. Matter},
  volume = {25},
  number = {15},
  pages = {152101},
  publisher = {IOP Publishing},
  issn = {0953-8984},
  doi = {10.1088/0953-8984/25/15/152101},
  url = {https://doi.org/10.1088/0953-8984/25/15/152101},
  urldate = {2026-09-16},
  langid = {english}
}

@article{reiher2017elucidating,
  title={Elucidating reaction mechanisms on quantum computers},
  author={Reiher, Markus and Wiebe, Nathan and Svore, Krysta M and Wecker, Dave and Troyer, Matthias},
  journal={Proceedings of the national academy of sciences},
  volume={114},
  number={29},
  pages={7555--7560},
  year={2017},
  publisher={National Academy of Sciences}
}

@article{kronenberger2026trotter,
  title={Trotter error and orbital transformations in quantum phase estimation},
  author={Kronenberger, Marvin and Erakovic, Mihael and Reiher, Markus},
  journal={Molecular Physics},
  pages={2681062},
  year={2026},
  publisher={Taylor \& Francis}
}

@article{aspuru-guzik_simulated_2005-1,
  title = {Simulated {{Quantum Computation}} of {{Molecular Energies}}},
  author = {Aspuru-Guzik, Alán and Dutoi, Anthony D. and Love, Peter J. and Head-Gordon, Martin},
  date = {2005-09-09},
  journal = {Science},
  volume = {309},
  number = {5741},
  pages = {1704--1707},
  publisher = {American Association for the Advancement of Science},
  doi = {10.1126/science.1113479},
  url = {https://www.science.org/doi/10.1126/science.1113479},
  urldate = {2026-09-18}
}

@article{riplinger_efficient_2013,
  title = {An Efficient and near Linear Scaling Pair Natural Orbital Based Local Coupled Cluster Method},
  author = {Riplinger, Christoph and Neese, Frank},
  date = {2013-01-17},
  journal = {The Journal of Chemical Physics},
  shortjournal = {J. Chem. Phys.},
  volume = {138},
  number = {3},
  pages = {034106},
  issn = {0021-9606},
  doi = {10.1063/1.4773581},
  url = {https://doi.org/10.1063/1.4773581},
  urldate = {2026-09-18}
}

@article{berry_rapid_2025,
  title = {Rapid {{Initial-State Preparation}} for the {{Quantum Simulation}} of {{Strongly Correlated Molecules}}},
  author = {Berry, Dominic W. and Tong, Yu and Khattar, Tanuj and White, Alec and Kim, Tae In and Low, Guang Hao and Boixo, Sergio and Ding, Zhiyan and Lin, Lin and Lee, Seunghoon and Chan, Garnet Kin-Lic and Babbush, Ryan and Rubin, Nicholas C.},
  date = {2025-05-08},
  journal = {PRX Quantum},
  shortjournal = {PRX Quantum},
  volume = {6},
  number = {2},
  pages = {020327},
  publisher = {American Physical Society},
  doi = {10.1103/PRXQuantum.6.020327},
  url = {https://link.aps.org/doi/10.1103/PRXQuantum.6.020327},
  urldate = {2026-09-18}
}

@article{elgart_note_2012,
  title = {A Note on the Switching Adiabatic Theorem},
  author = {Elgart, Alexander and Hagedorn, George A.},
  date = {2012-09-18},
  journal = {Journal of Mathematical Physics},
  shortjournal = {J. Math. Phys.},
  volume = {53},
  number = {10},
  pages = {102202},
  issn = {0022-2488},
  doi = {10.1063/1.4748968},
  url = {https://doi.org/10.1063/1.4748968},
  urldate = {2026-09-18}
}

@article{de_roeck_persistence_2019,
  title = {Persistence of {{Exponential Decay}} and {{Spectral Gaps}} for {{Interacting Fermions}}},
  author = {De Roeck, Wojciech and Salmhofer, Manfred},
  date = {2019-01-01},
  journal = {Communications in Mathematical Physics},
  shortjournal = {Commun. Math. Phys.},
  volume = {365},
  number = {2},
  pages = {773--796},
  issn = {1432-0916},
  doi = {10.1007/s00220-018-3211-z},
  url = {https://doi.org/10.1007/s00220-018-3211-z},
  urldate = {2026-09-18},
  langid = {english}
}

@article{koridon_orbital_2021,
  title = {Orbital Transformations to Reduce the 1-Norm of the Electronic Structure {{Hamiltonian}} for Quantum Computing Applications},
  author = {Koridon, Emiel and Yalouz, Saad and Senjean, Bruno and Buda, Francesco and O'Brien, Thomas E. and Visscher, Lucas},
  date = {2021-08-06},
  journal = {Physical Review Research},
  shortjournal = {Phys. Rev. Res.},
  volume = {3},
  number = {3},
  pages = {033127},
  publisher = {American Physical Society},
  doi = {10.1103/PhysRevResearch.3.033127},
  url = {https://link.aps.org/doi/10.1103/PhysRevResearch.3.033127},
  urldate = {2026-09-18}
}

@article{ding_even_2023,
  title = {Even {{Shorter Quantum Circuit}} for {{Phase Estimation}} on {{Early Fault-Tolerant Quantum Computers}} with {{Applications}} to {{Ground-State Energy Estimation}}},
  author = {Ding, Zhiyan and Lin, Lin},
  date = {2023-05-26},
  journal = {PRX Quantum},
  shortjournal = {PRX Quantum},
  volume = {4},
  number = {2},
  pages = {020331},
  publisher = {American Physical Society},
  doi = {10.1103/PRXQuantum.4.020331},
  url = {https://link.aps.org/doi/10.1103/PRXQuantum.4.020331},
  urldate = {2026-09-18}
}

@article{su_fault-tolerant_2021-1,
  title = {Fault-{{Tolerant Quantum Simulations}} of {{Chemistry}} in {{First Quantization}}},
  author = {Su, Yuan and Berry, Dominic W. and Wiebe, Nathan and Rubin, Nicholas and Babbush, Ryan},
  date = {2021-11-11},
  journal = {PRX Quantum},
  shortjournal = {PRX Quantum},
  volume = {2},
  number = {4},
  pages = {040332},
  publisher = {American Physical Society},
  doi = {10.1103/PRXQuantum.2.040332},
  url = {https://link.aps.org/doi/10.1103/PRXQuantum.2.040332},
  urldate = {2026-09-18}
}

@article{mizuta_theory_2026,
  title = {Theory of {{Trotter Errors}} for {{Time-Dependent Product}} and {{Multi-product Formulas}}},
  author = {Mizuta, Kaoru and Ikeda, Tatsuhiko N. and Fujii, Keisuke},
  date = {2026-09-21},
  journal = {PRX Quantum},
  shortjournal = {PRX Quantum},
  volume = {7},
  number = {3},
  pages = {033061},
  publisher = {American Physical Society},
  doi = {10.1103/ywv1-7fhf},
  url = {https://link.aps.org/doi/10.1103/ywv1-7fhf},
  urldate = {2026-09-24}
}

@article{chen_framework_2026,
  title = {Framework for {{Robust Quantum Speedups}} in {{Practical Correlated Electronic Structure}} and {{Dynamics}}},
  author = {Chen, Jielun and Chan, Garnet Kin-Lic},
  date = {2026-09-21},
  journal = {Physical Review Letters},
  shortjournal = {Phys. Rev. Lett.},
  volume = {137},
  number = {13},
  pages = {130601},
  publisher = {American Physical Society},
  doi = {10.1103/v2ms-wmz1},
  url = {https://link.aps.org/doi/10.1103/v2ms-wmz1},
  urldate = {2026-09-28}
}

@online{kitaev_quantum_1995,
  title = {Quantum Measurements and the {{Abelian Stabilizer Problem}}},
  author = {Kitaev, A. Yu},
  date = {1995-11-20},
  number = {arXiv:quant-ph/9511026},
  eprinttype = {arXiv},
  doi = {10.48550/arXiv.quant-ph/9511026},
  url = {http://arxiv.org/abs/quant-ph/9511026},
  urldate = {2026-09-28},
  pubstate = {prepublished}
}

@article{lee_even_2021-1,
  title = {Even {{More Efficient Quantum Computations}} of {{Chemistry Through Tensor Hypercontraction}}},
  author = {Lee, Joonho and Berry, Dominic W. and Gidney, Craig and Huggins, William J. and McClean, Jarrod R. and Wiebe, Nathan and Babbush, Ryan},
  date = {2021-07-08},
  journal = {PRX Quantum},
  shortjournal = {PRX Quantum},
  volume = {2},
  number = {3},
  pages = {030305},
  publisher = {American Physical Society},
  doi = {10.1103/PRXQuantum.2.030305},
  url = {https://link.aps.org/doi/10.1103/PRXQuantum.2.030305},
  urldate = {2026-09-28}
}

@article{von_burg_quantum_2021,
  title = {Quantum Computing Enhanced Computational Catalysis},
  author = {von Burg, Vera and Low, Guang Hao and Häner, Thomas and Steiger, Damian S. and Reiher, Markus and Roetteler, Martin and Troyer, Matthias},
  date = {2021-07-16},
  journal = {Physical Review Research},
  shortjournal = {Phys. Rev. Res.},
  volume = {3},
  number = {3},
  pages = {033055},
  publisher = {American Physical Society},
  doi = {10.1103/PhysRevResearch.3.033055},
  url = {https://link.aps.org/doi/10.1103/PhysRevResearch.3.033055},
  urldate = {2026-09-28}
}

@online{parrish_quantum_2019,
  title = {Quantum {{Filter Diagonalization}}: {{Quantum Eigendecomposition}} without {{Full Quantum Phase Estimation}}},
  shorttitle = {Quantum {{Filter Diagonalization}}},
  author = {Parrish, Robert M. and McMahon, Peter L.},
  date = {2019-09-19},
  number = {arXiv:1909.08925},
  eprinttype = {arXiv},
  eprintclass = {quant-ph},
  doi = {10.48550/arXiv.1909.08925},
  url = {http://arxiv.org/abs/1909.08925},
  urldate = {2026-09-28},
  pubstate = {prepublished}
}

@article{lee_evaluating_2023-2,
  title = {Evaluating the Evidence for Exponential Quantum Advantage in Ground-State Quantum Chemistry},
  author = {Lee, Seunghoon and Lee, Joonho and Zhai, Huanchen and Tong, Yu and Dalzell, Alexander M. and Kumar, Ashutosh and Helms, Phillip and Gray, Johnnie and Cui, Zhi-Hao and Liu, Wenyuan and Kastoryano, Michael and Babbush, Ryan and Preskill, John and Reichman, David R. and Campbell, Earl T. and Valeev, Edward F. and Lin, Lin and Chan, Garnet Kin-Lic},
  date = {2023-04-07},
  journal = {Nature Communications},
  shortjournal = {Nat Commun},
  volume = {14},
  number = {1},
  pages = {1952},
  publisher = {Nature Publishing Group},
  issn = {2041-1723},
  doi = {10.1038/s41467-023-37587-6},
  url = {https://www.nature.com/articles/s41467-023-37587-6},
  urldate = {2026-09-28},
  langid = {english}
}

@article{martinez2023assessment,
  title={Assessment of various Hamiltonian partitionings for the electronic structure problem on a quantum computer using the Trotter approximation},
  author={Mart{\'\i}nez-Mart{\'\i}nez, Luis A and Yen, Tzu-Ching and Izmaylov, Artur F},
  journal={Quantum},
  volume={7},
  pages={1086},
  year={2023},
  publisher={Verein zur F{\"o}rderung des Open Access Publizierens in den Quantenwissenschaften}
}

@article{mehendale2025estimating,
  title={Estimating Trotter approximation errors to optimize Hamiltonian partitioning for lower eigenvalue errors},
  author={Mehendale, Shashank G and Mart{\'\i}nez-Mart{\'\i}nez, Luis A and Kamath, Prathami Divakar and Izmaylov, Artur F},
  journal={Digit. Discov.},
  volume={4},
  number={12},
  pages={3540--3551},
  year={2025},
  publisher={Royal Society of Chemistry}
}

@article{liakos2020comprehensive,
  title={Comprehensive benchmark results for the domain based local pair natural orbital coupled cluster method (DLPNO-CCSD (T)) for closed-and open-shell systems},
  author={Liakos, Dimitrios G and Guo, Yang and Neese, Frank},
  journal={The Journal of Physical Chemistry A},
  volume={124},
  number={1},
  pages={90--100},
  year={2020},
  publisher={ACS Publications}
}

@inbook{doi:10.1137/1.9781611975949.ch19,

title = {Chapter 19. Clenshaw—Curtis and Gauss Quadrature},
booktitle = {Approximation Theory and Approximation Practice, Extended Edition},
author = {Trefethen, Lloyd N},
publisher = {SIAM},
year = {2018},
chapter = {19},
pages = {143-153},
doi = {10.1137/1.9781611975949.ch19},
URL = {https://epubs.siam.org/doi/abs/10.1137/1.9781611975949.ch19},
eprint = {https://epubs.siam.org/doi/pdf/10.1137/1.9781611975949.ch19}
}

@book{wiener_fourier_1934,
  title = {Fourier {{Transforms}} in the {{Complex Domain}}},
  author = {Wiener, N. and Paley, R.},
  date = {1934-12-31},
  series = {Colloquium {{Publications}}},
  volume = {19},
  publisher = {American Mathematical Society},
  location = {Providence, Rhode Island},
  doi = {10.1090/coll/019},
  url = {https://www.ams.org/coll/019},
  urldate = {2026-10-06},
  isbn = {978-0-8218-1019-4 978-0-8218-7456-1 978-1-4704-3167-9},
  langid = {english}
}

@article{sung2026ffsim,
  title={ffsim: Faster simulation of fermionic quantum circuits},
  author={Sung, Kevin J and Choi, Inho and Amico, Mirko and Andrews, Bartholomew and Ayantuna, Esra and Kawashima, Yukio and Lin, Wan-Hsuan and Omanovic, David and Piccinelli, Samuele and Moreno, Javier Robledo and others},
  journal={arXiv preprint arXiv:2605.03123},
  year={2026}
}

@article{sun2018pyscf,
  title={PySCF: the Python-based simulations of chemistry framework},
  author={Sun, Qiming and Berkelbach, Timothy C and Blunt, Nick S and Booth, George H and Guo, Sheng and Li, Zhendong and Liu, Junzi and McClain, James D and Sayfutyarova, Elvira R and Sharma, Sandeep and others},
  journal={Wiley Interdiscip. Rev. Comput. Mol. Sci.},
  volume={8},
  number={1},
  pages={e1340},
  year={2018},
  publisher={Wiley Online Library}
}

@online{tao_random_2010,
  title = {Random Matrices: {{Universality}} of Local Eigenvalue Statistics},
  shorttitle = {Random Matrices},
  author = {Tao, Terence and Vu, Van},
  date = {2010-06-29},
  number = {arXiv:0906.0510},
  eprinttype = {arXiv},
  eprintclass = {math.PR},
  doi = {10.48550/arXiv.0906.0510},
  url = {http://arxiv.org/abs/0906.0510},
  urldate = {2026-10-13},
  pubstate = {prepublished}
}

@article{kohn_nobel_1999,
  title = {Nobel {{Lecture}}: {{Electronic}} Structure of Matter---Wave Functions and Density Functionals},
  shorttitle = {Nobel {{Lecture}}},
  author = {Kohn, W.},
  date = {1999-10-01},
  journal = {Reviews of Modern Physics},
  shortjournal = {Rev. Mod. Phys.},
  volume = {71},
  number = {5},
  pages = {1253--1266},
  publisher = {American Physical Society},
  doi = {10.1103/RevModPhys.71.1253},
  url = {https://link.aps.org/doi/10.1103/RevModPhys.71.1253},
  urldate = {2026-10-14}
}

@online{tubman_postponing_2018,
  title = {Postponing the Orthogonality Catastrophe: Efficient State Preparation for Electronic Structure Simulations on Quantum Devices},
  shorttitle = {Postponing the Orthogonality Catastrophe},
  author = {Tubman, Norm M. and Mejuto-Zaera, Carlos and Epstein, Jeffrey M. and Hait, Diptarka and Levine, Daniel S. and Huggins, William and Jiang, Zhang and McClean, Jarrod R. and Babbush, Ryan and Head-Gordon, Martin and Whaley, K. Birgitta},
  date = {2018-09-14},
  number = {arXiv:1809.05523},
  eprinttype = {arXiv},
  eprintclass = {quant-ph},
  doi = {10.48550/arXiv.1809.05523},
  url = {http://arxiv.org/abs/1809.05523},
  urldate = {2026-10-14},
  pubstate = {prepublished}
}

@misc{OAI:QMA-hardness-of-continuum-Coulomb-energy-with-unit-nuclear-charges-September-24-2026,
  author = {{OpenAI}},
  title = {{QMA-hardness of continuum Coulomb energy with unit nuclear charges}},
  howpublished = {OpenAI Math Release preprint
                  \href{https://github.com/openai/math/blob/main/preprints/QMA-hardness-of-continuum-Coulomb-energy-with-unit-nuclear-charges-September-24-2026/QMA-hardness-of-continuum-Coulomb-energy-with-unit-nuclear-charges-September-24-2026.pdf}{OAI:QMA-hardness-of-continuum-Coulomb-energy-with-unit-nuclear-charges-September-24-2026}},
  year = {2026}
}

@article{rocca2024reducing,
  title={Reducing the runtime of fault-tolerant quantum simulations in chemistry through symmetry-compressed double factorization},
  author={Rocca, Dario and Cortes, Cristian L and Gonthier, J{\'e}r{\^o}me F and Ollitrault, Pauline J and Parrish, Robert M and Anselmetti, Gian-Luca and Degroote, Matthias and Moll, Nikolaj and Santagati, Raffaele and Streif, Michael},
  journal={J. Chem. Theory Comput.},
  volume={20},
  number={11},
  pages={4639},
  year={2024}
}

@article{ollitrault2025improving,
  title={Improving the runtime of quantum phase estimation for chemistry through basis set optimization},
  author={Ollitrault, Pauline J and Gonthier, J{\'e}r{\^o}me F and Rocca, Dario and Anselmetti, Gian-Luca and Degroote, Matthias and Moll, Nikolaj and Santagati, Raffaele and Streif, Michael},
  journal={J. Chem. Theory Comput.},
  volume={21},
  number={23},
  pages={12034},
  year={2025}
}

@article{wiebe_simulating_2011,
  title = {Simulating Quantum Dynamics on a Quantum Computer},
  author = {Wiebe, Nathan and Berry, Dominic W and Høyer, Peter and Sanders, Barry C},
  date = {2011-10},
  journal = {Journal of Physics A: Mathematical and Theoretical},
  shortjournal = {J. Phys. A: Math. Theor.},
  volume = {44},
  number = {44},
  pages = {445308},
  publisher = {IOP Publishing},
  issn = {1751-8121},
  doi = {10.1088/1751-8113/44/44/445308},
  url = {https://doi.org/10.1088/1751-8113/44/44/445308},
  urldate = {2026-10-20},
  langid = {english}
}

@article{motta_efficient_2019,
  title = {Efficient {{Ab Initio Auxiliary-Field Quantum Monte Carlo Calculations}} in {{Gaussian Bases}} via {{Low-Rank Tensor Decomposition}}},
  author = {Motta, Mario and Shee, James and Zhang, Shiwei and Chan, Garnet Kin-Lic},
  date = {2019-06-11},
  journal = {Journal of Chemical Theory and Computation},
  shortjournal = {J. Chem. Theory Comput.},
  volume = {15},
  number = {6},
  pages = {3510--3521},
  issn = {1549-9618},
  doi = {10.1021/acs.jctc.8b00996},
  url = {https://doi.org/10.1021/acs.jctc.8b00996},
  urldate = {2026-10-24}
}

@article{Prodan_2005,
   title={Nearsightedness of electronic matter},
   volume={102},
   ISSN={1091-6490},
   url={http://dx.doi.org/10.1073/pnas.0505436102},
   DOI={10.1073/pnas.0505436102},
   number={33},
   journal={Proceedings of the National Academy of Sciences},
   publisher={National Academy of Sciences},
   author={Prodan, E. and Kohn, W.},
   year={2005},
   month=Aug, pages={11635–11638} 
}

@article{kohn_density_1996-1,
  title = {Density {{Functional}} and {{Density Matrix Method Scaling Linearly}} with the {{Number}} of {{Atoms}}},
  author = {Kohn, W.},
  date = {1996-04-22},
  journal = {Physical Review Letters},
  shortjournal = {Phys. Rev. Lett.},
  volume = {76},
  number = {17},
  pages = {3168--3171},
  publisher = {American Physical Society},
  doi = {10.1103/PhysRevLett.76.3168},
  url = {https://link.aps.org/doi/10.1103/PhysRevLett.76.3168},
  urldate = {2026-10-25}
}

@article{saebo_local_1993-1,
  title = {Local {{Treatment}} of {{Electron Correlation}}},
  author = {Saebo, S and Pulay, P},
  date = {1993},
  journal = {Annual Review of Physical Chemistry},
  volume = {44},
  pages = {213--236},
  publisher = {Annual Reviews},
  issn = {1545-1593},
  doi = {10.1146/annurev.pc.44.100193.001241},
  url = {https://www.annualreviews.org/content/journals/10.1146/annurev.pc.44.100193.001241},
  urldate = {2026-10-25}
}

@article{kuwahara_area_2020,
  title = {Area Law of Noncritical Ground States in {{1D}} Long-Range Interacting Systems},
  author = {Kuwahara, Tomotaka and Saito, Keiji},
  date = {2020-09-08},
  journal = {Nature Communications},
  shortjournal = {Nat Commun},
  volume = {11},
  number = {1},
  pages = {4478},
  publisher = {Nature Publishing Group},
  issn = {2041-1723},
  doi = {10.1038/s41467-020-18055-x},
  url = {https://www.nature.com/articles/s41467-020-18055-x},
  urldate = {2026-10-25},
  langid = {english}
}

@article{poulin_trotter_2015,
  title = {The Trotter Step Size Required for Accurate Quantum Simulation of Quantum {{Chemistry}}},
  author = {Poulin, David and Hastings, Matthew B. and Wecker, Dave and Wiebe, Nathan and Doberty, Andrew C. and Troyer, Matthias},
  date = {2015-04-01},
  journal = {Quantum Information \& Computation},
  shortjournal = {Quantum Info. Comput.},
  volume = {15},
  number = {5--6},
  pages = {361--384},
  issn = {1533-7146}
}

@article{sahinoglu_hamiltonian_2021,
  title = {Hamiltonian Simulation in the Low-Energy Subspace},
  author = {Şahinoğlu, Burak and Somma, Rolando D.},
  date = {2021-07-27},
  journal = {npj Quantum Information},
  shortjournal = {npj Quantum Inf},
  volume = {7},
  number = {1},
  pages = {119},
  publisher = {Nature Publishing Group},
  issn = {2056-6387},
  doi = {10.1038/s41534-021-00451-w},
  url = {https://www.nature.com/articles/s41534-021-00451-w},
  urldate = {2026-10-25},
  langid = {english}
}

@article{low_hamiltonian_2019-2,
  title = {Hamiltonian {{Simulation}} by {{Qubitization}}},
  author = {Low, Guang Hao and Chuang, Isaac L.},
  date = {2019-07-12},
  journal = {Quantum},
  volume = {3},
  pages = {163},
  publisher = {Verein zur Förderung des Open Access Publizierens in den Quantenwissenschaften},
  doi = {10.22331/q-2019-07-12-163},
  url = {https://quantum-journal.org/papers/q-2019-07-12-163/},
  urldate = {2026-10-26},
  langid = {british}
}

@incollection{helgaker_second_2000,
  title = {Second {{Quantization}}},
  booktitle = {Molecular {{Electronic-Structure Theory}}},
  author = {Helgaker, Trygve and Jørgensen, Poul and Olsen, Jeppe},
  date = {2000},
  pages = {1--33},
  publisher = {John Wiley \& Sons, Ltd},
  doi = {10.1002/9781119019572.ch1},
  url = {http://onlinelibrary.wiley.com/doi/abs/10.1002/9781119019572.ch1},
  urldate = {2025-02-27},
  isbn = {978-1-119-01957-2},
  langid = {english}
}

@book{dym_fourier_1985,
  title = {Fourier {{Series}} and {{Integrals}}},
  author = {Dym, H. and McKean, H. P.},
  date = {1985-10-28},
  eprinttype = {googlebooks},
  publisher = {Elsevier Science},
  isbn = {978-0-12-226451-1},
  langid = {english},
  pagetotal = {316}
}

@article{landau_eigenvalue_1980,
  title = {Eigenvalue Distribution of Time and Frequency Limiting},
  author = {Landau, H. J and Widom, H},
  date = {1980-10-01},
  journal = {Journal of Mathematical Analysis and Applications},
  shortjournal = {Journal of Mathematical Analysis and Applications},
  volume = {77},
  number = {2},
  pages = {469--481},
  issn = {0022-247X},
  doi = {10.1016/0022-247X(80)90241-3},
  url = {https://www.sciencedirect.com/science/article/pii/0022247X80902413},
  urldate = {2026-10-26}
}

@book{akhiezer_theory_1956,
  title = {Theory of {{Approximation}}},
  author = {Akhiezer, Naum Ilʹich},
  date = {1956},
  eprinttype = {googlebooks},
  publisher = {F. Ungar Publishing Company},
  isbn = {978-0-8044-4019-6},
  langid = {english},
  pagetotal = {328}
}

@article{suzuki_general_1991,
  title = {General Theory of Fractal Path Integrals with Applications to Many‐body Theories and Statistical Physics},
  author = {Suzuki, Masuo},
  date = {1991-02},
  journal = {Journal of Mathematical Physics},
  volume = {32},
  number = {2},
  pages = {400--407},
  issn = {0022-2488},
  doi = {10.1063/1.529425},
  url = {https://doi.org/10.1063/1.529425}
}

@article{suzuki_general_1993,
  title = {General {{Decomposition Theory}} of {{Ordered Exponentials}}},
  author = {Suzuki, Masuo},
  date = {1993},
  journal = {Proceedings of the Japan Academy, Series B},
  volume = {69},
  number = {7},
  pages = {161--166},
  doi = {10.2183/pjab.69.161}
}

@article{wiebe_higher_2010,
  title = {Higher Order Decompositions of Ordered Operator Exponentials},
  author = {Wiebe, Nathan and Berry, Dominic and Høyer, Peter and Sanders, Barry C},
  date = {2010-01},
  journal = {Journal of Physics A: Mathematical and Theoretical},
  shortjournal = {J. Phys. A: Math. Theor.},
  volume = {43},
  number = {6},
  pages = {065203},
  issn = {1751-8121},
  doi = {10.1088/1751-8113/43/6/065203},
  url = {https://doi.org/10.1088/1751-8113/43/6/065203},
  urldate = {2026-09-24},
  langid = {english}
}

@book{demmel_applied_1997,
  title = {Applied {{Numerical Linear Algebra}}},
  author = {Demmel, James W.},
  date = {1997-08-01},
  eprinttype = {googlebooks},
  publisher = {SIAM},
  isbn = {978-0-89871-389-3},
  langid = {english},
  pagetotal = {426}
}

@online{gunther_sublinear-depth_2026,
  title = {Sublinear-Depth {{Quantum Simulation}} of {{Electrons}} with {{Atomic Orbitals}}},
  author = {Günther, Jakob and Harrow, Aram W.},
  date = {2026-10-01},
  number = {arXiv:2610.00950},
  eprinttype = {arXiv},
  eprintclass = {quant-ph},
  doi = {10.48550/arXiv.2610.00950},
  url = {http://arxiv.org/abs/2610.00950},
  urldate = {2026-10-28},
  pubstate = {prepublished}
}

@article{lin_dissipative_2025,
  title = {Dissipative Preparation of Many-Body Quantum States: {{Toward}} Practical Quantum Advantage},
  shorttitle = {Dissipative Preparation of Many-Body Quantum States},
  author = {Lin, Lin},
  date = {2025-09-10},
  journal = {APL Computational Physics},
  shortjournal = {APL Computational Physics},
  volume = {1},
  number = {1},
  pages = {010901},
  issn = {3066-0017},
  doi = {10.1063/5.0283315},
  url = {https://doi.org/10.1063/5.0283315},
  urldate = {2026-10-28}
}

@article{babbush2018encoding,
  title={Encoding electronic spectra in quantum circuits with linear T complexity},
  author={Babbush, Ryan and Gidney, Craig and Berry, Dominic W and Wiebe, Nathan and McClean, Jarrod and Paler, Alexandru and Fowler, Austin and Neven, Hartmut},
  journal={Phys. Rev. X},
  volume={8},
  number={4},
  pages={041015},
  year={2018},
  publisher={APS}
}

@article{poulin2018quantum,
  title={Quantum algorithm for spectral measurement with a lower gate count},
  author={Poulin, David and Kitaev, Alexei and Steiger, Damian S and Hastings, Matthew B and Troyer, Matthias},
  journal={Phys. Rev. Lett.},
  volume={121},
  number={1},
  pages={010501},
  year={2018},
  publisher={APS}
}

@article{chan_spiers_2024-1,
  title = {Spiers {{Memorial Lecture}}: {{Quantum}} Chemistry, Classical Heuristics, and Quantum Advantage},
  shorttitle = {Spiers {{Memorial Lecture}}},
  author = {Chan, Garnet Kin-Lic},
  date = {2024-11-01},
  journal = {Faraday Discussions},
  shortjournal = {Faraday Discuss.},
  volume = {254},
  pages = {11--52},
  issn = {1359-6640},
  doi = {10.1039/d4fd00141a},
  url = {https://doi.org/10.1039/d4fd00141a},
  urldate = {2026-10-28}
}

@article{zini_quantum_2023,
  title = {Quantum Simulation of Battery Materials Using Ionic Pseudopotentials},
  author = {Zini, Modjtaba Shokrian and Delgado, Alain and Reis, Roberto and Casares, Pablo Antonio Moreno and Mueller, Jonathan E. and Voigt, Arne-Christian and Arrazola, Juan Miguel},
  date = {2023-07-10},
  journal = {Quantum},
  volume = {7},
  pages = {1049},
  publisher = {Verein zur F\"orderung des Open Access Publizierens in den Quantenwissenschaften},
  doi = {10.22331/q-2023-07-10-1049}
}

@article{santagati_drug_2024,
  title = {Drug Design on Quantum Computers},
  author = {Santagati, Raffaele and Aspuru-Guzik, Alan and Babbush, Ryan and Degroote, Matthias and González, Leticia and Kyoseva, Elica and Moll, Nikolaj and Oppel, Markus and Parrish, Robert M. and Rubin, Nicholas C. and Streif, Michael and Tautermann, Christofer S. and Weiss, Horst and Wiebe, Nathan and Utschig-Utschig, Clemens},
  date = {2024-04},
  journal = {Nature Physics},
  shortjournal = {Nat. Phys.},
  volume = {20},
  number = {4},
  pages = {549--557},
  publisher = {Nature Publishing Group},
  issn = {1745-2481},
  doi = {10.1038/s41567-024-02411-5},
  url = {https://www.nature.com/articles/s41567-024-02411-5},
  urldate = {2026-10-28},
  langid = {english}
}

@article{kremenetski2021simulation,
  title={Simulation of adiabatic quantum computing for molecular ground states},
  author={Kremenetski, Vladimir and Mejuto-Zaera, Carlos and Cotton, Stephen J and Tubman, Norm M},
  journal={The Journal of Chemical Physics},
  volume={155},
  number={23},
  year={2021},
  publisher={AIP Publishing}
}

@article{sugisaki2022adiabatic,
  title={Adiabatic state preparation of correlated wave functions with nonlinear scheduling functions and broken-symmetry wave functions},
  author={Sugisaki, Kenji and Toyota, Kazuo and Sato, Kazunobu and Shiomi, Daisuke and Takui, Takeji},
  journal={Communications Chemistry},
  volume={5},
  number={1},
  pages={84},
  year={2022},
  publisher={Nature Publishing Group UK, London}
}

@article{ORCA5,
author = {Neese,F.},
title = {Software update: the ORCA program system, version 5.0},
journal = {WIRES Comput. Molec. Sci.},
volume = {12},
number = {1},
pages = {e1606},
DOI = {10.1002/wcms.1606},
year = {2022},
type = {journal Article}
}

@article{cpcm,
  title={Quantum calculation of molecular energies and energy gradients in solution by a conductor solvent model},
  author={Barone, Vincenzo and Cossi, Maurizio},
  journal={The Journal of Physical Chemistry A},
  volume={102},
  number={11},
  pages={1995--2001},
  year={1998},
  publisher={ACS Publications}
}

@article{rijcosx,
  title={An improved chain of spheres for exchange algorithm},
  author={Helmich-Paris, Benjamin and de Souza, Bernardo and Neese, Frank and Izs{\'a}k, R{\'o}bert},
  journal={The Journal of Chemical Physics},
  volume={155},
  number={10},
  year={2021},
  publisher={AIP Publishing}
}

@article{def2basis,
  title={Balanced basis sets of split valence, triple zeta valence and quadruple zeta valence quality for H to Rn: Design and assessment of accuracy},
  author={Weigend, Florian and Ahlrichs, Reinhart},
  journal={Physical chemistry chemical physics},
  volume={7},
  number={18},
  pages={3297--3305},
  year={2005},
  publisher={The Royal Society of Chemistry}
}

@article{dlpno-steom,
  title={Unveiling the photophysical properties of boron-dipyrromethene dyes using a new accurate excited state coupled cluster method},
  author={Berraud-Pache, Romain and Neese, Frank and Bistoni, Giovanni and Izs{\'a}k, R{\'o}bert},
  journal={Journal of chemical theory and computation},
  volume={16},
  number={1},
  pages={564--575},
  year={2020},
  publisher={ACS Publications}
}

@article{foster1960canonical,
  title={Canonical configurational interaction procedure},
  author={Foster, JM and Boys, SFi},
  journal={Rev. Mod. Phys.},
  volume={32},
  number={2},
  pages={300},
  year={1960},
  publisher={APS}
}

@article{virtanen2020scipy,
  title   = {{{SciPy} 1.0: Fundamental Algorithms for Scientific
            Computing in Python}},
  author  = {Virtanen, Pauli and Gommers, Ralf and Oliphant,      Travis E. and
            Haberland, Matt and Reddy, Tyler and Cournapeau, David and
            Burovski, Evgeni and Peterson, Pearu and Weckesser, Warren and
            Bright, Jonathan and {van der Walt}, St{\'e}fan J. and
            Brett, Matthew and Wilson, Joshua and Millman, K. Jarrod and
            Mayorov, Nikolay and Nelson, Andrew R. J. and Jones, Eric and
            Kern, Robert and Larson, Eric and Carey, C J and
            Polat, {\.I}lhan and Feng, Yu and Moore, Eric W. and
            {VanderPlas}, Jake and Laxalde, Denis and Perktold, Josef and
            Cimrman, Robert and Henriksen, Ian and Quintero, E. A. and
            Harris, Charles R. and Archibald, Anne M. and
            Ribeiro, Ant{\^o}nio H. and Pedregosa, Fabian and
            {van Mulbregt}, Paul and {SciPy 1.0 Contributors}},
  journal = {Nature Methods},
  year    = {2020},
  volume  = {17},
  pages   = {261--272},
  doi     = {10.1038/s41592-019-0686-2},
}

@article{litinski2019,
   title={A Game of Surface Codes: Large-Scale Quantum Computing with Lattice Surgery},
   volume={3},
   ISSN={2521-327X},
   url={http://dx.doi.org/10.22331/q-2019-03-05-128},
   DOI={10.22331/q-2019-03-05-128},
   journal={Quantum},
   publisher={Verein zur Forderung des Open Access Publizierens in den Quantenwissenschaften},
   author={Litinski, Daniel},
   year={2019},
   month=Mar, pages={128} }

@misc{litinski2022av,
      title={Active volume: An architecture for efficient fault-tolerant quantum computers with limited non-local connections}, 
      author={Daniel Litinski and Naomi Nickerson},
      year={2022},
      eprint={2211.15465},
      archivePrefix={arXiv},
      primaryClass={quant-ph},
      url={https://arxiv.org/abs/2211.15465}, 
}

@misc{ross2016,
      title={Optimal ancilla-free Clifford+T approximation of z-rotations}, 
      author={Neil J. Ross and Peter Selinger},
      year={2016},
      eprint={1403.2975},
      archivePrefix={arXiv},
      primaryClass={quant-ph},
      url={https://arxiv.org/abs/1403.2975}, 
}

@article{litinski2019ms,
  title={Magic state distillation: Not as costly as you think},
  author={Litinski, Daniel},
  journal={Quantum},
  volume={3},
  pages={205},
  year={2019},
  publisher={Verein zur F{\"o}rderung des Open Access Publizierens in den Quantenwissenschaften}
}

@misc{patil2026,
      title={A Platform-aware Compilation Framework for Fault-tolerant Quantum Computation}, 
      author={Srushti Patil and Susan X. Chen and Andreas Juul Bay-Smidt and Stefan Alaric Schäffer and Peter Krogstrup and Stefano Paesani and Gemma C. Solomon},
      year={2026},
      eprint={2609.08908},
      archivePrefix={arXiv},
      primaryClass={quant-ph},
      url={https://arxiv.org/abs/2609.08908}, 
}

\clearpage
\onecolumngrid

\setcounter{section}{0}
\setcounter{equation}{0}
\setcounter{figure}{0}
\setcounter{table}{0}
\setcounter{page}{1}

\renewcommand{\thesection}{SI \arabic{section}}
\renewcommand{\theequation}{S\arabic{equation}}
\renewcommand{\thefigure}{S\arabic{figure}}
\renewcommand{\thetable}{S\arabic{table}}
\renewcommand{\thepage}{S\arabic{page}}

\setcounter{theorem}{0}

\renewcommand{\thetheorem}{S\arabic{theorem}}
\renewcommand{\thelemma}{S\arabic{theorem}}
\renewcommand{\thecorollary}{S\arabic{theorem}}
\renewcommand{\theproposition}{S\arabic{theorem}}

\let\addcontentsline\oldaddcontentsline

\begin{center}
    {\large \bfseries Supplementary information for: \\ \vspace{0.2cm}
    ``Quantum algorithm for weakly correlated molecular electronic ground states with sub-quintic scaling in the system size''}
\end{center}
\vspace{0.5cm}

\tableofcontents

\vspace{0.8cm}

\section{Note on weak correlation in molecular electronic structure theory}
\label{app:weak_correlation}

Walter Kohn coined the term ``nearsightedness of electronic matter'' \cite{kohn_density_1996-1,Prodan_2005} to express the intuition that the important correlation effects between electrons in the ground states of neutral molecules and materials tend to be constrained by a finite correlation length. In weakly correlated systems, which comprise the majority of molecules at their equilibrium geometry, this localization effect is particularly pronounced \cite{chan_spiers_2024-1}. The flipside of this localization effect is that there is typically a significant energy penalty to excite an electron over the energetic barrier into the delocalized states, which generally manifests as a large spectral gap. A useful diagnostic for the spectral gap in weakly correlated systems is the difference in the highest occupied and lowest unoccupied single-particle states in the mean-field description (the HOMO-LUMO gap). For example, extended protein systems tend to have HOMO-LUMO gaps above $\sim 5$ eV \cite{lever_electrostatic_2013} regardless of the system size, which is further supported by our findings for alanine chains in Section \ref{sec:spectral_gap} in the main text. This finding is in line with rigorous results from many-body physics that equate area laws of entanglement, finite correlation lengths, and spectral gaps in the ground states of local Hamiltonians \cite{hastings_spectral_2006}, and which have been extended to fermionic systems \cite{de_roeck_persistence_2019} and to systems with power-law interactions \cite{kuwahara_area_2020}. 

For weakly correlated molecular electronic ground states in a second-quantized wavefunction description, the wavefunction is typically well-described by perturbative corrections on top of a single Slater determinant (the `reference state'), which can be reliably obtained by a mean-field Hartree-Fock calculation in this regime. This is because the HOMO-LUMO gap suppresses fluctuations around the reference Slater determinant, resulting in rapidly decaying support on the highly excited determinants in the CI expansion. This reduces the classical computer memory needed to store an approximate representation of the wavefunction, enabling relatively reliable convergence of perturbative `single-reference' methods such as coupled-cluster in weakly correlated systems, particularly when exploiting orbital localization \cite{saebo_local_1993-1,riplinger_efficient_2013}. While approximate solutions can often be obtained in this regime with low-order time complexity in the number of orbitals, $n$ (as low as $O(n)$ in some cases \cite{riplinger_efficient_2013}), high-accuracy solutions may require higher-order polynomial scalings (e.g., as required in thermochemistry benchmarks\,\cite{karton2006w4} or to describe large non-covalent $\pi$ interactions\,\cite{schafer2025understanding} or in near metallic systems\,\cite{neufeld2023highly}). For example, CCSDT scales as $O(n^8)$, while CCSDTQ scales as $O(n^{10})$.

\section{Computational details for the polyalanine spectral-gap study}
\label{sec:si_gap_methods}

The molecular models consisted of neutral, acetyl- and N-methylamide-capped poly-L-alanine chains, Ac-(Ala)$_N$-NHMe, constructed in an ideal $\alpha$-helical conformation. Cartesian coordinates were generated using a custom Python script with backbone torsion angles of $\phi=-57^\circ$, $\psi=-47^\circ$, and $\omega=180^\circ$. All calculations were performed at fixed geometries without further structural optimization.

Restricted Hartree-Fock (RHF) calculations were performed using the def2-SVP basis set \cite{def2basis} in ORCA \cite{ORCA5}. Gas-phase calculations were carried out for chains containing $N=1,2,4,8,16,32,$ and $64$ alanine residues. The RIJCOSX approximation \cite{rijcosx}, and \texttt{TightSCF} convergence settings were employed. SCF stability analysis was performed for all systems, and the resulting RHF solutions were found to be stable within the implemented stability test. HOMO-LUMO gaps were obtained directly from the converged orbital energies.
To assess the influence of an aqueous environment, additional RHF/def2-SVP calculations were performed for $N=4,8,16,32,$ and $64$ using the conductor-like polarizable continuum model (CPCM) \cite{cpcm} with water as the solvent. These calculations retained the same molecular geometries as the gas-phase calculations. 

The instantaneous spectral gaps along the adiabatic interpolation were calculated using PySCF \cite{sun2018pyscf}. Canonical molecular orbitals obtained from an RHF calculation were kept fixed throughout the interpolation.
Complete active spaces were constructed from the frontier orbitals, so that there are equal numbers of occupied and virtual RHF orbitals. Orbitals below the active space were kept doubly occupied, while higher virtual orbitals were excluded. The initial Hamiltonian was constructed from the projected one-body RHF Fock operator, and the final Hamiltonian corresponded to the interacting electronic Hamiltonian within the same active space. For the single-residue system, five active spaces ranging from CAS(4,4) to CAS(12,12) were investigated, using a uniform interpolation interval of $\Delta s=0.01$.

Correlated vertical excitation energies were calculated for the one- and two-residue systems using DLPNO-STEOM-CCSD \cite{dlpno-steom} in ORCA. The calculations employed RHF references, and the def2-SVP orbital basis. The frozen core approximation was applied. Four singlet and four triplet excitation energies were calculated for each system. Both coupled-cluster calculations and the requested excited-state roots converged. However, the lowest triplet states showed active-space convergence warnings, with active characters of 95.68\% and 91.78\% for one and two residues, respectively. The corresponding excitation energies should therefore be regarded as preliminary estimates.

\section{Adiabatic state preparation using a smoothed Gevrey schedule}
\label{app:adiabatic}

\subsection{Switching adiabatic theorem with a Gevrey schedule}
\label{app:adiabatic_theorem}

The quantum adiabatic theorem is a foundational result in quantum mechanics, which states, informally, that if a system is prepared in its ground state at time $t=0$, and is then evolved under a smoothly varying time-dependent Hamiltonian that has a robust spectral gap at all times, then the state of the system at time $t$ is approximately the ground state of the Hamiltonian at time $t$. Various rigorous formulations of the quantum adiabatic theorem have been derived for the purpose of state preparation by efficient (i.e., polynomial-sized) quantum circuits \cite{lidar_adiabatic_2009,boixo_necessary_2010,elgart_note_2012,jansen_bounds_2007}.

Refs. \cite{elgart_note_2012,jansen_bounds_2007} established near-optimal bounds for the runtime of state preparation methods based on the quantum adiabatic theorem, using an interpolation function from the Gevrey class $G^\alpha$. For our purposes, we construct a schedule function $s(\tau)\in[0,1]$, which is a monotonically increasing function of the dimensionless time parameter, $\tau=t/T_\text{ASP}$, and is contained in the class $G^2$. We first define a \emph{weight function}
\begin{align}
w(u) = \exp\left( -\frac{1}{u(1-u)}\right),
\end{align}
where $u\in(0,1)$. We then define the integral of this function from zero to one,
\begin{align}
I_\text{tot} = \int_{0}^1 w(u) du \approx 7.03 \times 10^{-3},
\end{align}
and we define the schedule function by
\begin{align}
s(\tau) :=\frac{1}{I_\text{tot}}\int_{0}^\tau w(u) du \in[0,1].
\label{eq:gevrey_schedule}
\end{align}
The value of $s(\tau)$ for any point $\tau\in[0,1]$ can be easily computed to numerical precision by fast numerical quadrature algorithms \cite{doi:10.1137/1.9781611975949.ch19}.

This schedule function has two important properties. The first is that the derivatives vanish at the end-points, that is, for any integer $k\geq 1$,
\begin{align}
\frac{d^k s}{d\tau^k}\bigg|_{\tau=0} = 0, \qquad \frac{d^k s}{d\tau^k}\bigg|_{\tau=1} = 0.
\end{align}
The second important property is that all $k$-th order derivatives at all intermediate times $\tau\in(0,1)$ are compactly supported on the interval $(0,1)$. 

Using this smooth Gevrey schedule, Ref. \cite{elgart_note_2012} proved the following result:

\begin{lemma}[Switching adiabatic theorem for $G^2$ Hamiltonians \cite{elgart_note_2012}]
Let $H^{(s)}$ be the interpolated Hamiltonian
\begin{align}
H^{(s)} = (1-s)H^{(0)} + sH^{(1)},
\end{align}
where $s(\tau)$ is defined in Eq \ref{eq:gevrey_schedule}, and for all values of $s\in[0,1]$, the spectral gap is bounded below by
\begin{align}
\Delta^{(s)} := E_1^{(s)} - E_0^{(s)} \geq \Delta,
\end{align}
where $\Delta$ is a constant. Let $P^{(s)}_\perp$ be the projector onto the orthogonal complement of $\ket{\psi^{(s)}_0}$, and let $\ket{\Phi}$ be the time-evolved state,
\begin{align}
\ket{\Phi} = \mathcal{T} \exp\left( -i T_\text{ASP} \int_{0}^1 H^{(s(\tau))} \, d\tau \right)\ket{\psi^{(0)}_0}.
\end{align}
Then the projection onto the orthogonal complement of $\ket{\psi^{(1)}_0}$ is bounded by
\begin{align}
\|P^{(1)}_\perp\ket{\Phi}\| \leq \mathcal{C}_1 \exp\left( -\mathcal{C}_2 \frac{\Delta^2 T_\emph{ASP}}{\|H^{(1)}-H^{(0)}\|} \right),
\label{eq:adiabatic_suppression}
\end{align}
for some constants $\mathcal{C}_1,\mathcal{C}_2 = \emph{polylog}(1/\Delta)$.
\end{lemma}

Rearranging Eq. \ref{eq:adiabatic_suppression} for the minimal runtime to ensure $\|P^{(1)}_\perp\ket{\Phi}\|\leq \epsilon_\perp$, we obtain
\begin{align}
T_\text{ASP} = O\left( \frac{\mathcal{C}_2\|H^{(1)}-H^{(0)}\|}{\Delta^2} \ln \left(\frac{\mathcal{C}_1}{\epsilon_\perp}\right) \right).
\label{eq:T_asp_bound}
\end{align}
Under the size-extensivity assumption, $\|H^{(1)}-H^{(0)}\|=\Theta(n)$, this simplifies to
\begin{align}
T_\text{ASP} = \tilde{O}\left(\frac{n}{\Delta^2}\right),
\end{align}
proving Proposition \ref{prop:adiabatic_sim_time} in the main text.

\subsection{Time-dependent Suzuki product formulae}
\label{app:adiabatic_product_formulae}

This section discusses the theory of standard time-dependent product formulae \cite{mizuta_theory_2026,wiebe_higher_2010,suzuki_general_1993}. The complete time evolution from $0$ to $T$ can be written as
\begin{align}
    U(T,0) = \mathcal{T} \exp\left( -iT\int_{0}^1 H^{(s)} d\tau \right),
\end{align}
where $\mathcal{T}$ is the time-ordering superoperator. We can also split this exactly into a product of shorter time evolutions over smaller timesteps,
\begin{align}
t_\nu = (\nu-1)\delta t, \qquad \delta t = \frac{T}{N},
\end{align}
then
\begin{align}
U(T,0) = \prod_{\nu=1}^N U(t_{\nu+1}, t_\nu).
\end{align}
It remains to decompose $U(t_{\nu+1}, t_\nu)$ by a Trotter product formula. With a slight abuse of notation, let
\begin{align}
H^{(t)} = H^{(s(\tau))} = H^{(s(t/T))}.
\end{align}
Then, following Refs. \cite{suzuki_general_1993,wiebe_higher_2010}, we define the $p$-th order time-dependent Suzuki product formula, $S_p(t_{\nu+1},t_\nu)$ as follows:
\begin{align}
S_p(t_{\nu+1}, t_\nu) = \prod_{l=1}^{\mathcal{L}_p} \exp\left( -i \, a_l \delta t \, H_{\pi_p(l)}^{(t_\nu + c_l \delta t)} \right),
\end{align}
where the sum from $l=1,\ldots,\mathcal{L}_p$ picks out different Hamiltonian terms according to the \emph{channel function}, $\pi_p(l)$. For $p=1$, $\mathcal{L}_1=L$ and $\pi_1(l) = l$, while for $p=2$, $\mathcal{L}_2=2L$, and the channel function $\pi_2(l)$ is given by
\begin{align}
\pi_2(l) = \begin{cases}
l & l\in\{1,\ldots,L\}, \\
2L-l+1 & l\in\{L+1,\ldots,2L\}
\end{cases}.
\end{align}
Higher order product formulae are defined analogously to the recursive time-independent Lie-Trotter-Suzuki constructions. The coefficients $a_l,c_l\in[0,1]$ are detailed in Refs. \cite{suzuki_general_1993,wiebe_higher_2010}.

It remains to bound the Trotter error from the time-dependent product formula. Now let $\Xi_k^{(t)}$ be a commutator map that acts on any operator $A$ to give
\begin{align}
\Xi_k^{(t)} A = [H_k^{(t)}, A].
\end{align}
The current state-of-the-art in commutator norm bounds for time-dependent Trotter product formulae has been worked out in Ref. \cite{mizuta_theory_2026}, which provides the following lemma.
\begin{lemma}[Nested commutator bound for standard time-dependent Trotter product formulae \cite{mizuta_theory_2026}]
Let $U$ and $S_p$ be defined as above. Then the Trotter error for a single timestep from $t_\nu$ to $t_{\nu+1}=t_\nu+\delta t$ is bounded by
\begin{align}
\|U(t_{\nu+1},t_\nu) - S_p(t_{\nu+1},t_\nu)\| = O\left( (\delta t)^{p+1} \max_{t\in[t_\nu,t_{\nu+1}]} \aleph_p^{(t)} \right),
\end{align}
where $\aleph_p^{(t)}$ is the time-dependent commutator sum,
\begin{align}
\aleph_p^{(t)} = \sum_{k_0=1}^L \sum_{k_1, \ldots, k_p=1}^{L+1} \left\| \bigg( \prod_{q=p}^1 \mathcal{D}^{(t)}_{k_q} \bigg) H_{k_0}^{(t)} \right\|,
\end{align}
with
\begin{align}
\mathcal{D}^{(t)}_k := \begin{cases}
\Xi_k^{(t)} & k\in\{1,\ldots,L\}, \\
L\frac{d}{dt} & k=L+1. 
\end{cases}
\end{align}
\label{lem:mizuta_bound}
\end{lemma}
This expresses the Trotter error as a sum over nested commutators and their time derivatives.

\subsection{Nested time-derivative commutator bound}
\label{app:adiabatic_commutator_bound}

 Using the smooth Gevrey interpolation schedule, which have compactly supported derivatives, and assuming size-extensivity of the Hamiltonian norm, we can now asymptotically bound the commutator error for time-dependent Trotterization by the following lemma.
\begin{lemma}
Let 
\begin{align}
H^{(s)} = \sum_{k=1}^L H_k^{(s)} ,\qquad H_k^{(s)} = H_k^{(0)} + s\bar{H}_k,
\end{align}
and
\begin{align}
\lambda^{(s)} = \sum_{k} \| H_k^{(s)}\| \leq \lambda_\emph{max}.
\end{align}
Let the interpolation function $s(\tau)=s(t/T)$ be the $G^2$ Gevrey function defined in Eq. \ref{eq:gevrey_schedule}. Then for all $t \in[0,T]$, the time-dependent commutator sum for the $p$-th order Lie-Trotter-Suzuki formula is asymptotically bounded by 
\begin{align}
\aleph_p^{(t)} = O(\lambda_\emph{max}^{p+1}),
\end{align}
provided that we choose
\begin{align}
T = \Omega\left(\frac{L}{\lambda_\emph{max}}\right).
\end{align}
\label{lem:td_comm_bound}
\end{lemma}

\begin{proof}
By the triangle inequality, the nested commutator and derivative sum can be separated into distinct sums containing $m$ time derivatives:
\begin{align}
\aleph_p^{(t)} \leq \sum_{m=0}^p \aleph_{p;m}^{(t)}.
\end{align}
The zero-derivatives term is identical to the standard nested commutator sum for time-independent product formulae,
\begin{align}
\aleph_{p;0}^{(t)} &= \sum_{k_0, \ldots, k_p=1}^{L} \left\| \bigg( \prod_{q=p}^1 \Xi_{k_q}^{(t)} \bigg) H_{k_0}^{(t)} \right\| \\
&\leq \left( \sum_{k=1}^L \|H_k^{(t)}\|\right)^{p+1} \\
& = O(\lambda_\text{max}^{p+1}).
\end{align}
We first bound the single-derivative terms. Let $r$ represent a position in the product where a time derivative could be inserted, then we can bound
\begin{align}
\aleph_{p;1}^{(t)} \leq L \sum_{r=1}^{p} \sum_{\{k_q | q\neq r\}=1}^L \left\| \bigg(\prod_{q=p}^{r+1} \Xi_{k_q}^{(t)} \bigg) \frac{d}{dt}\bigg(\prod_{q=r-1}^{1} \Xi_{k_q}^{(t)} \bigg) H^{(t)}_{k_0} \right\|.
\end{align}
Let the index $l$ run from $r-1$ down to $0$. Then define the compact notation for the $(r-1)$-th order nested commutator,
\begin{align}
[H^{(t)}_{k_{r-1}},\ldots,H^{(t)}_{k_{l}},\ldots,H^{(t)}_{k_{0}}] := \bigg(\prod_{q=r-1}^{1} \Xi_{k_q}^{(t)} \bigg) H^{(t)}_{k_0}.
\end{align}
Then the time derivative of this nested commutator is obtained by the Leibniz rule,
\begin{align}
\frac{d}{dt} [H^{(t)}_{k_{r-1}},\ldots,H^{(t)}_{k_{l}},\ldots,H^{(t)}_{k_{0}}] = \sum_{l=0}^{r-1}\, [H^{(t)}_{k_{r-1}}\ldots,\frac{d}{dt}H^{(t)}_{k_{l}},\ldots,H^{(t)}_{k_{0}}].
\end{align}
The time derivative of each Hamiltonian term evaluates to
\begin{align}
\frac{d}{dt}H_{k}^{(t)} = \frac{1}{T}\frac{d s(\tau)}{d\tau}\bar{H}_k.
\end{align}
The derivative of the Gevrey function $s(\tau)$ is simply 
\begin{align}
\frac{d s(\tau)}{d\tau} = \exp\left( -\frac{1}{\tau(1-\tau)}\right),
\end{align}
which is maximized at $\tau=1/2$, therefore $\frac{d s(\tau)}{d\tau} \leq e^{-1/4}$. We can pull the prefactor of $1/T$ out of the nested commutator, resulting in
\begin{align}
\aleph_{p;1}^{(t)} = O\left( \frac{L}{T}\sum_{r=1}^p \sum_{l=0}^{r-1} \sum_{\{k_q | q\neq r\}=1}^L \left\| [H^{(t)}_{k_p},\ldots, H^{(t)}_{k_{r+1}}, H^{(t)}_{k_{r-1}}\ldots,\bar{H}_{k_{l}},\ldots,H^{(t)}_{k_{0}}] \right\|\right).
\end{align}
This commutator sum is then easy to bound as
\begin{align}
\aleph_{p;1}^{(t)} = O\left( \frac{L}{T} \sum_{r=1}^p \sum_{l=0}^{r-1} \bigg( \sum_{k=1}^L \|H^{(t)}_{k}\| \bigg)^{p-1} \bigg( \sum_{k=1}^L \|\bar{H}_{k}\| \bigg) \right).
\end{align}
Finally, we can use the triangle inequality to bound $\bar{H}_k$ in terms of the decomposition of the Hamiltonian at $s=0$ and $s=1$,
\begin{align}
\sum_{k=1}^L \|\bar{H}_{k}\| &= \sum_{k=1}^L \|H^{(1)}_{k} - H^{(0)}_{k}\| \\
&\leq  \sum_{k=1}^L \|H^{(1)}_{k}\| + \sum_{k=1}^L \|H^{(0)}_{k}\| \\
&= O(\lambda_\text{max}).
\end{align}
This yields
\begin{align}
\aleph_{p;1}^{(t)} = O\left( \frac{p^2L \lambda_\text{max}^p}{T}\right).
\end{align}
We can generalize the exact same mathematical procedure to the terms with $m$ time derivatives, which results in
\begin{align}
\aleph_{p;m}^{(t)} = O\left( \frac{p^{2m} L^m \lambda_\text{max}^{p+1-m}}{T^m}\right).
\end{align}
This implies that 
\begin{align}
\aleph_{p}^{(t)} = O(\lambda_\text{max}^{p+1}),
\end{align}
provided that we choose $T$ to satisfy
\begin{align}
T = \Omega\left(\frac{L}{\lambda_\text{max}}\right).
\end{align}

\end{proof}

This lemma allows us to asymptotically bound the number of Trotter steps needed to achieve a target error $\epsilon_\perp^\text{trot}$ in exactly the same way as for time-independent product formulae.

\begin{corollary}[Proposition \ref{prop:adiabatic_trotter_stepcount} in the main text]
Let the approximate evolved state $\ket{\tilde{\Phi}}$ be obtained by substituting Trotterized time evolution for the time-dependent evolution operator $U(T_\emph{ASP},0)$, defined by interpolating between $H^{(0)}$ and $H^{(1)}$ using a smoothed $G^2$ Gevrey schedule, with $T_\emph{ASP}=\Omega(L/\lambda_\emph{max})$. Then using a number of Trotter steps
\begin{align}
N = O\left( \frac{(T_\emph{ASP} \lambda_\emph{max})^{1+1/p}}{(\epsilon_\perp^{\emph{trot}})^{1/p}} \right)
\end{align} 
is sufficient to bound the total Trotter error for the adiabatic state preparation by $\| \ket{\tilde{\Phi}} - \ket{\Phi} \| \leq \epsilon_\perp^{\emph{trot}}$.
\label{cor:adiabatic_trotter_stepcount}
\end{corollary}

\begin{proof}
Lemma \ref{lem:mizuta_bound} bounds the Trotter error $\epsilon_\perp^\text{trot}$ for a single Trotter step, and the accumulated error bound for $N$ Trotter steps simply picks up a factor of $N$. Substituting the one-norm bound on $\aleph^{(t)}_p$ from Lemma \ref{lem:td_comm_bound} and $\delta t = T_\text{ASP}/N$, then rearranging for $N$, yields the desired bound on the Trotter step count.
\end{proof}

\section{Eigenvalue estimation by quantum prolate diagonalization}
\label{app:prolate}

\subsection{Prolate spheroidal functions and the Paley-Wiener space}
\label{app:prolate__theory}

We begin by introducing the theory of prolate spheroidal functions in a self-contained manner. The theory was first developed at Bell labs in the 1960's by Slepian, Landau, and Pollak \cite{slepian_prolate_1961}, where they investigated the problem of reconstructing a time signal, $C(t)$, which has a Fourier transform $\hat{C}(\omega)$ that has zero support outside of a finite frequency band, $\omega \in[-W,W]$, by evaluating the function at a discrete set of points over a finite time domain, $t\in[-T,T]$.

Mathematically, we say that $C(t)$ belongs to the \emph{Paley-Wiener space}, $B_W$ \cite{wiener_fourier_1934}. Because $C(t)$ could be a periodic function, it may not be square-integrable over the whole real line. We therefore define the Paley-Wiener space over the space of \emph{tempered distributions}, $\mathcal{S}'(\mathbb{R})$, which is an extension of the Schwartz space that includes all continuous square-integrable functions on $\mathbb{R}$, as well as delta functions and plane waves.

\begin{definition}[Paley-Wiener space, $B_W$]
    \begin{align}
    B_W := \left\{ \phi \in \mathcal{S}'(\mathbb{R}) \,\,\big|\,\, \emph{supp}(\hat{\phi}) \subseteq [-W,W] \right\}.
    \end{align}
\end{definition}

The Paley-Wiener space can also be extended to distributions in two or more time coordinates, such that the Fourier transform has support constrained to $[-W,W]$ in each frequency coordinate. For a function $C(t,t')$ defined over two time coordinates, we will write this as $C \in B_{W,W}$.

We now introduce some notation to distinguish integrals over the whole real line from integrals over $[-T,T]$. For some arbitrary distributions $\phi,\psi\in\mathcal{S}'(\mathbb{R})$, we define the inner products
\begin{align}
\braket{\phi, \psi}_\infty := \int_{-\infty}^\infty \phi^*(t) \psi(t) dt, \qquad \braket{\phi, \psi}_T := \int_{-T}^T \phi^*(t) \psi(t) dt,
\label{eq:int_notation}
\end{align}
which provide the $L^2$-norm definitions $\|\phi\|_{2} = \sqrt{\braket{\phi,\phi}_\infty}$ and $\|\phi\|_{2,T} = \sqrt{\braket{\phi,\phi}_T}$. We also define the $L^1$-norms,
\begin{align}
\|\phi\|_{1} = \int_{-\infty}^{\infty} |\phi(t)| dt, \qquad \|\phi\|_{1,T} = \int_{-T}^{T} |\phi(t)| dt,
\end{align}
This notation allows us to compactly write the useful inequality $\|\phi\|_{1,T} \leq \sqrt{2T}\|\phi\|_{2,T}$, for example.

A useful property of the Paley-Weiner space is that the absolute value of any continuous square-integrable function $\phi\in B_W$ satisfies the reproducing kernel norm bound on $B_W$ \cite{dym_fourier_1985}, 
\begin{align}
\phi \in B_W \implies |\phi(t)|\leq \sqrt{\frac{W}{\pi}} \|\phi\|_2 \qquad \forall\, t\in(-\infty,\infty).
\end{align}

Prolate spheroidal functions form an orthonormal basis for all functions in $B_W$, and can be defined as follows.

\begin{definition}[Prolate spheroidal functions] Consider a finite time window $[-T,T]$ and a finite frequency band $[-W,W]$, and let $\phi(t)\in B_W$ be any continuous function belonging to the Paley-Wiener space. Then the zeroth-order prolate spheroidal function is given by
\begin{align}
f_0(t):=\underset{\phi \in B_W, \, \|\phi\|_2 = 1}{\mathrm{argmax}}\braket{\phi,\phi}_T.
\end{align}
The higher-order prolates can then be defined recursively as
\begin{align}
f_p(t):=\underset{\phi \in B_W, \, \|\phi\|_2 = 1}{\mathrm{argmax}}\left[\braket{\phi,\phi}_T \,\middle|\, \braket{\phi,f_q}_\infty=0\quad \forall q<p \right].
\end{align}
\end{definition}

In words, the zeroth-order prolate spheroidal wavefunction is the normalized function in the Paley-Wiener space that is maximally concentrated within the finite time window $[-T,T]$. The next higher-order prolate is then defined recursively as the maximally concentrated normalized function in the Paley-Wiener space that is orthogonal to the previously obtained set consisting of all the lower-order prolates. These functions can also be obtained as the solutions to an integral equation or as the solutions to a differential equation, and in practice they are approximated by a convergent expansion in a basis of Legendre polynomials \cite{slepian_prolate_1961,walter_sampling_2003}.

The prolate functions provide a basis for the Paley-Wiener space that satisfies an intriguing double-orthogonality property:

\begin{proposition}[Double-orthogonality]
The set of prolate spheroidal functions $\{f_p(t)\}_{p=0}^\infty$ forms an orthogonal basis for $B_W$ under both inner products simultaneously:
\begin{align}
\braket{ f_p, f_q }_\infty &= \delta_{pq}, \\
\braket{ f_p, f_q }_T &= \gamma_p \delta_{pq},
\end{align}
where $\gamma_p:=\braket{f_p,f_p}_T \in (0,1)$ is the $p$-th Slepian eigenvalue.
\end{proposition}

This double-orthogonality property is very useful because it implies that the prolate functions form an orthogonal basis for band-limited functions over both the domain $(-\infty,\infty)$ and the restricted domain $[-T,T]$, leading to the following identity:

\begin{proposition}
For any $C\in B_W$, we have $\braket{C,f_p}_T = \gamma_p\braket{C,f_p}_\infty$.
\end{proposition}

This fact leads directly to the \emph{prolate expansion formula}, which can be used to reconstruct any band-limited function using only information obtained within the finite time window:

\begin{proposition}[Prolate expansion formula \cite{slepian_prolate_1961}]
For any $C\in B_W$,
\begin{align}
C(t) = \sum_{p=0}^\infty \frac{\braket{f_p,C}_T}{\gamma_p} f_p(t).
\end{align}
\label{prop:prolate_expansion}
\end{proposition}

\subsection{Tail sum bound for the prolate expansion formula}
\label{app:prolate_tail_bound}

We now derive a pointwise exponential convergence bound for the prolate expansion formula in Proposition \ref{prop:prolate_expansion} on the domain $[-T,T]$, when the function $C(t)=C_0e^{i\omega t}$ is a single plane wave with frequency $\omega\in[-W,W]$ weighted by $|C_0| \leq 1$, and when truncated to $M$ prolates, provided that the time-bandwidth product satisfies $TW\geq \pi$, and the number of prolate functions is chosen above the \emph{Nyquist-Shannon number},
\begin{align}
M_0 =\frac{2WT}{\pi}.
\end{align}

The Landau-Widom counting theorem \cite{landau_eigenvalue_1980} proves that the Slepian eigenvalues $\gamma_p$ decay exponentially to zero past $M_0$. To state this formally,

\begin{lemma} For $p>M_0+\ln(WT)$, where $M_0$ is the Nyquist-Shannon number and $WT\geq \pi$, the Slepian eigenvalues $\gamma_p$ are bounded from above and from below by
\begin{align}
\gamma_p = \Theta\left(e^{-\alpha(p-M_0)}\right), \qquad \alpha = \frac{\pi^2}{\ln(WT)}.
\end{align}
\label{lem:slepian_bounds}
\end{lemma}
\begin{proof}
The asymptotic upper and lower bounds follow immediately from the Landau-Widom counting theorem \cite{landau_eigenvalue_1980}, which states that the number of Slepian eigenvalues greater than $\delta\in[0,1]$ is given by
\begin{align}
\mathcal{N}_\delta = M_0 + \frac{1}{\pi^2}\left(\ln \left(\frac{1-\delta}{\delta}\right)\right)\ln(WT) + O(\ln(WT)).
\end{align}
We can bound this from above by setting the last term to be $+\mathcal{C}\ln(WT)$ for some constant $\mathcal{C}$, and from below by setting this term to be $-\mathcal{C}\ln(WT)$. Choosing $\mathcal{N}_{\gamma_p}=p$ and rearranging for $\gamma_p$ yields the desired asymptotic upper and lower bounds.
\end{proof}

We also prove a helpful lemma that bounds the maximum value of $|f_p(t)|$ on the domain $[-T,T]$ in terms of $W$, $T$, and $\gamma_p$.

\begin{lemma}
For $TW>\frac{1}{2}$,
\begin{align}
f^{(\text{max})}_p := \max_{t\in[-T,T]} |f_p(t)| \leq 2eW(1+WT)\sqrt{2T\gamma_p}.
\label{eq:f_max_bound}
\end{align}
\label{lem:f_max_bound}
\end{lemma}
\begin{proof}
Because $f_p(t)\in B_W$, we apply a local Markov-Bernstein inequality \cite{akhiezer_theory_1956} to bound the magnitude of the first time derivative, 
\begin{align}
\max_{t\in[-T,T]}|\frac{d}{dt}f_p(t)|\leq eW(1+WT)f^{(\text{max})}_p.
\end{align}
Now consider that the maximum value $f^{(\text{max})}_p$ occurs at some point $t_\text{max}\in[-T,T]$, and draw straight lines from $(t_\text{max},f^{(\text{max})}_p)$ with slope $\pm eW(1+WT)f^{(\text{max})}_p$ down to their points of intersection with the $t$-axis. Then the area under this triangle that falls within the domain $[-T,T]$ provides a lower bound for the integral of $|f_p(t)|$ over the time domain. Assuming $TW >\frac{1}{2}$, the smallest possible value for this area occurs when half the triangle is outside the window, at $t_\text{max}=\pm T$. Then we have that
\begin{align}
\frac{f^{(\text{max})}_p}{2 eWT(1+WT)} \leq \|f_p\|_{1,T} \leq \sqrt{2T}\|f_p\|_{2,T} = \sqrt{2T \gamma_p}.
\end{align}
Rearranging for $f_p^{(\text{max})}$ yields the bound in Eq. \ref{eq:f_max_bound}.
\end{proof}

Lemmas \ref{lem:slepian_bounds} and \ref{lem:f_max_bound} now allow us to asymptotically bound the truncation error in the prolate expansion formula for a single plane wave restricted to $M$ prolates,

\begin{corollary} Choose $M>M_0+\ln(WT)$. For $C(t)=C_0e^{i\omega t}$, with $\omega\in[-W,W]$ and $|C_0|\leq 1$, and for any $t\in[-T,T]$, we have
\begin{align}
\left|\sum_{p=M}^\infty \frac{\braket{f_p,C}_T}{\gamma_p} f_p(t)\right| = O\left((W T)^{\frac{7}{2}} e^{-\frac{\alpha}{2}(M-M_0)} \right).
\end{align}
\label{cor:tail_sum_bound}
\end{corollary}

\begin{proof}
Given $C(t)=C_0e^{i\omega t}$, the coefficient $\braket{f_p,C}_T$ is the finite Fourier transform of $f_p$ on the domain $[-T,T]$. By Slepian's Fourier transform identity \cite{slepian_prolate_1961}, this is equal to
\begin{align}
\braket{f_p,C}_T = C_0 i^p \sqrt{\frac{2\pi T\gamma_p}{W}} \; f_p\left( \frac{\omega T}{W} \right).
\end{align}
We can then substitute the bound on $f_p^{(\text{max})}$ from Lemma \ref{lem:f_max_bound}, which gives
\begin{align}
|\braket{f_p,C}_T| &\leq 4e\sqrt{\pi}|C_0| T\sqrt{W} (1+WT) \gamma_p \\
&= O(T^2W^\frac{3}{2} \gamma_p).
\end{align}
Substituting this into the tail sum expression, we get
\begin{align}
\left|\sum_{p=M}^\infty \frac{\braket{f_p,C}_T}{\gamma_p} f_p(t)\right| &= O\left(T^2 W^\frac{3}{2} \sum_{p=M}^\infty f_p^{(\text{max})}\right) \\
&= O\left((WT)^\frac{7}{2}\sum_{p=M}^\infty \sqrt{\gamma_p}\right),
\end{align}
which is absolutely summable, and is asymptotically bounded above by $O(e^{-\frac{\alpha}{2}(M-M_0)})$ by Lemma \ref{lem:slepian_bounds}.
\end{proof}

\subsection{Gauss-Legendre quadrature error bounds for 1d and 2d integrals of band-limited functions}
\label{app:prolate_quadrature}

We now consider discretizing 1d and 2d overlap integrals for functions in the Paley-Wiener space. In particular, we will make use of the Paley-Wiener theorem, which extends functions of real continuous variables in $B_W$ to entire functions in the complex plane.

\begin{theorem}[Paley-Wiener theorem \cite{wiener_fourier_1934}]
Any $f\in B_W$ extends to an entire function on $\mathbb{C}$ obeying the exponential growth bound,
\begin{align}
|f(t+i\sigma)|\leq f_\emph{max} e^{ W |\sigma|}, \qquad f_\emph{max} = \sup_{t\in(-\infty,\infty)} |f(t)|.
\end{align}
\label{thm:paley_wiener}
\end{theorem}

This fact enables tight error bounds for numerical quadrature of overlap integrals of the form $\braket{f,g}_T$ for any $f,g\in B_W$. In the following lemma, we prove super-exponential convergence of the quadrature error in the number of grid-points $K$, when the timesteps and the associated quadrature weights, $\{t_k,w_k\}_{k=1}^K$, are chosen according to a standard Gauss-Legendre quadrature rule (see e.g. Chapter 19 in Ref. \cite{doi:10.1137/1.9781611975949.ch19}).

\begin{lemma} Consider a quadrature grid of $K$ grid-points, with timesteps and quadrature weights $\{t_k,w_k\}_{k=1}^K$ chosen according to a Gauss-Legendre quadrature rule on $[-T,T]$. For any $f,g\in B_W$, with $|f(t)|,|g(t)|\leq \emph{poly}(W,T)$ on the domain $[-T,T]$, we have
\begin{align}
\left|\braket{f,g}_T - \sum_{k=1}^K w_k f^*(t_k) g(t_k) \right| = O^*\left(e^{-2K\ln(\beta K)} \right), \qquad \beta = \frac{2}{eWT},
\label{eq:quadrature_bound}
\end{align}
where the asymptotic notation $O^*(\cdot)$ drops polynomial factors in $T$ and $W$.
\label{lem:1d_quadrature_bound}
\end{lemma}

\begin{proof}
Because $f,g\in B_W$, their product $h(\tau):=f(\tau T)g(\tau T)\in B_{2W}$, where we have susbstituted the renormalized time parameter $\tau=t/T \in[-1,1]$. By Theorem \ref{thm:paley_wiener}, $h(\tau)$ extends to an entire function on $\mathds{C}$ obeying the exponential growth bound,
\begin{align}
|h(\tau+i\sigma)|\leq h_\text{max} e^{2 W T |\sigma|}, \qquad h_\text{max} = \max_{\tau\in[-1,1]}|h(\tau)|.
\end{align}
Now consider the Bernstein ellipse $E_\rho \subset \mathds{C}$ (that is, an elliptical disc in the complex plane with radius $\rho$, and foci fixed at $-1$ and $1$). For any point $z\in E_\rho$, the imaginary part $|y|<\rho/2$, which bounds
\begin{align}
|h(z)|\leq h_\text{max} e^{W T \rho}.
\end{align}
Now, applying Trefethen's error bound on Bernstein ellipses (Eq. 19.3 in Ref. \cite{doi:10.1137/1.9781611975949.ch19}), then inserting this bound and rearranging, we get
\begin{align}
\left|\int_{-1}^1 h(\tau) d\tau - \sum_{k=1}^K \frac{w_k}{T} h(\tau_k) \right| =O \left(h_\text{max} e^{WT\rho - 2K\ln \rho}\right).
\end{align}
The exponent is minimized by finding $\rho>1$ such that $\frac{d}{d\rho}(WT\rho - 2K \ln \rho)=0$, which is obtained at $\rho = 2K/(WT)$. Substituting this into the exponent, rescaling $\tau \rightarrow t$, and bounding $h_\text{max}\leq f_\text{max}g_\text{max}$, yields Eq. \ref{eq:quadrature_bound}.
\end{proof}

We now extend the overlap integral notation in Eq. \ref{eq:int_notation} to integrals over two time coordinates, $t$ and $t'$. For example, for two single-coordinate functions $f,p\in B_W$ and a two-coordinate function $C \in B_{W,W}$, we will use the compact 2d integral notation
\begin{align}
\braket{\braket{f,C}^*_T,g}_T = \iint_{-T}^T f^*(t) C(t,t') g(t') dt\,dt',
\end{align}
where it is understood that the inner integral is over the $t$ coordinate, and produces a function on the second coordinate, i.e., $h(t') := \braket{f,C}_T$. We can then bound the 2d quadrature error as follows.

\begin{lemma}
Let $f,g\in B_W$, with $|f(t)|,|g(t)|\leq \emph{poly}(W,T)$ on the domain $t\in[-T,T]$, and let $C \in B_{W,W}$, with $|C(t,t')|\leq \emph{poly}(W,T)$ on the domain $[-T,T]^2$. Let $\vec{x}$ and $\vec{y}$ be $K$-dimensional vectors, with element-wise definitions
\begin{align}
x_{k} =w_k f(t_k), \qquad y_{k} = w_k g(t_k),
\end{align}
where $\{t_k,w_k\}_{k=1}^K$ are chosen according to a Gauss-Legendre quadrature rule on $[-T,T]$, and let $\mathbf{A}$ be the $K\times K$ matrix with elements $a_{kl}=C(t_k-t_l)$. Then 
\begin{align}
|\braket{\braket{f, C}^*_T, g}_T - \vec{x}^\dag \mathbf{A} \vec{y}\,| = O^*(e^{-2K\ln(\beta K)}).
\end{align}
\label{lem:2d_quadrature_bound}
\end{lemma}

\begin{proof}
First, bound the error in the inner integral. Let $\vec{a}(t')$ be the vector with elements $a_k(t')=C(t_k-t')$, then taking the inner integral over $t$, for any constant $t'$, by Lemma \ref{lem:1d_quadrature_bound} we have
\begin{align}
\braket{f,C}_T  = \vec{x}^{\,\dag}\vec{a}(t') + O^*( e^{-2K \ln(\beta K)}).
\end{align}
We can now define the function $h(t') = \vec{x}^{\,\dag}\vec{a}(t')$, which is a sum over band-limited functions, therefore $h\in B_{W}$. Plugging this into the outer integral over $t'$, we get
\begin{align}
\braket{\braket{f, C}^*_T, g}_T = \braket{h^*, g}_T + O^*(e^{-2K \ln(\beta K)}),
\end{align}
since integrating the error term over $[-T,T]$ only incurs a factor of $T$. Finally, let $\vec{z}$ be the vector with elements $z_l = h(t_l)$. Because $h\in B_{W}$, we can again apply Lemma \ref{lem:1d_quadrature_bound} to give
\begin{align}
\braket{h^*,g}_T = \vec{z}^{\,T} \vec{y} + O^*(e^{-2K \ln(\beta K)}).
\end{align}
Unpacking the notation, we see that
\begin{align}
\vec{z}^{\,T} \vec{y} = \sum_l z_l y_l = \sum_{kl} x^*_k a_{kl} y_l = \vec{x}^\dag \mathbf{A} \vec{y},
\end{align}
which completes the proof.
\end{proof}

\subsection{Quantum prolate diagonalization with a single in-band eigenphase}
\label{app:qpd_outline}

The core idea behind quantum prolate diagonalization (QPD), introduced by Stroschein \emph{et al.} \cite{stroschein_ground_2025}, is to approximate the eigenvalues of the second-quantized molecular electronic Hamiltonian in an active space of $n$ orbitals,
\begin{align}
H = \sum_{pq}^n h_{pq} \hat{a}_p^\dag \hat{a}_q + \sum_{pqrs}^n h_{pqrs} \hat{a}_p^\dag \hat{a}_q^\dag \hat{a}_r\hat{a}_s,
\label{eq:el_ham_app1}
\end{align}
by embedding it within a subspace of reference states $\{\ket{v_0},\ldots,\ket{v_{M-1}}\}$ given by
\begin{align}
\ket{v_p} = F_{p}\ket{\Phi},
\label{eq:prolate_basis_state}
\end{align}
where $\ket{\Phi}$ is an input state that can be efficiently prepared, meaning by a quantum circuit with a gate count that scales polynomially in the system size, and $F_p$ is a spectral transformation obtained by the Fourier transformation of the prolate spheroidal wavefunction $f_p$, weighted by the Slepian eigenvalue $\gamma_p$,
\begin{align}
F_p = \gamma_p^\frac{1}{2}\int_{t= -\infty}^{\infty} f_p(t) e^{-iH t} dt.
\label{eq:prolate_filter_integral}
\end{align} 
Note that we work in atomic units, so we equate energies with frequencies, all expressed in units of Ha. Because the prolate functions belong to the Paley-Wiener space, $F_p$ projects the state onto the energy band $[-W,W]$. The energies within this band are then obtained by solving a generalized eigenvalue problem
\begin{align}
\textbf{H}\textbf{C} = \textbf{S}\textbf{C}\textbf{E},
\label{eq:geneig_exact}
\end{align}
where the columns of the $M\times M$ solution matrix $\textbf{C}$ encode the generalized eigenvectors, subject to the normalization constraint
\begin{align}
\textbf{C}^\dag \textbf{S} \textbf{C} = \mathds{1},
\end{align}
and $\textbf{E}$ is a diagonal matrix, where the diagonal elements are the generalized eigenvalues, $E_1\leq \cdots \leq E_M$. The matrix elements of  the $M\times M$ subspace matrices $\textbf{H}$ and $\textbf{S}$ are given by
\begin{align}
h_{pq} = \braket{v_p|H|v_q}, \qquad s_{pq} = \braket{v_p|v_q},
\label{eq:matels}
\end{align}

By substituting the expressions for $\ket{v_p}$ and $\ket{v_q}$ into Eqs. \ref{eq:matels}, one obtains
\begin{align}
h_{pq} &= -i (\gamma_{p} \gamma_{q})^\frac{1}{2}\iint_{-\infty}^{\infty} f_{p}^*(t) \left[\frac{\partial}{\partial \tau}C(t - t')\right] f_{q}(t') dt dt', \label{eq:h_pq_integral}\\
s_{pq} &= (\gamma_{p} \gamma_{q})^\frac{1}{2} \iint_{-\infty}^{\infty} f_{p}^*(t) C(t - t') f_{q}(t') dt dt',
\label{eq:s_pq_integral}
\end{align}
where $C(t-t')$ is the two-point autocorrelation function,
\begin{align}
C(t) = \braket{\Phi|e^{i H (t-t')}|\Phi}.
\end{align}
The autocorrelation function decomposes into in-band and out-of-band components, 
\begin{align}
C(t)=C_W(t)+C_\perp(t),
\end{align}
with $C_W\in B_W$, and $|C_\perp(t)| \leq \epsilon_\perp^2$. We can also write the in-band component of the two-point autocorrelation function as $C_W(t-t')$, where now $C_W \in B_{W,W}$. Stroschein \emph{et al.} \cite{stroschein_ground_2025} used the properties of the prolate basis to prove that these integrals can be expressed exactly over the domain $[-T,T]$, resulting in
\begin{align}
h_{pq} &= -i (\gamma_{p} \gamma_{q})^{-\frac{1}{2}}\iint_{-T}^{T} f_{p}^*(t) \left[\frac{\partial}{\partial t}C_W(t - t')\right] f_{q}(t') dt d t', \label{eq:h_pq_integral}\\
s_{pq} &= (\gamma_{p} \gamma_{q})^{-\frac{1}{2}}\iint_{-T}^{T} f_{p}^*(t) C_W(t - t') f_{q}(t') dt dt'.
\label{eq:s_pq_integral}
\end{align}
Solving the generalized eigenavalue problem then yields exact expressions for all energies in the band-limited signal, using a time window $T$ that scales inversely with the smallest separaton between in-band phases, and in the limit $M\rightarrow \infty$.

Note that the $F_p$ are non-unitary operators, thus the states $\ket{v_p}$ are not guaranteed to be normalized, and in general are not expected to be. The generalized eigenvalue problem can therefore become ill-conditioned if there are eigenstates within the band that have small support on $\ket{\Phi}$, meaning that if one substitutes approximate expressions for the matrix elements up to some additive error, the error will be scaled by the condition number of the overlap matrix, $\kappa(\mathbf{S})$. 

As discussed in the main text, the condition number problem is completely resolved if one assumes that there is only a single eigenstate within the band. Under this assumption, $C_W\in B_W$ is simply a single phase oscillation,
\begin{align}
C_W(t) = |\phi_0|^2 e^{iE_0 t},
\end{align}
and the purpose of QPD is then simply to efficiently read off the phase $E_0$. Projecting onto a single band-limited prolate function is then sufficient to compute the phase,
\begin{align}
E_0 = \frac{\braket{v_0|H|v_0}}{\braket{v_0|v_0}}.
\end{align}
The problem with this approach is that the denominator $\braket{v_0|v_0}$ will depend on the positioning of $E_0$ within the band. For example, the Fourier transform of the lowest order prolate function $f_0(t)$ is heavily peaked around the center of the frequency band, and decays rapidly approaching the edges at $-W$ or $W$. However, given an obtained set of autocorrelation measurements, it is very computationally cheap to extend the classical post-processing to $M$ prolates, where $M$ need only be slightly larger than the Nyquist-Shannon number $M_0=2WT/\pi$ to ensure that the largest eigenvalue of the overlap matrix converges exponentially to $\|\textbf{S}\|\rightarrow 2T|\phi_0|^2$, which we state formally as follows.

\begin{lemma}
Assume a single in-band eigenphase $E_0$, and assume $WT\geq \pi$, then for $M>M_0+\ln(WT)$ prolate basis functions, we have
\begin{align}
\|\mathbf{S}\| = 2T|\phi_0|^2 - O^*(\epsilon_\emph{trunc}),
\end{align}
where
\begin{align}
\epsilon_\emph{trunc} = O^*\left(e^{-\frac{\alpha}{2}(M-M_0)}\right), \qquad \alpha = \frac{\pi^2}{\ln(WT)}.
\end{align}
\label{lem:S_norm_trunc_bound}
\end{lemma}

\begin{proof}
Here we use the compact integral notation introduced in \ref{app:prolate__theory}. 
The basis functions project out the excited states, so they have the form
\begin{align}
\ket{v_p} &= \gamma_p^{\frac{1}{2}} \braket{f_p,C_W}_\infty \ket{\psi_0} \\ 
&= \gamma_p^{-\frac{1}{2}} \braket{f_p,C_W}_T \ket{\psi_0}.
\end{align}
Because all basis states are parallel vectors, and $\ket{\psi_0}$ is normalized by definition, then the nonzero eigenvalue of the overlap matrix is simply the sum of squares of the coefficients,
\begin{align}
\|\mathbf{S}\| = \sum_{p=0}^{M-1}\frac{1}{\gamma_p} |\braket{f_p,C_W}_T|^2.
\end{align}
Now consider integrating the squared magnitude of the in-band autocorrelation function over the domain $[-T,T]$:
\begin{align}
\int_{-T}^T|C_W(t)|^2 dt = 2T|\phi_0|^2.
\end{align}
We can also expand $C_W(t)$ in the basis of $M$ prolates according to Proposition \ref{prop:prolate_expansion} and Corollary \ref{cor:tail_sum_bound}, which gives
\begin{align}
C_W(t) = \sum_{p=0}^{M-1}\frac{\braket{f_p,C_W}_T}{\gamma_p}f_p(t) + O^*(e^{-\frac{\alpha}{2}(M-M_0)}).
\end{align}
Inserting this into the integral expression yields
\begin{align}
\int_{-T}^T|C_W(t)|^2 dt &= \int_{-T}^T \left(\sum_{pq=0}^{M-1} \frac{\braket{f_p,C_W}_T^*\cdot\braket{f_q,C_W}_T}{\gamma_p\gamma_q}\right)f_p^*(t) f_q(t) dt + O^*(e^{-\frac{\alpha}{2}(M-M_0)})\\
&= \sum_{pq=0}^{M-1} \frac{\braket{f_p,C_W}_T^*\cdot\braket{f_q,C_W}_T}{\gamma_p\gamma_q} \braket{f_p,f_q}_T + O^*(e^{-\frac{\alpha}{2}(M-M_0)})\\
&= \sum_{p=0}^{M-1} \frac{1}{\gamma_p} |\braket{f_p,C_W}_T|^2 + O^*(e^{-\frac{\alpha}{2}(M-M_0)}).
\end{align}
By comparison, we see that 
\begin{align}
\|\mathbf{S}\|= 2T|\phi_0|^2 - O^*(e^{-\frac{\alpha}{2}(M-M_0)}),
\end{align}

which completes the proof.
\end{proof}
In the case of a single eigenphase, the overlap matrix $\textbf{S}$ is a rank-1 matrix, and the eigenphase can be robustly estimated by diagonalizing
\begin{align}
\textbf{S} = \textbf{U} \boldsymbol{\Lambda} \textbf{U}^\dag,
\label{eq:S_diag}
\end{align}
then extracting the first column $\vec{u}$ from the eigenvector matrix, and projecting the Hamiltonian matrix directly onto this vector to obtain the ground state energy estimate:
\begin{align}
E_0 = \frac{\vec{u}^\dag \textbf{H} \vec{u} }{\|\mathbf{S}\|}.
\label{eq:E_estimator}
\end{align}

Exact error bounds are easily obtained in this case, following a similar line of argumentation as laid out for more general cases in Ref. \cite{stroschein_ground_2025}.

\begin{proposition}
Let $(\mathbf{H},\mathbf{S})$ be the exact matrix pencil with matrix elements obtained by Eqs. \ref{eq:h_pq_integral} and \ref{eq:s_pq_integral}, where $C_W(t)=|\phi_0|^2e^{i E_0 t}$ and $E_0\in[-W,W]$, and let
\begin{align}
\tilde{\mathbf{H}} = \mathbf{H} + \delta \mathbf{H}, \qquad \tilde{\mathbf{S}} = \mathbf{S} + \delta \mathbf{S}.
\end{align}
Then substituting the perturbed matrix pencil $(\tilde{\mathbf{H}},\tilde{\mathbf{S}})$ into Eqs. \ref{eq:S_diag} and $\ref{eq:E_estimator}$ produces an energy estimate $\tilde{E}_0$ that satisfies the asymptotic bound
\begin{align}
|\tilde{E}_0 - E_0| = O\left( \frac{\|\delta \mathbf{H}\| + W \|\delta \mathbf{S}\|}{2T|\phi_0|^2 - O^*(\epsilon_\emph{trunc}) - \|\delta \mathbf{S}\|}\right), \qquad \epsilon_\emph{trunc} = O^*(e^{-\frac{\alpha}{2}(M-M_0)}).
\label{eq:energy_bound}
\end{align}
\label{prop:energy_perturbation_theory}
\end{proposition}
\begin{proof}
By the Davis-Kahan $\sin \theta$ theorem, and because the unperturbed overlap matrix has a single nonzero eigenvalue, the principal eigenvector of $\tilde{\mathbf{S}}$, $\vec{u}+\delta \vec{u}$, is perturbed as
\begin{align}
\|\delta \vec{u}\| \leq \|\delta\mathbf{S}\|\cdot\|\mathbf{S}\|^{-1},
\end{align}
where we use the standard definition $\vec{u}^\dag \delta \vec{u}=0$, with the intermediate normalization convention $\|\vec{u}+\delta \vec{u}\|=1+O(\|\delta \vec{u}\|^2)$. Substituting into $\tilde{E}_0-E_0$, and then using the fact that every vector in the subspace is an eigenvector of $\mathbf{H}$ with eigenvalue $E_0$, we have
\begin{align}
\tilde{E}_0-E_0 &= (\vec{u} + \delta\vec{u})^\dag (\mathbf{H} + \delta \mathbf{H}) (\vec{u} + \delta\vec{u})\cdot \|\tilde{\mathbf{S}}\|^{-1} - \vec{u}^\dag \mathbf{H} \vec{u}\cdot \|\mathbf{S}\|^{-1} \\
&\leq \|\delta \mathbf{H}\|\cdot\|\tilde{\mathbf{S}}\|^{-1} + |E_0|\cdot \|\tilde{\mathbf{S}}\|^{-1}\cdot\left( \|\delta \mathbf{S}\| + \|\delta \mathbf{S}\|^2\cdot\|\mathbf{S}\|^{-1} \right).
\end{align}
We can then use Weyl's theorem to bound
\begin{align}
\|\tilde{\mathbf{S}}\| \geq \|\mathbf{S}\| - \|\delta\mathbf{S}\|.
\end{align}
Substituting $|E_0|\leq W$, keeping only the leading $\|\delta \mathbf{S}\|$ term in the parentheses, and substituting Lemma \ref{lem:S_norm_trunc_bound} yields the asymptotic bound in Eq. \ref{eq:energy_bound}.
\end{proof}

\subsection{Prior work on QPD and suppression of the out-of-band error}
\label{app:out_of_band}

Section~\ref{app:qpd_outline} uses QPD to reconstruct the signal. With $M>M_0$ prolates, the in-band signal $C_W$ is recovered on $[-T,T]$, and the out-of-band component $C_\perp$ enters as additive noise.
This is justified, as adiabatic state preparation keeps $C_\perp$  small (\ref{app:adiabatic}).
The original formulation of QPD highlights a complementary regime~\cite{stroschein_ground_2025}. With $M<M_0$, the prolates act as filters, and the out-of-band contribution is suppressed by an error factor that decays exponentially in $WT$ instead of entering as noise.
The resource analysis below does not need this regime, but the numerical benchmark in Sec.~\ref{sec:benchmark_main} uses this variant of QPD.
We therefore summarize it for the ground state of a gapped system and state the corresponding error bound from Ref.~\cite{stroschein_ground_2025}.

\begin{definition}[Section II.D \cite{stroschein_ground_2025}]
    Given the signal $C(t)=\sum_n |\phi_n|^2 e^{iE_n t}$, with out-of-band weight $\epsilon_\perp^2=\sum_{n>0}|\phi_n|^2$, for $t\in[-2T,2T]$, a frequency band $B_{f}=[-W, W]$ of interest, and an expansion dimension $M$,
\emph{Prolate Filter Diagonalization} returns the $M \times M$ matrices $\mathbf{H}$ and $\mathbf{S}$ of a generalized eigenvalue problem with matrix elements:
\begin{align}
    \mathbf{H}_{sl} & = - i \int_{-T}^{T} \int_{-T}^{T} f_s(\tau) \partial_\tau C(\tau - t) f_l(t) d t d\tau   \label{eq:VHV_sl}\\
    \mathbf{S}_{sl} & = \int_{-T}^{T} \int_{-T}^{T} f_s(\tau) C(\tau - t) f_l(t) d t d\tau \label{eq:VV_sl}
\end{align}
where $s,l = 0, \dots , M-1$ and $\{f_l(t)\}_l$ is the sequence of prolate spheroidal wave functions band-limited to $B_f$ and maximally concentrated on $[-T,T]$.
\end{definition}

In practice, the matrices are computed from a finite number of noisy samples of an approximately propagated signal. As in Proposition~\ref{prop:energy_perturbation_theory}, we write 
$\tilde{\mathbf H}=\mathbf H+\delta\mathbf H$ and $\tilde{\mathbf S}=\mathbf S+\delta\mathbf S$, where $\delta\mathbf H,\delta\mathbf S$ collect discretization, shot-noise and time-propagation errors.

To estimate the ground-state energy, we place the filter window such that it contains $E_0$ but not $E_1$. 
We then project onto the leading eigenvector $\vec u$ of $\tilde{\mathbf{S}}$ and take $\tilde E_0=\vec u^\dagger\tilde{\mathbf H}\vec u/\|\tilde{\mathbf S}\|$.

Specialising Theorem~1 of Ref.~\cite{stroschein_ground_2025} to a single in-band frequency, gives
\begin{align}
-\frac{\|\delta\mathbf H\|+ W \|\delta\mathbf S\|}{\|\tilde{\mathbf S}\|-\epsilon^{\mathrm{Prlt}}_M\epsilon_\perp^2-\|\delta\mathbf S\|}
\le \tilde E_0-E_0 \le
\frac{\epsilon^{\mathrm{Prlt}}_M\sum_{n>0}(E_n-\tilde E_0)|\phi_n|^2+\|\delta\mathbf H\|+ W \|\delta\mathbf S\|}{\|\tilde{\mathbf S}\|-\epsilon^{\mathrm{Prlt}}_M\epsilon_\perp^2-\|\delta\mathbf S\|},
\end{align}
The noise terms $\|\delta\mathbf H\|+W\|\delta\mathbf S\|$ enter as in Theorem~\ref{thm:qpd_error_single_eigenphase}.
Here, however, the out-of-band weight $\epsilon_\perp^2$, which entered linearly there, is multiplied by $\epsilon^{\mathrm{Prlt}}_M$.
The price is an energy-weighted prefactor that grows with the spectral range of the input state.
It can be estimated from the signal itself via the mean energy,
\begin{align*}
    -i C'(0) = \sum_{n} E_n |\phi_n|^2 .
\end{align*}
The error factor $\epsilon^{\mathrm{Prlt}}_M$ is defined in Eq.~(29) of Ref.~\cite{stroschein_ground_2025}. With $c=WT$, for fixed $M$ and large $c$,
\begin{align}
\epsilon^{\mathrm{Prlt}}_M < MT\frac{\pi^{3/2} 2^{3M} c^{M+1/2} e^{-2c}}{(M-1)!}[1+O(1/c)].
\end{align}
It shares the same sharp transition as the prolate leakage factors $1-\gamma_{M-1}(c)$. It remains exponentially close to zero as long as  $M$ stays below $M_0=2WT/\pi$ by a margin logarithmic in the target accuracy
but sharply increases around $M_0$. 
This error bound implies a threshold on the necessary observation time. 
To estimate a single frequency, it requires $2TW\gtrsim\pi$.
For our gapped system this means
\begin{align}
    W\lesssim \Delta \quad \text{and} \quad 2T \gtrsim \frac{\pi}{W} > \frac{\pi}{\Delta}.
\end{align}
The constraint on $W$ is necessary to ensure that the band indeed only contains the ground state energy, and the constraint on $T$ 
that off-band frequencies are suppressed and the matrix problem is stable with respect to other error sources.

It remains to discretise the matrix elements such that only finitely many samples of $C$ on $[-2T,2T]$ are needed, while keeping $\|\delta\mathbf H\|+W\|\delta\mathbf S\|$ small.
Ref.~\cite{stroschein_ground_2025} uses the prolate sampling formula on the equidistant grid $t_k=\pi k/W_s$. It requieres a sampling rate $W_s\ge W_C+W$, where $W_C$ is the bandwidth of $C$ (Theorem~2 of Ref.~\cite{stroschein_ground_2025}).
The resulting sampling error is governed by the same leakage factors $1-\gamma_l(c)$ as $\epsilon^{\mathrm{Prlt}}_M$. So that subspace and discretisation error are coupled and 
share the threshold above.

For the resource analysis we follow a different route. We return to the reconstruction regime of Section~\ref{app:qpd_outline}, treat $C_\perp$ as additive noise, and discretise by Gauss--Legendre quadrature (Section~\ref{app:numerical_quad}).
A systematic comparison of both discretisations is left to future work.

\subsection{Subspace matrices from numerical quadrature}
\label{app:numerical_quad}

We now prove the following lemma.

\begin{lemma}
Let $(\mathbf{H},\mathbf{S})$ be the exact matrix pencil with the element-wise definition in Eqs. \ref{eq:h_pq_integral} and \ref{eq:s_pq_integral}, then the subspace matrices are obtained by the matrix expressions
\begin{align}
\mathbf{S} &= \mathbf{X}\mathbf{A}\mathbf{X}^\dag + O^*(\epsilon_\emph{quad}) \\
\mathbf{H} &= \frac{i}{2}\left(\mathbf{Y}\mathbf{A}\mathbf{X}^\dag - \mathbf{X}\mathbf{A}\mathbf{Y}^\dag\right) + O^*(\epsilon_\emph{quad}),
\end{align}
where the $M\times K$ prolate transformation matrices $\mathbf{X}$ and $\mathbf{Y}$ have the element-wise definitions
\begin{align}
x_{pk} &= \gamma_p^{-\frac{1}{2}}w_kf_p(t_k), \\ 
y_{pk} &= \gamma_p^{-\frac{1}{2}} \left( w_k f'_p(t_k) + \delta_{k1} f_p(t_1) + \delta_{kK}f_p(t_K) \right),
\end{align}
where $\delta_{kk'}$ is the Kronecker delta, and obey the spectral norm bounds
\begin{align}
\|\mathbf{X}\| &\leq \sqrt{\frac{\pi T}{K}} + O^*(\epsilon_\emph{quad}), \\
\|\mathbf{Y}\| &\leq W\sqrt{\frac{\pi T}{K}} + 2eW\sqrt{2MT}(1+WT) + O^*(\epsilon_\emph{quad}),
\end{align}
with 
\begin{align}
\epsilon_\emph{quad} = O^*(e^{-2K \ln(\beta K)}), \qquad \beta = \frac{2}{eWT}.
\end{align}
The asymptotic notation $O^*(\cdot)$ drops polynomial factors in $W,T,M,$ and $K$.
\label{lem:subspace_quadrature}
\end{lemma}

\begin{proof}
For readability we define the renormalized prolate functions 
\begin{align}
g_p(t) = \gamma_p^{-\frac{1}{2}}f_p(t),
\end{align}
which are normalized on the domain $[-T,T]$, i.e., $\braket{g_p,g_q}_T=\delta_{pq}$. By Lemma \ref{lem:f_max_bound}, we then have $|g_p(t)|\leq eW(1+WT)\sqrt{2T}$ on the domain $[-T,T]$. Using the compact integral notation from \ref{app:prolate__theory}, the exact overlap matrix elements are given by
\begin{align}
s_{pq} = \braket{\braket{g_p,C}^*_T,g_q}_T,
\end{align}
so by constructing the $M\times K$ prolate transformation matrix $\mathbf{X}$, with matrix elements $x_{pk} = w_k g_p(t_k)$, and applying the 2d quadrature bound in Lemma \ref{lem:2d_quadrature_bound}, we get
\begin{align}
\mathbf{S} = \mathbf{X}\mathbf{A}\mathbf{X}^\dag + O^*(\epsilon_\text{quad}).
\end{align}

We can bound the norm of the $\mathbf{X}$ matrix by noting that 
\begin{align}
[\mathbf{X} \mathbf{X}^\dag]_{pq} &= \sum_{k=1}^K w_k^2 g_p(t_k) g_q(t_k) \\
&\leq \frac{\pi T}{K} \sum_{k} w_k g_p(t_k) g_q(t_k) \\
&= \frac{\pi T}{K} \left(\braket{g_p,g_q}_T + \epsilon_\text{quad}\right) \\
&= \frac{\pi T}{K}(\delta_{pq}+\epsilon_\text{quad}),
\end{align}
where we used the upper bound on the quadrature weights using a Gauss-Legendre quadrature grid, $w_k \leq \pi T /K$. We therefore have that
\begin{align}
\|\mathbf{X}\| \leq \sqrt{\frac{\pi T}{K}} + O^*(\epsilon_\text{quad}).
\end{align}
Obtaining the Hamiltonian subspace matrix takes some more work, because of the time derivative of the autocorrelation function. To simplify the presentation, we define the compact notation for the partial derivative operator with respect to a variable $u$, 
\begin{align}
\partial_u := \frac{\partial}{\partial u}.
\end{align}
In order to ensure Hermiticity of the Hamiltonian matrix, we can average the partial time derivative over both coordinates,
\begin{align}
h_{pq} = \frac{i}{2} \Big(\braket{\braket{g_p, \partial_t C}^*_T, g_q}_T - \braket{\braket{g_p, \partial_{t'} C}^*_T, g_q}_T  \Big).
\label{eq:h_pq_formula}
\end{align}
We now apply integration by parts to the inner integral over the $t$ variable in the first term (or to the integral over $t'$ in the second term), which separates the integral into bulk and boundary contributions, and shifts the time derivative onto the prolate functions in the bulk term. Applying integration by parts to the inner integral in $t$ in the first term gives
\begin{align}
\braket{g_p, \partial_t C}_T = \chi_p(t') - \braket{g'_p, C}_T,
\end{align}
where $g'_p(t) = \frac{d}{dt} g_p(t)$, and
\begin{align}
\chi_p(t') := g_{p}(T) C(T - t') - g_{p}(-T) C(-T - t').
\end{align}
We obtain a similar expression for the integral in $t'$ in the second term. Plugging these into Eq. \ref{eq:h_pq_formula} results in a sum of bulk and edge terms:
\begin{align}
h_{pq} &= \frac{i}{2}\left(h_{pq}^{\text{(bulk)}} + h_{pq}^{\text{(edge)}}\right), \label{eq:h_pq_bulk_bound}\\
h_{pq}^{\text{(bulk)}} &= \braket{\braket{g'_p, C}^*_T, g_q}_T - \braket{\braket{g_p, C}^*_T, g'_q}_T, \\
h_{pq}^{\text{(edge)}} &= \braket{\chi_p^*,g_q}_T - \braket{g_p,\chi^*_q}_T.
\end{align}

Because the time derivatives of the renormalized prolate functions $ g'_p \in B_W$, we can immediately apply Lemma \ref{lem:2d_quadrature_bound} to the bulk terms, resulting in the matrix expression
\begin{align}
\mathbf{H}^{\text{(bulk)}} = \bar{\mathbf{Y}}\mathbf{A}\mathbf{X}^\dag - \mathbf{X}\mathbf{A}\bar{\mathbf{Y}}^\dag + O^*(\epsilon_\text{quad}),
\end{align}
where the prolate derivative transformation matrix $\bar{\mathbf{Y}}$ has elements
\begin{align}
y_{pk} = w_k g'_{p}(t_k).
\end{align}
We can now use the same tricks as for the $\mathbf{X}$ matrix, i.e., computing the matrix elements of $\bar{\mathbf{Y}}\bar{\mathbf{Y}}^\dag$, to bound
\begin{align}
\|\bar{\mathbf{Y}}\| \leq W\sqrt{\frac{\pi T}{K}} + O^*(\epsilon_\text{quad}),
\end{align}
where we have used the fact that $\braket{g'_p,g'_p}_T \leq W^2\delta_{pq}$.

We also have that $\chi_p\in B_W$, therefore we can use Lemma \ref{lem:1d_quadrature_bound} to write
\begin{align}
h_{pq}^{\text{(edge)}} = \sum_{k} \left(\chi_p(t_k) x_{qk} - x_{pk} \chi^*_q(t_k) \right) + O^*(\epsilon_\text{quad}).
\end{align}
Expanding the definition of $\chi_p$, and grouping the sums over matrix elements, we see that we can write
\begin{align}
\mathbf{H}^{(\text{edge})} = \mathbf{Z}\mathbf{A}\mathbf{X}^\dag - \mathbf{X}\mathbf{A}\mathbf{Z}^\dag + O^*(\epsilon_\text{quad}),
\end{align}
where 
\begin{align}
\mathbf{Z} = w_K^{-1}\vec{x}_K\vec{e}_K^{\,\dag} - w_1^{-1}\vec{x}_1\vec{e}_1^{\,\dag},
\end{align}
with $\vec{x}_k$ is the $k$'th column of $\textbf{X}$, and $\vec{e}_k$ the $k$'th single-element unit vector. The norm of this matrix is bounded by
\begin{align}
\|\mathbf{Z}\| &\leq w_1^{-1}\|\vec{x}_1\| + w_K^{-1}\|\vec{x}_K\| \\
&\leq 2 \bigg(\sum_{p=0}^{M-1} |g_p(T)|^2 \bigg)^{\frac{1}{2}}. \\
&\leq 2eW\sqrt{2MT}(1+WT),
\end{align}
where we have used the pointwise bound on $|g_p(t)|$ on the domain $[-T,T]$.

Combining the bulk and boundary matrices,
\begin{align}
\mathbf{Y} = \bar{\mathbf{Y}} + \mathbf{Z},
\end{align}
yields the final expression
\begin{align}
\mathbf{H} = \frac{i}{2}\left( \mathbf{Y}\mathbf{A}\mathbf{X}^\dag - \mathbf{X}\mathbf{A}\mathbf{Y}^\dag\right) + O^*(\epsilon_\text{quad}),
\end{align}
where now 
\begin{align}
\|\mathbf{Y}\| \leq W\sqrt{\frac{\pi T}{K}} + 2eW\sqrt{2MT}(1+WT) + O^*(\epsilon_\text{quad}).
\end{align}

\end{proof}

\subsection{Positioning the spectral window using a level-shift}\label{app:level-shift}

To uniformly shift all the frequencies in the autocorrelation matrix by some constant $\omega_0$, we simply transform
\begin{align}
\bar{C}(t-t') = e^{i\omega_0 (t-t')}C(t-t').
\end{align}
This amounts to a trivial classical post-processing operation on the obtained autocorrelation matrix. Defining the $K\times K$ diagonal phase matrix, $\mathbf{B}$, with matrix elements
\begin{align}
b_{kk'} = e^{i \omega_0 t_k}\delta_{kk'},
\end{align}
then the level-shifted autocorrelation matrix is given by
\begin{align}
\bar{\mathbf{A}} = \mathbf{B}\tilde{\mathbf{A}}\mathbf{B}^\dag.
\end{align}
Because the phase matrix is unitary, this does not affect the quadrature error or the magnitude of any propagated errors from noisy autocorrelation measurements. After solving the generalized eigenvalue problem with the level-shifted matrix pencil $(\bar{\mathbf{H}},\bar{\mathbf{S}})$, the ground state energy estimate is simply
\begin{align}
E_0=\bar{E}_0 - \omega_0.
\end{align}
In order to locate the ground state inside the window without knowing the true value of $E_0$ beforehand, in the worst case one would have to `sweep' the window across the entire Hamiltonian spectrum, which incurs a classical post-processing cost of at worst $O(n/W)$, assuming a size-extensive Hamiltonian norm. In the following, we will assume that the spectral window has been positioned appropriately such that the ground state is isolated within the band $[-W,W]$, and the excited states are all excluded from this frequency band.

\subsection{Error propagation from noisy autocorrelation measurements}
\label{app:prolate_error_prop}

Having demonstrated that the numerical quadrature error is easily controlled, the main source of error that affects the energy estimator is due to the noise in the autocorrelation matrix. By measuring the autocorrelation function on a quantum computer, we obtain
\begin{align}
\tilde{\textbf{A}} = \textbf{A} + \delta\textbf{A},
\end{align}
where both terms are $K\times K$ Hermitian matrices. We can further decompose the error matrix into
\begin{align}
\delta \textbf{A} = \delta\mathbf{A}_\text{stoch} + \delta\mathbf{A}_\text{coh} + \delta\mathbf{A}_\perp,
\end{align}
where the first error matrix $\delta\mathbf{A}_\text{stoch}$ represents the \emph{stochastic} or uncorrelated noise, arising from shot noise and stochastic gate erorrs. Importantly, the elements of this error matrix are independent random variables, thus assuming that each random matrix element is normally distributed, and has zero mean and variance bounded by
\begin{align}
\text{Var}\left[\delta a_{kl}^\text{(stoch)}\right] \leq \epsilon^2_\text{stoch},
\end{align}
then we can model $\delta\textbf{A}_\text{stoch}$ as a random Hermitian Wigner matrix. by a standard result in random matrix theory (see e.g. Ref. \cite{tao_random_2010}), the spectral norm of $\delta\textbf{A}_\text{stoch}$ is bounded almost surely in the asymptotic limit by
\begin{align}
\|\delta\textbf{A}_\text{stoch}\| = O(\sqrt{K}\epsilon_\text{stoch}).
\end{align}

The second error matrix $\delta\textbf{A}^\text{(coh)}$ contains all error contributions which cannot be assumed to be independently distributed, including Trotter error and coherent gate errors. Assuming that we can bound
\begin{align}
|\delta a^\text{(coh)}_{kl}| \leq \epsilon_\text{coh},
\end{align}
Then the worst-case norm bound for this error matrix will be
\begin{align}
\|\delta\textbf{A}_\text{coh}\| = O(K\epsilon_\text{coh}).
\end{align}
The exact same logic applies to the out-of-band noise in the autocorrelation matrix,
\begin{align}
\|\delta\textbf{A}_\perp\| = O(K\epsilon^2_\perp),
\end{align}
where we used that $|C_\perp(t)|\leq \epsilon^2_\perp$. In the following we assume that $K$ is sufficiently large such that $\epsilon_\text{quad}$ is negligible compared to other sources of error, and can be safely dropped from the analysis. 

We now prove the final error bound theorem for QPD with a single in-band eigenphase.

\begin{theorem}[Theorem \ref{thm:qpd_error_single_eigenphase} in the main text]
Let $C(t)$ be the autocorrelation function obtained from the input state $\ket{\Phi}$, with an in-band component $C_W(t)=|\phi_0|^2e^{iE_0 t}$, where $E_0\in[-W,W]$, and an out-of-band component $C_\perp(t)$ bounded by $|C_\perp(t)|\leq\epsilon^2_\perp$. Let $\tilde{E}_0$ be the energy estimate obtained by 
\begin{align}
\tilde{E}_0 = \frac{\vec{u}^\dag \tilde{\mathbf{H}} \vec{u}}{\|\tilde{\mathbf{S}}\|},
\end{align}
where the $\vec{u}$ is the principal eigenvector of $\tilde{\mathbf{S}}$, and the noisy matrix pencil $(\tilde{\mathbf{H}},\tilde{\mathbf{S}})$ is obtained according to Lemma \ref{lem:subspace_quadrature}, using two-point autocorrelation function measurements $C(t-t')$ at a series of grid-points $\{t_k,w_k\}_{k=1}^K$ selected according to a Gauss-Legendre quadrature rule, with a suitable choice of $K,M>M_0$ such that $\epsilon_\emph{quad}$ and $\epsilon_\emph{trunc}$ are negligible compared to other sources of error. Assume that each autocorrelation measurement is obtained with a stochastic error term of variance $\epsilon_\emph{stoch}^2$, and a coherent error term of magnitude at most $\epsilon_\emph{coh}$. Then the eigenvalue estimate $\tilde{E}_0$ satisfies the asymptotic bound
\begin{align}
|\tilde{E}_0-E_0| = O\left( W(1+\mathcal{C}) \left( \frac{\epsilon_\emph{stoch}}{\sqrt{K}} + \epsilon_\emph{coh} + \epsilon_\perp^2 \right) \right),
\end{align}
where $\mathcal{C} = \sqrt{MK}(1+TW) = \tilde{O}(1)$ by choosing $TW=O(1)$.
\end{theorem}

\begin{proof}
By Proposition \ref{prop:energy_perturbation_theory}, and assuming $\epsilon_\text{trunc}$ is negligible, the energy estimate is bounded by
\begin{align}
|\tilde{E}_0 - E_0| = O\left( \frac{\|\delta \mathbf{H}\| + W \|\delta \mathbf{S}\|}{2T(1-\epsilon_\perp^2) - \|\delta \mathbf{S}\|}\right).
\label{eq:energy_bound_rep}
\end{align}
Using Lemma \ref{lem:subspace_quadrature}, we obtain bounds on the error in the Hamiltonian and overlap matrices,
\begin{align}
\|\delta \textbf{S}\| &\leq \|\textbf{X}\|^2\cdot \|\delta \textbf{A}\| =O \left(T\left( \frac{\epsilon_\text{stoch}}{\sqrt{K}} + \epsilon_\text{coh} + \epsilon_\perp^2 \right)  \right), \\
\|\delta \textbf{H}\| &\leq \|\textbf{Y}\|\cdot\|\textbf{X}\|\cdot \|\delta \textbf{A}\| = O\left( WT\left(1+\sqrt{MK}(1+WT)\right)\left( \frac{\epsilon_\text{stoch}}{\sqrt{K}} + \epsilon_\text{coh} + \epsilon_\perp^2 \right)  \right).
\end{align}
In the asymptotic limit where we suppress $\epsilon_\perp,\|\delta\mathbf{S}\|\rightarrow 0$, the denominator in Eq. \ref{eq:energy_bound} simplifies to $2T$, which proves the theorem.
\end{proof}

As an immediate corollary, if we assume ideal (i.e., noiseless) quantum hardware, then the stochastic error term is entirely due to shot noise, and can be suppressed as
\begin{align}
\label{eq:shot_noise}
\epsilon_{\text{stoch}} = O(N_\text{shot}^{-1/2}),
\end{align}
where $N_\text{shots}$ is the number of shots per autocorrelation evaulation.

\begin{corollary}[Corollary \ref{cor:qpd_error_T_scaling} in the main text]
Assume that the stochastic error is entirely due to shot noise, and is suppressed by taking a number of measurement $N_\emph{shot}$ per grid-point evaluation of $C(t)$. Choose a spectral bandwidth of $W=O(1/T)$, and allow $T$ as a free parameter. Then holding $\epsilon_\emph{coh}$ and $\epsilon_\perp$ constant, the energy error is asymptotically bounded by
\begin{align}
|\tilde{E}_0-E_0| = O\left(\frac{1}{T}\left(1+\frac{1}{\sqrt{K N_\emph{shot}}}\right)\right).
\end{align}
\end{corollary}

This reveals a fundamental tradeoff between the circuit depth and the shot count: by choosing $T=O(1)$, to suppress $|\tilde{E}_0-E_0|\leq \varepsilon$ requires a total number of measurements $\propto K^2 N_\text{shot}=\Omega(1/\epsilon^2)$, resulting in a total runtime of $\Omega(1/\varepsilon^2)$. However, by choosing $T=\Omega(1/\varepsilon)$, and shrinking the width of the spectral band as $W=\pi/T$, one can resolve the eigenphase to any error $\epsilon$ by essentially setting the width to $W\propto \varepsilon$, then requiring a total number of measurements $\propto K^2 N_\text{shot}=O(1)$. In other words, the total number of circuit runs is then at worst polylogarithmic in $1/\varepsilon$, and the total quantum runtime is $\tilde{O}(1/\varepsilon)$, thus achieving true Heisenberg-limited scaling. Note that this incurs a classical post-processing cost of $O(n/\varepsilon)$ to locate the ground state energy, by `sliding' the narrow frequency band using the level shift across the range of the Hamiltonian spectrum (which grows as $O(n)$ under the size-extensivity assumption), until the ground state is detected inside the band.

\section{Double-factorized Hamiltonian decomposition}

\subsection{Low-rank factorization of the Hamiltonian}
\label{sec:circuit_lowrank}

As described in Eq. \ref{eq:el_ham} of the manuscript,
the molecular electronic Hamiltonian has the form
\begin{align}
H = \sum_{pq}^n h_{pq} \hat{a}_p^\dag \hat{a}_q + \sum_{pqrs}^n h_{pqrs} \hat{a}_p^\dag \hat{a}_q^\dag \hat{a}_r \hat{a}_s \ .
\end{align}
By a low-rank double factorization, the Hamiltonian \cite{motta_low_2021-1} is re-expressesd in the form of mutually commuting sets of number operators expressed in different orbital bases,
\begin{align}\label{eq:df_ham}
H = \sum_{k=1}^L H_k = H_1 + \sum_{k=2}^{L} G_k A_k^2 G_k^\dag,
\end{align}
where $H_1$ represents the one-body terms in $H$, the $G_k$ are unitary orbital rotation operators, and the $A_k$ are diagonal one-body fermionic Hamiltonians,
\begin{align}
A_k = \sum_{p=1}^{\rho_k} h^{(k)}_{p} \hat{n}_p.
\end{align}
The number of terms in distinct orbital bases is at most $L=O(n^2)$, however one frequently observes a practical scaling of $L=O(n)$ by truncating small singular values, resulting in errors that are below the chemical accuracy threshold \cite{motta_low_2021-1}. We present additional numerical results supporting the observed $L=O(n)$ scaling in \ref{app:df_error_vs_L}. Furthermore, in the asymptotic limit $n\rightarrow \infty$, the number of terms in each diagonal operator $A_k$ tends to $\rho_k=O(1)$, which is observed in practice for very large numbers of orbitals \cite{motta_low_2021-1}. The most costly step in the classical factorization can be performed exactly in time $O(n^6)$ by reshaping the two-electron repulsion integral tensor into an $n^2 \times n^2$ real-valued s.p.d. matrix which is then diagonalized or factorized into its Cholesky decomposition \cite{motta_low_2021-1}. However, since only $L=O(n)$ eigenvalue/vector pairs are typically needed in practice, this step can be performed in time $O(Ln^4)=O(n^5)$ using iterative solvers such as the Lanczos algorithm \cite{demmel_applied_1997}.

\subsection{Analytic gate count for a single Trotter step}
\label{app:df_analytic_gc}

Here we summarize the analysis in Ref. \cite{motta_low_2021-1} of the gate count to implement a single Trotter step, decomposing into a gate set of Clifford + $Z$-rotations, which we will later compare against our numerical results. We break down the gate count to implement a single Trotter step using the first order product formula $S_1$ defined in Eq. \ref{eq:first_order_pf}, since this forms the repeat-unit circuit block for the higher order product formulae in the Lie-Suzuki-Trotter construction.

Time evolution under the one-body term ($e^{iH_1 t}$) can be directly implemented by a linear depth circuit $G_1$ of exactly $n \choose 2$ Givens rotation gates under the Jordan-Wigner transformation \cite{kivlichan_quantum_2018}. By writing the exponential function as a Taylor expansion and grouping terms, we see that for $k\geq 2$ we have
\begin{align}
e^{iH_k t} = G_k \exp(it A_k^2) G_k^\dag,
\end{align}
where each orbital rotation $G_k$ can also be implemented by the same Givens rotation circuits \cite{kivlichan_quantum_2018}. The exponentiated diagonal two-body operator decomposes as
\begin{align}
D^{(t)}_k &= \exp(it A_k^2) \\
&= \exp\left( i\sum_{pq}^{\rho_k} \theta^{(k)}_{pq}\hat{n}_p \hat{n}_q \right), \qquad \theta^{(k)}_{pq} := h^{(k)}_{p} h^{(k)}_{q} t \\
&= \prod_{pq}^{\rho_k} \exp \left(i \theta_{pq}^{(k)} \hat{n}_p \hat{n}_q \right).
\label{eq:tb_decomp}
\end{align}
Under the Jordan-Wigner transformation, each double number excitation in Eq. \ref{eq:tb_decomp} corresponds to a controlled-$Z$ rotation with angle $\theta_{pq}^{(k)}$. Now define the grouped orbital rotations between terms $k-1$ and $k$, for all $k\geq 2$,
\begin{align}
\tilde{G}_k := G_{k-1}^\dag G_{k}, \qquad G_1^\dag = e^{iH_1 t}. 
\end{align}
We then have the following expression for the first order product formula,
\begin{align}
S_1(t) = \left(\prod_{k=2}^L \tilde{G}_k D_k^{(t)} \right) G_L^\dag.
\label{eq:pf_df_first_order}
\end{align}

A simple counting exercise then yields
\begin{align}
N_\text{gate}^\text{step} = L{n \choose 2} + \sum_{k=2}^L \rho_k^2.
\end{align}
However, the more detailed analysis in Ref. \cite{motta_low_2021-1} also exploited partial basis rotations, $z$-spin symmetry, and fermionic swap networks to reduce the prefactors in the above gate count. The gates can be further decomposed into Clifford gates and single-qubit rotations (i.e., Pauli $Z$-rotations), where each Givens rotation gate requires two single-qubit rotations, and each controlled-$Z$ rotation requires only one. The end result of this analysis in Ref. \cite{motta_low_2021-1} was an exact single-qubit rotation gate count of 
\begin{align}
N_{R_z}^\text{step} = \left(\frac{n}{2}-2\right)\sum_{k=1}^L \rho_k.
\end{align}
Note that adding controls to these gates, as required by phase estimation circuits, requires only CNOT gates, which are Clifford gates, and thus do not factor into the dominant T-gate cost for fault-tolerant quantum algorithms.
For a $p$-th order product formula, we can use the prefactor from the recursive Lie-Trotter-Suzuki formula \cite{suzuki_general_1991} to bound the total gate count by $2\cdot 5^{p/2-1}\cdot N_{R_z}^\text{step}$.

\subsection{Fermionic seminorm bound for $\lambda_\text{DF}$}
\label{app:fermionic_seminorm}

Starting from the double factorized Hamiltonian in Eq. \ref{eq:df_ham}, we first absorb the one-body terms into the two-body terms for simplification, resulting in the expression

\begin{align}\label{eq:df_ham_diag_rotated_obt}
H = \sum_{k=1}^L H_k = \sum_{k=1}^{L} G_k A_k^2 G_k^\dag \ .
\end{align}
This is possible, since for a set of one body operators, one can always find an orbital basis in which they are diagonal \cite{martinez2023assessment}, and since number operators are idempotent.

Note that the double number excitations can be further factorized, yielding
\begin{align}
\label{double_factorization}
H_k = G_k A_k^2 G_k^\dag, \qquad A_k = \sum_{p=1}^{\rho_k} h^{(k)}_p \hat{n}_p.
\end{align}
We can define the induced one-norm as 
\begin{align}\label{eq:df_ham_onorm}
\lambda_\text{DF} = \sum_k\|H_k\|,
\end{align}
where $\|\cdot\|$ denotes the spectral norm (i.e., the operator 2-norm). However, because Trotter error arises from non-canceling nested commutator terms which all conserve particle number, we can get a tighter bound using the \emph{fermionic seminorm} \cite{su_nearly_2021}, which measures the largest transition amplitude between states of fixed electron number, $\eta$. Let $\ket{\phi_\eta}$ denote any state in the $\eta$-electron Fock space. For any Hermitian operator $A$, this reduces to an operator norm definition,
\begin{align}\label{eq:fermi_seminorm_def}
\|A\|_\eta = \max_{\ket{\phi_\eta}} \frac{\|A \ket{\phi_\eta}\|}{\|\ket{\phi_\eta}\|}, \qquad A = A^\dag.
\end{align}
We can write the induced one-norm in the $\eta$-electron subspace as
\begin{align}
\lambda^{(\eta)}_\text{DF} = \sum_{k=1}^L \|H_k\|_\eta,
\end{align}
then using the properties of the fermionic seminorm given in Proposition 4 of Ref. \cite{su_nearly_2021}, we get
\begin{align}
\lambda^{(\eta)}_\text{DF} = \sum_{k=1}^L \|A_k\|^2_\eta.
\end{align}
Because $A_k$ is Hermitian and is composed of a sum of number operators, the extremal eigenvectors of $A_k$ are simply $\eta$-electron Slater determinants. Order the coefficients in descending order, $h^{(k)}_1 \geq h^{(k)}_2 \geq \cdots \geq h^{(k)}_n$. We can then write

\begin{align}\label{A_k_seminorm_from_sd_coeff_1}
\|A_k\|_\eta = \max\left( \left|\sum_{p=1}^\eta h^{(k)}_p\right|, \left|\sum_{p=n-\eta+1}^n h^{(k)}_p\right| \right),
\end{align}
which yields the final expression
\begin{align}
\lambda^{(\eta)}_\text{DF} = \sum_{k=1}^L \max\left( \left|\sum_{p=1}^\eta h^{(k)}_p\right|, \left|\sum_{p=n-\eta+1}^n h^{(k)}_p\right| \right)^2.
\end{align}

Note that the coefficients $h_i^{(k)}$ can have a positive or negative
value, and hence we need to consider both the largest positive sign, as well as the largest negative sign coefficient sums in
Eq. \ref{A_k_seminorm_from_sd_coeff_1}.

\section{Physical resource estimation}
\label{app:resources}
This  details the assumptions and the pipeline behind the physical resource estimates of Section~\ref{sec:physical_resource_estimate}.
Every estimate is for one Trotter step of one circuit.
A logical qubit is encoded in a rotated surface code patch of $2d^{2}$ physical qubits, where $d$ is the code distance; the code cycle is one round of syndrome extraction, taken to last $1\,\mu$s on both superconducting and photonics platform, and a logical cycle is then defined as $d$ rounds of syndrome extraction.

\subsection{Noise model and distance calculation}
\label{app:noise}
We assume standard circuit-level noise on the surface code with an effective threshold of $1\%$, under which the logical error probability per logical qubit per code cycle is approximated as
\begin{equation}
    p_L(p,d) = 0.1\,(100p)^{(d+1)/2},
\label{circ-noise-model}
\end{equation}
where $p$ is the physical error rate.
The code distance is the smallest odd $d$ for which the probability of at least one logical error over the whole computation stays below the algorithmic failure budget $\epsilon=0.01$.
For a computation that occupies $n_Q$ logical tiles for $n_L$ logical cycles, i.e.\ $n_L\,d$ code cycles, this condition reads
\begin{equation}
    n_Q \cdot n_L \cdot d \cdot p_L(p,d) < \epsilon .
\label{eq:distance-criterion}
\end{equation}
Since $n_L$ itself depends on $d$ as well as magic state supply (~\ref{app:factory}), Eq.~\ref{eq:distance-criterion} is evaluated for each candidate factory seperately.
We report results for a physical error rate of $p=10^{-4}$.

\subsection{Logical circuits and $T$ counts}
\label{app:tcount}
The Trotter circuits are given in the $\{R_z,\sqrt{X},X,\mathrm{CX}\}$
gate set.
Every $R_z(\theta)$ whose angle is not a multiple of $\pi/4$ is
synthesised into a Clifford$+T$ sequence with GridSynth
\cite{ross2016} at precision $\epsilon_{\rm gs}=10^{-4}/N_{\rm rot}$,
where $N_{\rm rot}$ is the number of such rotations in the step
(Section~\ref{sec:physical_resource_estimate}); rotations by odd
multiples of $\pi/4$ are a single $T$ gate with a Clifford correction
and cost one $T$ state, and rotations by multiples of $\pi/2$ are
Clifford.
This synthesis budget is separate from the failure budget $\epsilon$
of Eq.~\ref{eq:distance-criterion}.
For the CAS($n$,$n$) series at the $10^{-3}$ cutoff, $N_{\rm rot}$ grows
from 192 to $1.4\times10^{5}$, the mean cost per rotation from 64 to 95
$T$ gates, and the $T$ count per step, denoted $T$ below, from
$1.2\times10^{4}$ to $1.3\times10^{7}$.

\subsection{Data block architectures}
\label{app:blocks}
Three of the four architectures are the data blocks of Ref.~\cite{litinski2019}, which execute a Clifford$+T$ circuit as a sequence of $\pi/8$ Pauli product rotations via lattice surgery between neighbouring patches, consuming one $T$ state per rotation:
\begin{itemize}
  \item \emph{Baseline}: $4n$ tiles for $n$ logical qubits (two tiles per
  qubit plus an equal routing region), consuming up to one $T$ state per
  logical cycle.
  \item \emph{Compact}: $1.5n+3$ tiles, consumption of one $T$ state  taking upto 9 logical cycles.
  \item \emph{Fast}: $2n+\sqrt{8n}+1$ tiles, consuming one $T$ state per
  logical cycle.
\end{itemize}
Writing $c$ for the number of logical cycles per $T$ state ($c=1$ for baseline and fast, $c=9$ for compact), a data block needs $n_L=T\,c$ logical cycles unless magic state production is slower (~\ref{app:factory}).
The fourth architecture, active volume \cite{litinski2022av}, assumes
non-local connections between logical qubits and is described in
~\ref{app:av}.

\subsection{Magic state factories}
\label{app:factory}
In any fault tolerant computation, we need to produce high fidelity magic states that can be consumed to perform T gates. 
For lattice surgery based computation, various protocols exist in the literature which takes noisy magic states and produce a high fidelity T state.
In practice, with the considered architectures, the protocol is run at a logical level via lattice surgery at finite code distance, and the output error also contains the logical errors of the distillation circuit itself; we therefore use the full error model and cost model of Ref.~\cite{litinski2019ms}, in which a factory is specified by the two spacial code distances $(d_X,d_Z)$ of the surface code patch taking part in distillation and a temporal distance $d_m$.
Within this model we consider four families of factories.
A 15-to-1 distillation, which consumes 15 noisy $T$ states of error $p_{\rm in}$ and outputs one state of error $\approx35\,p_{\rm in}^{3}$ when the Clifford operations of the protocol are perfect \cite{litinski2019ms}.
At logical level, its output error is $35p^{3}$ plus the logical errors accumulated at $(d_X,d_Z,d_m)$ during the protocol.
The small-footprint 15-to-1 variant serialises the rotations,
trading a longer duration per output state for fewer tiles.
When the error target lies below what one level can reach, the outputs of several low-distance first-level units are distilled again by a second unit at higher distance: in the $(15\text{-to-}1)\times(15\text{-to-}1)$ factory the second level is another 15-to-1 unit, giving an output error of order $35\,(35p^{3})^{3}$, and in the $(15\text{-to-}1)\times(20\text{-to-}4)$ factory it is a 20-to-4 protocol, which turns 20 inputs into four outputs of error $\approx22\,p_{\rm in}^{3}$ \cite{litinski2019ms}.
Specifying these parameters for a given computation gives us physical qubit count, the time required $t_{\rm fac}$ in code cycles to produce one high fidelity magic state, and the output error $p_{\rm out}$ of the factory.
Every output state is consumed by a $T$ gate of the computation, so we
require
\begin{equation}
  p_{\rm out} \le \epsilon / T ,
\label{eq:factory-target}
\end{equation}
which keeps the total probability of a faulty $T$ state below the
failure budget.
For each circuit and architecture we select, among single-level 15-to-1
factories, their small-footprint variants and the two-level
$(15\text{-to-}1)\times(15\text{-to-}1)$ and
$(15\text{-to-}1)\times(20\text{-to-}4)$ protocols, the factory with the
smallest space-time cost per $T$ state that satisfies
Eq.~\ref{eq:factory-target}, with the output distance $d_X$ equal to the
code distance $d$ of the data block.
At $p=10^{-4}$ a single-level factory suffices throughout:
$(d_X,d_Z,d_m)=(d,3,3)$ with 810 to 1530 physical qubits and
$t_{\rm fac}=18.1$ code cycles per $T$ state, except for the
40-spin-orbital circuits, where the target $7.8\times10^{-10}$ requires
$(11,5,3)$ (2058 qubits) or, for the compact block at $d=13$, $(13,3,5)$
(1970 qubits, $t_{\rm fac}=30.1$).
With $f$ factories the supply is $f$ $T$ states per $t_{\rm fac}$ code
cycles, while the data block can consume one per $c\,d$ code cycles.
If the supply is slower the computation is factory bound and runs for
$n_L = T\,t_{\rm fac}/(f\,d)$ logical cycles instead of $T\,c$.
The smallest number of factories at which the data block becomes the
bottleneck, is
\begin{equation}
  f_{\rm limit} = \lceil t_{\rm fac}/(c\,d) \rceil :
\end{equation}
three factories at $d=7,9$ and two at $d=11$ for the baseline and fast
blocks, and one for the compact block, whose consumption of one $T$ state
per nine logical cycles is always slower than the factory.
Figure~\ref{fig:runtime-vs-orbitals-floor} uses $f=f_{\rm react}$ for
every architecture.

\subsection{Qubit count and runtime}
\label{app:qubits-runtime}
With $d$ from Eq.~\ref{eq:distance-criterion}, the physical qubit count
is $2d^{2}\,n_Q$ for the data block plus the qubits of the $f$
factories, and the runtime is $n_L\,d$ code cycles of $1\,\mu$s.
For the CAS($n$,$n$) series this gives $d=7$ at 4 spin orbitals, $d=9$ up
to 16 spin orbitals and $d=11$ above, with $d=13$ for the compact block
at 40 spin orbitals.

\subsection{Active-volume estimate}
\label{app:av}
For the active-volume architecture the Clifford$+T$ circuit is compiled
into Pauli product rotations, the Clifford rotations are commuted to the
end of the circuit and absorbed into a Clifford frame, and the remaining
$T$ $\pi/8$ rotations and the final measurements are each specified by a
Pauli string on the $n$ qubits.
Each such operation is assigned an active volume of
$\lceil 1.5\,w_X\rceil+\lceil 1.5\,w_Z\rceil+1$ logical blocks, where
$w_X$ and $w_Z$ count the qubits on which the string acts as $X$ or $Z$,
with $Y$ counted in both and one extra block of each type when the
number of $Y$ is odd.
Because the Clifford frame conjugates every rotation, the strings are
dense: the mean volume per $T$ gate is $\approx1.5n$ blocks, with a
largest block of 17 to 98 across the series.
The operations are scheduled in circuit order into logical cycles of
workspace capacity $W$ blocks, an operation starting a new cycle when it
does not fit into the remainder of the current one.
We take $W$ as the smallest capacity, not below the larger of $2n$ and
the largest block, at which the schedule executes at least two $T$ gates
per logical cycle, matching the output of one factory; this gives
$W=23$ to $150$ and $2.00$ to $2.07$ $T$ gates per logical cycle for all
circuits.
The data footprint is $2W$ tiles, $W$ of workspace and $W$ of memory,
i.e.\ $n_Q=2W$ in Eq.~\ref{eq:distance-criterion} and $4Wd^{2}$
physical qubits.
Magic states come from a 15-to-1 factory realised in the same
architecture as 35 logical blocks at distance $d$, producing two $T$
states per logical cycle and occupying $70d^{2}$ physical qubits, with
output error
\begin{equation}
  p_{\rm out} \approx 35\,[\,4\,p_L(p,d/4)+p\,]^{3} + 2\,p_L(p,d),
\label{eq:av-factory}
\end{equation}
where the bracket is the error of an input $T$ state after the
distance-$d/4$ Clifford operations of the protocol and the last term the
logical error of the output patch.
Eq.~\ref{eq:factory-target} holds at the data distance for every
circuit at $p=10^{-4}$, with a margin of 4 to 56.
The schedule consumes at most $2.07$ $T$ states per logical cycle, so a
single factory suffices for the runtime and $n_L$ is the number of
scheduled logical cycles; the runtime is $n_L\,d$ code cycles as before.

\section{Numerical results}\label{app:numerical_results}

\subsection{Details on compilation of quantum circuits of the double factorized Hamiltonian}\label{app:compilation_details}

We employ the PySCF package \cite{sun2018pyscf} to prepare second quantized embedded QM/MM CASCI Hamiltonians of natural orbital basis active spaces of increasing size for a structure of an arginine-tryptophane dimer, embedded in a water solvent, as described in Ref.\,\cite{Castaldo2026ChemicallyAccurate}.
The ffsim library \cite{sung2026ffsim} was subsequently employed for the preparation of the double factorized Hamiltonian, where a slight modification was made to the ffsim code, to separate the Hamiltonian
simulation of the one-body term $H_1$ from the first orbital rotation, which are otherwise contracted by default.
The ffsim library already allows the compilation of Trotterized Hamiltonian simulation qiskit circuits from the double factorized Hamiltonian. We compiled our circuits by employing a first-order product formula, as defined in Eq. \ref{eq:first_order_pf}. We transpiled these qiskit circuits to a gate set of $\{ R_z(\theta), X, SX, CNOT \}$, using an optimization level of 2. Increasing the optimization level to 3 only made a negligible difference in terms of the number of $R_z$ gates.

For consistency with our implementation, and subsequent numerical evaluation of the Hamiltonian 1-norm scaling, we re-express the double factorized molecular Hamiltonian for this section of the SI as
\begin{align}
\label{eq:df_ham_ffsim}
H = \sum_{k=0}^L H_k & = \sum_{pq} h_{pq} a^{\dagger}_p a_q + \sum_{k=1}^{L} G_k J_k G_k^\dagger \\
J_k & = \sum_{pq}^{\rho_k} h^{(k)}_{pq} \hat{n}_p \hat{n}_q,
\end{align}
where $G_k$ denote orbital rotation unitaries, as previously,
and operators $J_k$ are two-body (number) operators.

The Trotterized time evolution operator for a first-order product
formula hence has the structure
\begin{equation}\label{eq:df_trotter_one_ffsim}
S_1(t) = U_1 \prod_{k=1}^{L} G_k D_k(t) G^\dagger_k,
\end{equation}
with

\begin{equation}
D_k(t) = e^{- it J_k} \ ,
\end{equation}
and

\begin{equation}
\prod_{p>q} e^{-it h_{pq} (a^{\dagger}_p a_q + a^\dagger_q a_p)}.
\end{equation}

The construction of $S_2(t)$, and other higher-order product formula then follows trivially from Eqs. \ref{eq:first_order_pf} and \ref{eq:sec_order_pf}.

Our code subsequently constructs controlled time evolution
circuits of Eq. \ref{eq:df_trotter_one_ffsim},
as is required for the Hadamard test.
By exploiting the identity
\begin{equation}
c(U D U^\dagger) = U cD U^\dagger,
\end{equation}
only the one-body unitary  $U_1$,
and the unitaries $D_k(t)$ need to be controlled on the ancilla qubit.

Note however that as mentioned in \ref{app:df_analytic_gc}, switching from uncontrolled to controlled simulation should not affect the number of $R_z$ gates, and hence not affect the total number of $T$ gates in the compiled circuits.

We briefly point out the possibility that rotating the one-body operator to a diagonal basis and implementing it as a two-body number operator, as discussed in \ref{app:fermionic_seminorm} might lead to slight improvements in terms of the $R_z$ gate counts. As $n$ increases, however, the quantum resource cost mainly comes from the double factorization of the ERI tensor, due to the $O(Ln^2) \approx O(n^3)$ scaling compared to the $O(n^2)$ scaling of the one body term $U_1$. Consequently, we expect the savings in $R_z$ gates as a function of $n$ to be negligible as $n$ increases.

\subsection{Energy error due to truncation of $L$}\label{app:df_error_vs_L}

To support the assumption that the double factorization parameter $L$
in Eq. \ref{eq:df_ham} grows linearly with the system size,
we performed the following numerical experiment.

First, we create the double factorized Hamiltonian with different double factorization thresholds, employing the modified Cholesky double factorization routine of the ffsim package \cite{sung2026ffsim}.
We re-factorize the obtained double factorized Hamiltonian,
to reconstruct the now approximate one- and two-body integrals,
i.e. rewrite the molecular electronic Hamiltonian in the form
of Eq. \ref{eq:el_ham}.

\begin{figure}[h]
    \centering
    \includegraphics[width=0.85\linewidth]{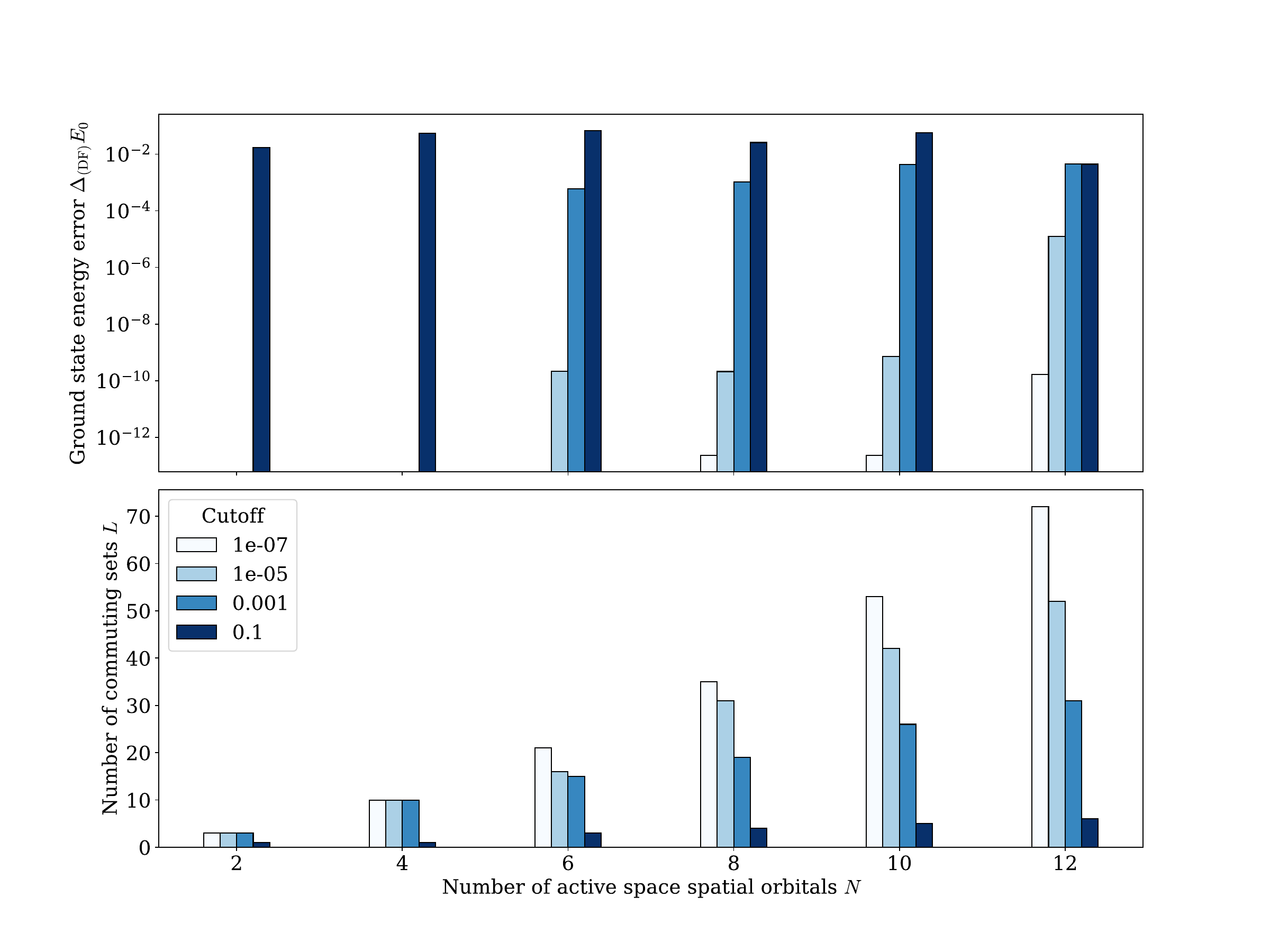}
    \caption{Ground state energy error $\Delta_{\mathrm{(DF)}}E_0$ (top)
    and number of commuting sets $L$ (bottom)
    in the double factorized Hamiltonian
    as a function of the number of active space orbitals $n$
    for different double factorization thresholds as defined
    by ffsim's modified Cholesky routine.}
    \label{fig:df_err_growth_vs_n}
\end{figure}

We then compare the difference in the CASCI energies between
exact (prior to double factorization), and reconstructed (after double factorization) electronic Hamiltonian,
which we denote as $\Delta_{(\mathrm{DF})}E_0$.
The results are depicted in Fig. \ref{fig:df_err_growth_vs_n}.

Although the range of values for $n$ is small, it appears that
the assumption of linear scaling, i.e. $L(n) = O(n)$, for a fixed
double factorization threshold, is sensible.

\subsection{Scaling of the double factorized Hamiltonian 1-norm}\label{app:numerics_df_onorm_scaling}

\begin{figure}[h]
    \centering
    \includegraphics[width=1.0\linewidth]{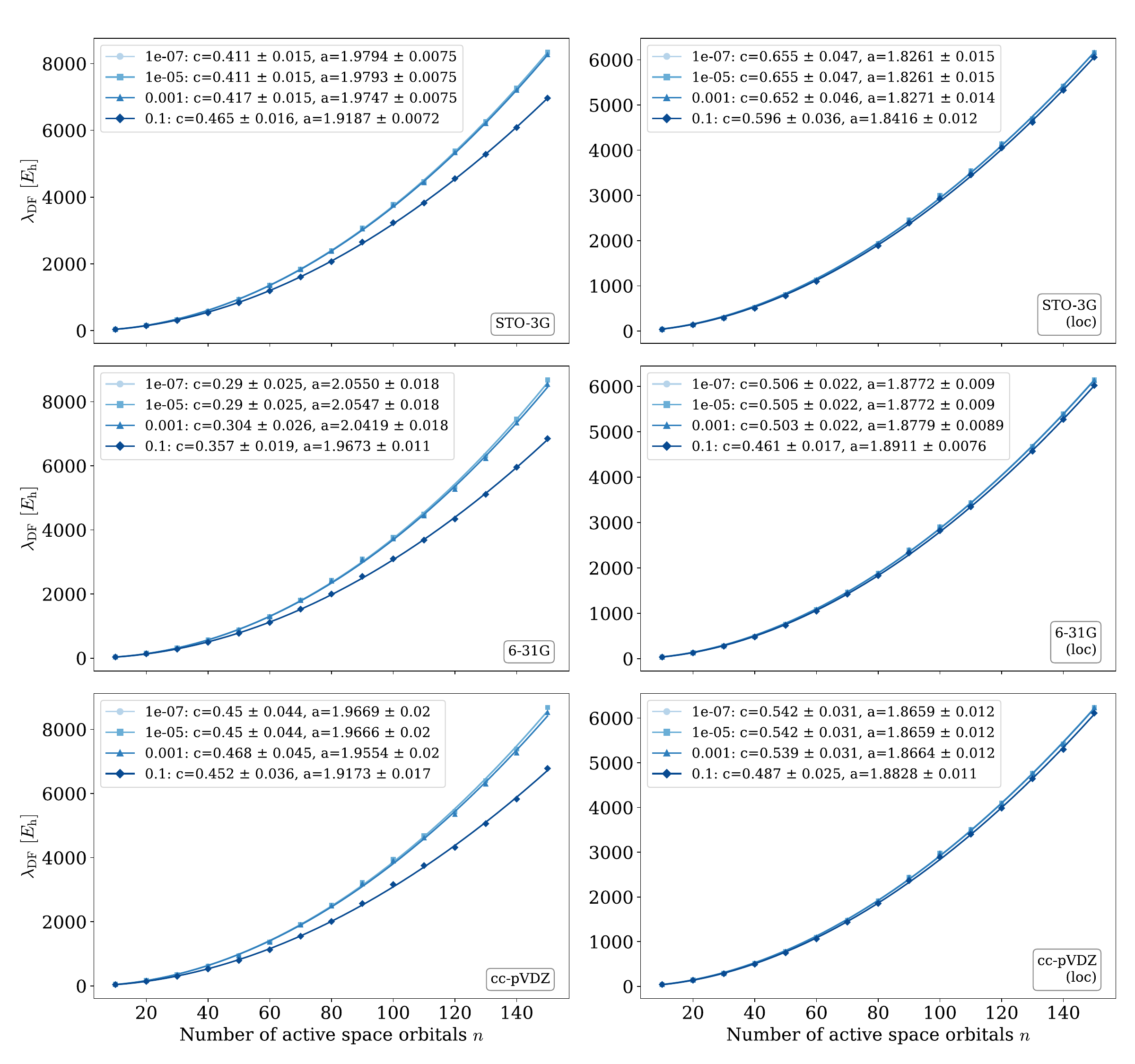}
    \caption{Plots of the induced Hamiltonian 1-norm $\lambda_{\mathrm{DF}}$ as a function of the number of spatial orbitals in the active space.
    The number of active space electrons equals the number of
    active space spatial orbitals.
    Different chemical basis sets were
    deployed: STO-3G (top), 6-31G (middle), cc-pVDZ (bottom).
    Additionally, we show data for MP2 natural orbitals (left),
    as well as an orbital basis obtained by subsequent orbital
    localization of the MP2 natural orbitals by the Foster-Boys
    localization scheme \cite{foster1960canonical} (right).
    Raw data points are shown by various dot symbols.
    Fits of the function $\lambda(n) = cn^a$ for the different
    double factorization thresholds, and
    computed by the scipy library \cite{virtanen2020scipy},
    are plotted as solid lines.
    The parameters for the prefactor $c$ and the polynomial
    coefficient $a$ obtained by the fitting are provided
    in the legend of each sub-figure.}
    \label{fig:df_onorm_scaling}
\end{figure}

We here present additional numerical data on the calculations
of the induced 1-norms of the double factorized Hamiltonian
$\lambda_{\mathrm{DF}}$, where we employed basis sets of
increasing size.
The results are depicted in Fig. \ref{fig:df_onorm_scaling}.

We included some smaller orbital bases compared to the cc-pVDZ basis, in particular the STO-3G basis to potentially model the scenario, where very high-lying virtual orbitals with negligible contribution to the system, are added, more accurately.
The idea was that for these small orbital bases,
one would approach the scaling of $\lambda_{\mathrm{DF}}(n)$
in the the asymptotic limit more quickly.

\begin{figure}[h]
    \centering
    \includegraphics[width=1.0\linewidth]{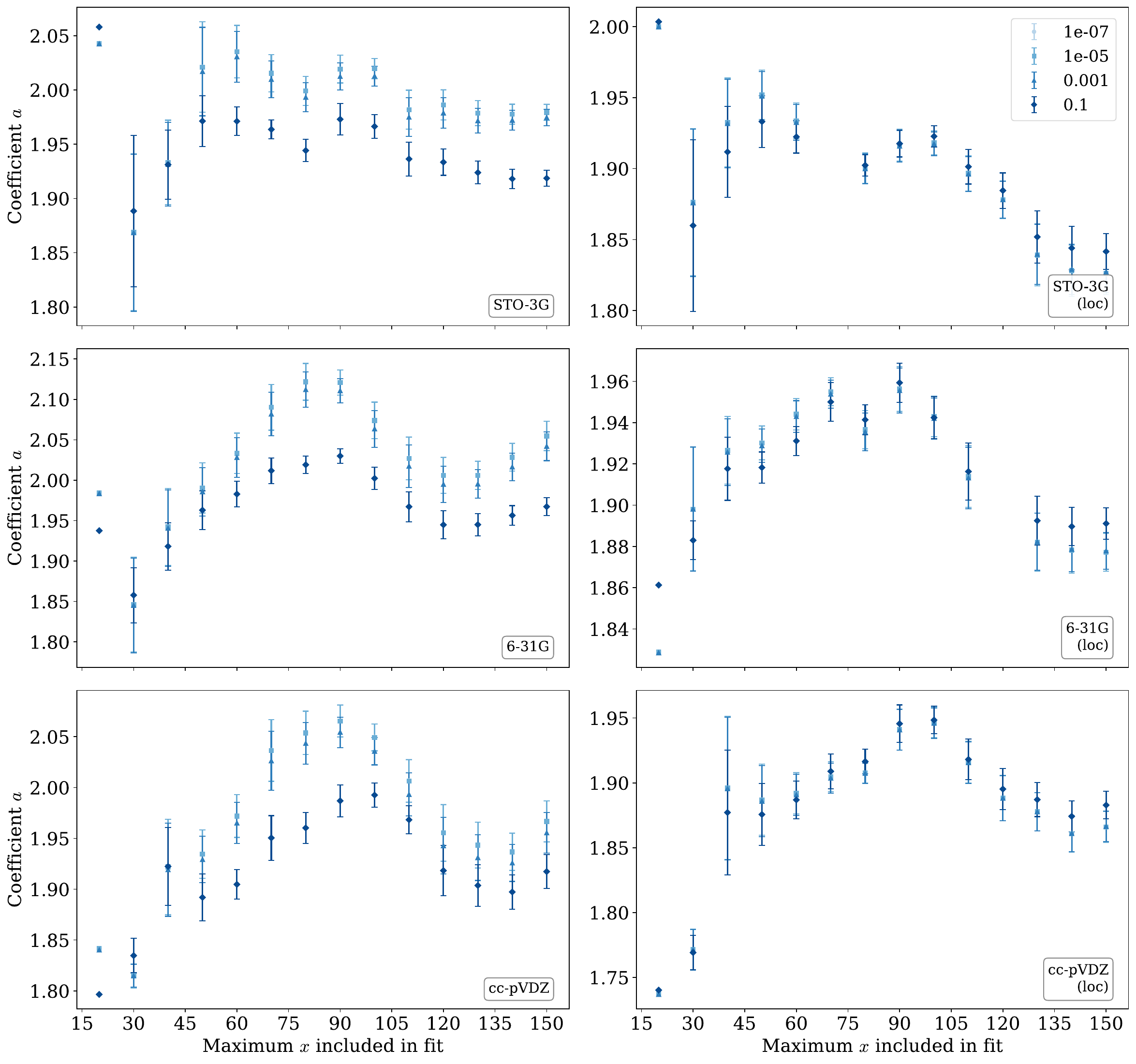}
    \caption{Plots of the change in the polynomial
    fit coefficients of the
    function $\lambda_{\mathrm{DF}}(n) = c n^a$,
    as a larger number of active orbital space data
    points from Fig. \ref{fig:df_onorm_scaling} are considered for the fitting.
    Details about the orbital bases, different DF truncation thresholds, and fit procedure
    are provided in the caption of Fig. \ref{fig:df_onorm_scaling}.}
    \label{fig:df_onorm_fit_coeff_convergence}
\end{figure}

To assess the convergence of the coefficient $c$,
we plot its change, as a larger number of data points
from Fig. \ref{fig:df_onorm_scaling} are considered
for the fitting. The results are shown in Fig. \ref{fig:df_onorm_fit_coeff_convergence}.

\subsection{Compiled circuit depths}\label{app:comp_circuit_depths}

\begin{figure}
    \centering
    \includegraphics[width=0.85\linewidth]{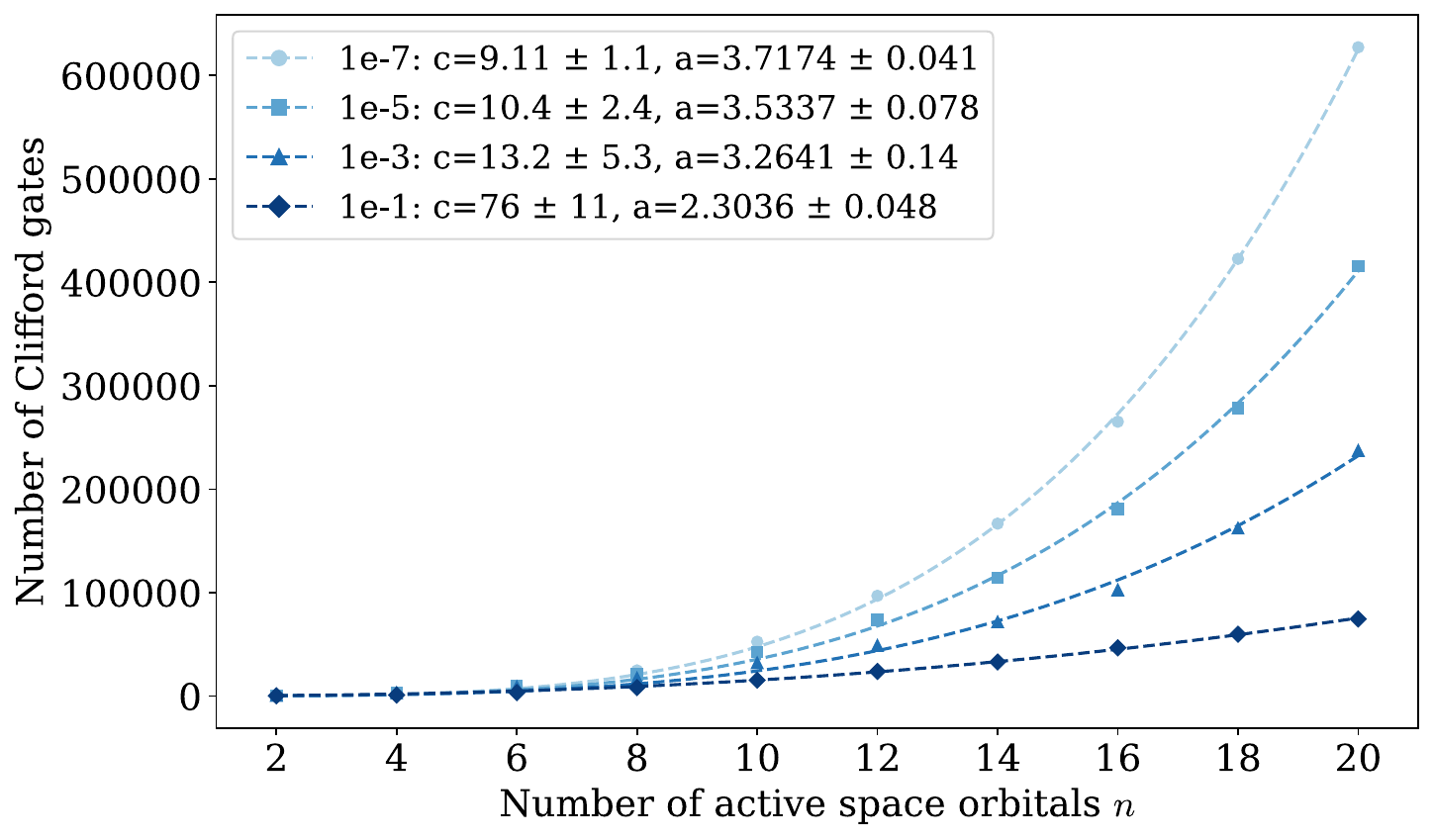}
    \caption{Total number of Clifford, i.e. non-$R_z$ gates for quantum circuits of a single Trotter step with a
    double factorized Hamiltonian with different double factorization truncation thresholds.
    Raw data points are shown by various dot symbols.
    Fits of the function $N_{\mathrm{Clifford}} = cn^a$ for the different
    double factorization thresholds, and
    computed by the scipy library \cite{virtanen2020scipy},
    are plotted as solid lines.
    The parameters for the prefactor $c$ and the polynomial
    coefficient $a$ obtained by the fitting are provided
    in the figure legend.}
    \label{fig:clifford_growth_with_n}
\end{figure}

For the software pipeline described in \ref{app:compilation_details},
we have compiled quantum circuits for a single Trotter step
to evaluate the scaling of the number of $R_z$ and Clifford
gates with the system size.
Recall that the quantum circuits were compiled to the following
gate set: $\{ R_z, X, SX, CNOT \}$.
$R_z$ gates have already been reported in
Fig. \ref{fig:rz_growth_with_n} of the manuscript.
In Fig. \ref{fig:clifford_growth_with_n},
we additionally report on the number of Clifford
gates in these circuits.

We observe approximately cubic scaling for the circuit depth
in terms of the number of $R_z$ and Clifford gates as a function
of the number of active space orbitals $n$.
A cubic scaling behavior would have been expected,
due to the DF circuit depth scaling as $O(L n^2)$,
where $L$ is known to scale as $O(n)$ from numerical
studies on chemical systems \cite{motta_low_2021-1}.

In the asymptotic limit, where a scaling of $\rho = O(1)$
has been suggested previously \cite{motta_efficient_2019, motta_low_2021-1}, it would follow from a light-cone argument
that only $O(n)$ pairwise orbital rotations, i.e. Givens
rotations would be required
for each operator $G_k$, as opposed to the $O(n^2)$ Givens rotations
required for arbitrary orbital rotations
Under these assumptions one would then expect
the circuit depths of the DF Hamiltonian to scale as $O(Ln)$,
i.e. $O(n^2)$.
In the range of 2-20 spatial orbitals, we cannot expect to observe
the asymptotic limit scaling of $\rho = O(1)$, however.
Based on the numerical study in Ref. \cite{motta_efficient_2019},
the $\rho = O(1)$-scaling is expected to only be reached
for thousands or even tens of thousands of
orbitals for three dimensional chemical systems.

From our results in figs. \ref{fig:rz_growth_with_n} and
\ref{fig:clifford_growth_with_n}, we estimate the prefactor
constant of the $R_z$ and Clifford gate count scaling to
be $C=10$, which is observed for the DF truncation thresholds
of $10^{-3}, 10^{-5}$, and $10^{-7}$.
We disregard the results for the truncation threshold
of $10^{-1}$, due to the extremely small values of $L$
resulting from this truncation threshold
(see Fig. \ref{fig:df_err_growth_vs_n}).

\section{Related work on quantum phase estimation}
\label{app:methods_phase_estimation}

Here, we summarize the theory underlying the methods against which we have compared our QPD implementation. All these single-ancilla methods can be described within the same cost model, as they differ only in the classical postprocessing applied to the same quantity measured on the quantum computer, namely, the autocorrelation function $C(t)$.

As discussed in the main text, the focus of this work is on the problem of estimating the ground state energy for the molecular Hamiltonian (see Eq.\,\ref{eq:el_ham}-). Given this starting point, all the protocols that we have compared consider an initial reference state $|\Phi\rangle$, with overlaps with the molecular Hamiltonian eigenstates

\begin{equation}
  p_k=\left|\braket{\psi_k|\Phi}\right|^2,\qquad \sum_k p_k=1 .
\end{equation}

Due to the symmetry conservation properties of the molecular Hamiltonian the autocorrelation signal $C(t)$ contains only contribution from states with $p_k > 0$, so in the following we will denote the ground state energy as the lowest energy in the symmetry sector spanned by the initial state $|\Phi\rangle$. We write $p_0$ for the ground-state overlap, $p_{\mathrm{tail}}=1-p_0$ for the residual weight, and
\begin{equation}
  \Delta=\min\{E_k-E_0:\ p_k>0,\ E_k>E_0\}
\end{equation}
for the gap to the first populated excited level. The task is to return $\hat E_0$ with $|\hat E_0-E_0|\le\epsilon$; the target accuracy tipically considered in the context of \textit{ab initio} quantum chemistry is $\epsilon^\ast=1~\mathrm{mHa}$.

Some of the methods below are formulated for a set $\mathcal D$ of ``dominant'' eigenvalues~\cite{ding2023simultaneous, ding2024quantum}. For ground-state estimation $\mathcal D=\{0\}$; in their notation $p_{\min}=p_0$ and the residual overlap is $R=p_{\mathrm{tail}}$.

Before going into the details of each method we summarize the main quantities and features that the different algorithms have in common. We refer to the original works of Ref.\,\cite{ding2023simultaneous, ding2024quantum, shen2025estimating, klymko2022real} for a more detailed account.

\subsection{Autocorrelation function from Hadamard test measurements and cost model}

All single-ancilla methods consume samples of the autocorrelation function
\begin{equation}
  C(t)=\braket{\Phi|e^{-iHt}|\Phi}=\sum_k p_k\,e^{-iE_k t},
  \qquad C(0)=1,\quad C(-t)=C(t)^\ast,\quad |C(t)|\le 1 .
  \label{eq:Z}
\end{equation}
Its Fourier transform corresponds to a distribution in the frequency domain induced by the overlaps $p_k$ as $\mathcal{F}[C(t)] = \int C(t)\,e^{i\omega t}\,dt=\sum_k p_k\,\delta(\omega-E_k)$. Every method is therefore an estimator of the position of the lowest-lying peak of this distribution from finitely many noisy samples of $C(t)$ on a finite time window obtained executing the Hadamard test circuit

\begin{equation}\label{eq:htest_circ}
    \Re/ \Im [ \langle \Psi | U(t_k) | \Psi \rangle ]  = \Qcircuit @C=1em @R=1em {
 & \qw & \gate{H} & \ctrl{1} & \gate{W} & \gate{H} & \meter \\
 & {/}^{\otimes^n} \qw & \gate{P} & \multigate{0}{U(t_k)} & \qw & \qw & \qw
}
\end{equation}
A single shot returns $x\in\{\pm1\}$ for $W=\mathbb{I}$ and $y\in\{\pm 1\}$ for $W=S^\dagger$, with
\begin{equation}
  P(x=\pm1)=\tfrac12\bigl(1\pm\Re \{C(t)]\}\bigr),\qquad
  P(y=\pm1)=\tfrac12\bigl(1\pm\Im \{C(t)\}\bigr).
  \label{binomial_distribution}
\end{equation}

Given access to this quantum subroutine, we specify the cost metrics of the following phase estimation protocols according to the their sets of $N_t$ distinct evolution times $\{t_n\}_{n=1}^{N_t}$ which can be either deterministic or random. Moreover, in order to acquire sufficient statistics, some methods require to accumulate multiple measurements per each distinct evolution times. Following the main text, we indicate $N_\text{shots}^n$ the number of repetitions of each Hamadard circuit at $t_n$. Finally, we also need to take into account that some protocol measure both the real and imaginary part. For these reasons it is customary to characterize the quantum computational cost of these algorithm using the following metrics:

\begin{align}
  T_{max} &= \max_{n}|t_n| ,\label{eq:Tmax}\\
  T_{runtime} &= \nu\sum_{n=1}^{N_t} N_\text{shots}^n\,|t_n| ,\label{eq:Trun}\\
  N_{meas}  &= \nu\sum_{n=1}^{N_t} N_\text{shots}^n .\label{eq:Nsh}
\end{align}
Where $\nu\in\{1,2\}$: $\nu=2$ when both $\Re \{C(t)\}$ and $\Im \{C(t)\}$ are measured, $\nu=1$ for real-part-only protocols.

\subsection{QPE: Quantum Phase Estimation}

\begin{figure}[h!]
    \centering
\begin{tikzpicture}

\draw (0,0) node[anchor=east] {\Qcircuit @C=1em @R=.7em {
            \lstick{\ket{0}} & \qw &  \gate{H} & \ctrl{4} & \qw & \dots &  & \qw & \qw & \multigate{3}{QFT^{-1}} & \meter \\
            \lstick{\ket{0}} & \qw &  \gate{H} & \qw & \ctrl{3} & \dots &  & \qw & \qw & \ghost{QFT^{-1}} & \meter \\
            & \vdots & & & & & \\
            \lstick{\ket{0}} & \qw &  \gate{H} &  \qw & \qw & \dots & & \ctrl{1} & \qw & \ghost{QFT^{-1}} & \meter \\
            \lstick{\ket{0}} & {/}^{\otimes^n} \qw &  \gate{P} & \gate{U} & \gate{U^2} & \dots & & \gate{U^{2^t}} & \qw & \qw & \qw \\
        }};
\end{tikzpicture}
    \caption{Schematic representation of the Quantum Phase Estimation circuit. The readout register consists of $t$ qubits, each initialized in the state $|0\rangle$, while the system register consists of $n$ qubits to which a generic state preparation routine $P$ is applied, preparing the state $\psi = \sum_u c_u |u\rangle$.}
    \label{fig:qpe_circuit}
\end{figure}
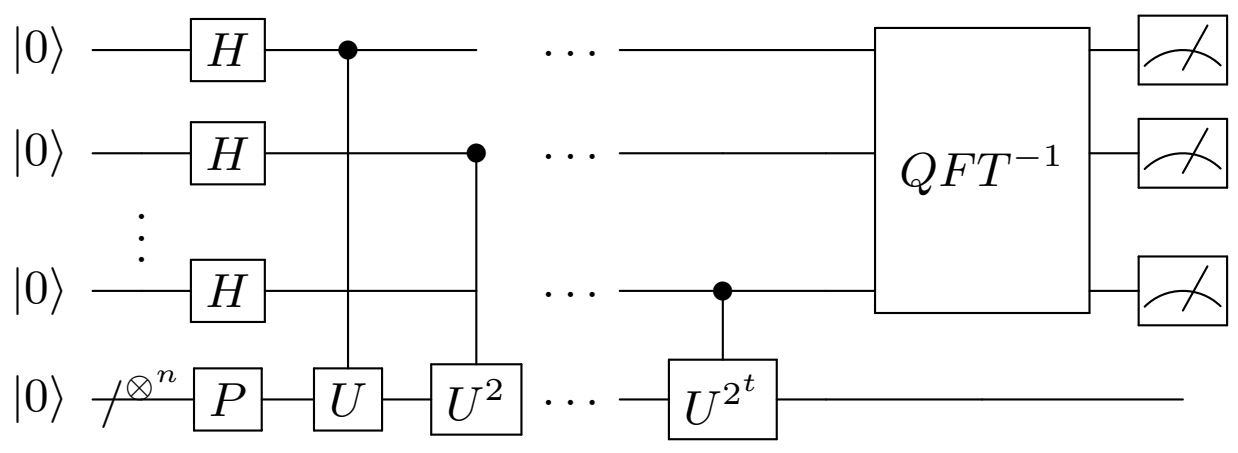

Here we summarize the textbook Quantum Phase Estimation (QPE) algorithm\cite{nielsen_quantum_2010} and show that can also be casted under the same cost metrics that we just discussed. Indeed, even though it relies on a different circuit we can interpret the QPE circuit as a black-box that allows to sample from the distribution $\mathcal{F}[C(t)] = \int C(t)\,e^{i\omega t}\,dt=\sum_k p_k\,\delta(\omega-E_k)$ upon measurement of the ancilla register (also called readout register) in the computational basis. The number of ancilla-qubits used defines the resolution with which we represent $\mathcal{F}[C(t)]$ with its values encoded in the amplitudes of the readout qubits' wavefunction (see Fig.\,\ref{fig:qpe_circuit}).

To better understand the previous statement we write the wavefunction of the QPU after applying the controlled evolutions as

\begin{equation}
    \sum_kc_k\Bigl(\frac{1}{\sqrt{N_t}}\sum_{x=0}^{N_t-1}e^{ix\theta_k}\ket x\Bigr)\otimes\ket{\psi_k}
    =\sum_kc_k\Bigl(\bigotimes_{j=1}^{t}\frac{\ket0+e^{i2^j\theta_k}\ket1}{\sqrt2}\Bigr)\otimes\ket{\psi_k}.
    \label{eq:qpe_kickback}
\end{equation}
where $\theta_k=\tilde{E}_k\Delta t\ (\mathrm{mod}\ 2\pi)$, $\Delta t$ is the evolution time of a single controlled application of U repeated by the QPE circuit, and $\tilde{E}_k$ is the shifted and rescaled Hamiltonian energy for the eigenstate $|\psi_k\rangle$.

As we can see, the amplitudes of the readout register in  Eq.\,\ref{eq:qpe_kickback} encode exactly $C(t)$ discretized on a grid of $N_t = 2^t$ points. Application of the inverse QFT on this wavefunction allows to sample from the distribution in frequency domain that reads

\begin{equation}
    P(y)=\sum_kp_k\,K_{N_T}(\theta_k-\omega_y),\qquad
    K_N(\theta)=|A_N(\theta)|^2=\frac{\sin^2(N\theta/2)}{N^2\sin^2(\theta/2)} .
    \label{eq:fejer}
\end{equation}
where $\omega_y$ are the frequency values in the interval [0, 1] represented on the grid defined by the readout register.

Finally, the energy estimate is obtained from the measured samples of Eq.\,\ref{eq:fejer}. Particularly, it is possible to define the QPE ground state energy estimator as: (i) the minimum sampled phase\,\cite{ding2024quantum}, (ii) the mode of the measurements\,\cite{nielsen_quantum_2010} or (iii) their circular average\cite{cruz2020optimizing, castaldo2024differentiable}. In this work we have used for our numerical calculations the mode of the sampled bitstrings as our energy estimator (also referred to as majority vote estimator), then we can write 

\begin{align}
  T_{max}^{QPE} &= 2^t,\\
  T_{runtime}^{QPE} &= N_\text{shots} \sum_{j=1}^{t}2^j,\\
  N_{meas}^{QPE}  &= N_\text{shots}.
\end{align}
Tipically, by choosing $N_\text{shots} = \mathcal{O}(p_o^{-1})$ it is possible to achieve an error scaling $T_{runtime}\propto T_{max}\propto 1/\epsilon$.

As already mentioned in the discussion, the scalings obtained w.r.t. these metrics can be further improved by applying window functions in the time domain. This can be achieved by appending unitary transformations to the first QFT application, before the controlled time evolutions\,\cite{rendon2022effects,loaiza2025nonlinear,patel2026optimal}.

\subsection{MM-QCELS: Multimodal-Multilevel Quantum Complex Exponentials Least Squares}

Here we summarize the multimodal, multilevel quantum complex exponential least squares (MM-QCELS) algorithm of Ding and Lin\,\cite{ding2023simultaneous}, which generalizes single-mode QCELS\,\cite{ding_even_2023} to the simultaneous estimation of the $K=|\mathcal D|$ dominant eigenvalues. Instead of sampling $C(t)$ on a grid, MM-QCELS draws $N$ evolution times independently from the truncated Gaussian
\begin{equation}
  a_T(t)=\frac{1}{A_\gamma\sqrt{2\pi}\,T}\,e^{-t^2/(2T^2)}\,\mathbf 1_{[-\gamma T,\gamma T]}(t),\qquad A_\gamma=\operatorname{erf}(\gamma/\sqrt2),
  \label{eq:aT_qcels}
\end{equation}
and runs each Hadamard-test circuit once per time ($N_\text{shots}^n=1$, $\nu=2$), so that each data point is the single-shot estimate $\hat C_n=x_n+i\,y_n$ of Eq.\,\ref{binomial_distribution}. The truncation guarantees $|t_n|\le\gamma T$.

The eigenvalues are estimated by fitting $K$ complex exponentials to these data, i.e. by minimizing the loss function
\begin{equation}
  L_K\bigl(\{r_k\},\{\theta_k\}\bigr)=\frac1N\sum_{n=1}^N\Bigl|\hat C_n-\sum_{k=1}^K r_k\,e^{-i\theta_k t_n}\Bigr|^2
  \label{eq:qcels_loss}
\end{equation}
over $r_k\in\mathbb C$ and $\theta_k\in\mathbb R$. Since $\mathbb E[\hat C_n]=C(t_n)$ and $t_n\sim a_T$, for $N\gg1$ the minimizers of Eq.\,\ref{eq:qcels_loss} approach those of the ideal loss $\int a_T(t)\,|C(t)-\sum_kr_k\,e^{-i\theta_kt}|^2\,dt$.

Considering the case $K=1$, relevant for the ground state, the authors show that this optimization problem corresponds to finding the maximum of the function\,\cite[Eqs.~(6)--(7)]{ding2023simultaneous}
\begin{equation}
  G(\theta)=\Bigl|\frac1N\sum_{n=1}^N\hat C_n\,e^{i\theta t_n}\Bigr| ,
  \label{eq:qmegs_G}
\end{equation}
i.e.\ the fitted energy is $\theta^\ast=\arg\max_\theta G(\theta)$. In expectation,
\begin{equation}
  \mathbb E\Bigl[\frac1N\sum_{n=1}^N\hat C_n\,e^{i\theta t_n}\Bigr]=\int a_T(t)\,C(t)\,e^{i\theta t}\,dt=\sum_kp_k\,F(\theta-E_k),\qquad
  F(x)=\int a_T(t)\,e^{ixt}\,dt ,
  \label{eq:gauss_filter}
\end{equation}
where $F(x)\to e^{-T^2x^2/2}$ for $\gamma\to\infty$. $G$ is thus a Monte Carlo estimate of the spectral distribution $\mathcal F[C(t)]$ smoothed by a Gaussian of width $\sim1/T$ (cf.\ Eqs.\,(6)--(7) of Ref.\,\cite{ding2023simultaneous}).

For large $T$, $G$ has many local maxima of width $\sim1/T$, and a fit at the final $T$ needs a good starting point. MM-QCELS therefore proceeds in levels $j=0,\dots,l$, each with a fresh set of $N_j$ times drawn from $a_{T_j}$ and $T_j=2^jT_0$. Level~0 searches $\theta_k\in[-\pi,\pi]$; level $j\ge1$ restricts $\theta_k$ to $\theta_k^\ast\pm\pi/T_{j-1}$, with $\theta_k^\ast$ the optimum of level $j-1$. Theorem~1 of Ref.\,\cite{ding2023simultaneous} shows that, with $\gamma$, $T_0$, $N_0$ and $N_{j\ge1}$ independent of $\epsilon$ and $l=\max\{\lceil\log_2(q/(\epsilon T_0))\rceil,1\}$, all dominant eigenvalues are found within $\epsilon$ with probability $1-\eta$ and $T_{max}=\delta/\epsilon$, where $\delta=\tilde\Theta(q\log q^{-1})$ and $q=\Theta(R/p_{\min})$. The depth constant $\delta$ can thus be much smaller than the $\pi$ required by QPE when $p_{\mathrm{tail}}\ll p_0$.

In this work we used $K=1$ and the level schedule ($T_0$, $l$, $\{N_j\}$, $\gamma$) reported in Tab.\,\ref{tab:settingsA}. Counting the samples of all levels, we can write
\begin{align}
  T_{max}^{\text{MM-QCELS}} &= \max_{j,n}\bigl|t_n^{(j)}\bigr|\le\gamma T_l,\\
  T_{runtime}^{\text{MM-QCELS}} &= 2\sum_{j=0}^{l}\sum_{n=1}^{N_j}\bigl|t_n^{(j)}\bigr|,\qquad
  \mathbb E\bigl[T_{runtime}^{\text{MM-QCELS}}\bigr]=2c_\gamma\sum_{j=0}^{l}N_jT_j,\\
  N_{meas}^{\text{MM-QCELS}} &= 2\sum_{j=0}^{l}N_j,
\end{align}
where $c_\gamma=\mathbb E|t|/T=\sqrt{2/\pi}\,(1-e^{-\gamma^2/2})/A_\gamma$. Since $T_j$ doubles at every level while $N_{j\ge1}$ is fixed, the sum is dominated by the last levels and $T_{runtime}\propto T_l\propto T_{max}$, which with $T_{max}=\delta/\epsilon$ gives the Heisenberg scaling $T_{runtime}\propto T_{max}\propto1/\epsilon$\,\cite[Thm.~1]{ding2023simultaneous}.

\subsection{QMEGS: Quantum Multiple Eigenvalue Gaussian filtered Search}

Here we summarize the quantum multiple eigenvalue Gaussian filtered search (QMEGS) of Ding et al.\,\cite{ding2024quantum}. QMEGS uses the same type of data as MM-QCELS, single-shot estimates $\hat C_n=x_n+i\,y_n$ ($N_\text{shots}^n=1$, $\nu=2$) at $N$ times drawn independently from a truncated Gaussian of width $T$ and truncation parameter $\gamma$ ($\sigma$ in Ref.\,\cite{ding2024quantum}), but replaces the nonlinear fit with a direct evaluation of the filtered spectral density $G(\theta)$ of Eq.\,\ref{eq:qmegs_G} on the grid
\begin{equation}
  \theta_j=-\pi+jq/T,\qquad 0\le j\le\lfloor2\pi T/q\rfloor ,
\end{equation}
whose spacing is set by the search parameter $q$. By Eq.\,\ref{eq:gauss_filter}, $G$ estimates $|\sum_kp_kF(\theta-E_k)|$, the spectral distribution $\mathcal F[C(t)]$ convolved with a Gaussian of width $\sim1/T$, playing for QMEGS the role that the distribution of Eq.\,\ref{eq:fejer} plays for QPE, with the Gaussian $F$ in place of the Fej\'er kernel $K_{N_T}$.

The $K$ dominant eigenvalues are taken as the $K$ largest values of $G(\theta_j)$, each selected outside the open intervals $(\theta_{k'}-\alpha/T,\theta_{k'}+\alpha/T)$ around the maxima already selected, so that the same peak is not selected twice (block parameter $\alpha$). For the ground state ($K=1$) this reduces to $\hat E_0=\arg\max_jG(\theta_j)$. For $K=1$, QMEGS and a single level of MM-QCELS therefore maximize the same function of the same type of data (Eq.\,\ref{eq:qmegs_G}); QMEGS uses a single $T$ and a grid search over $[-\pi,\pi)$ instead of a continuous optimization refined over several levels.

Theorem~3.1 of Ref.\,\cite{ding2024quantum} requires only $p_{\min}>p_{\mathrm{tail}}$, i.e.\ $p_0>1/2$ for the ground state, and no spectral gap. For $\alpha,\gamma=\Omega\bigl(\log^{1/2}[1/(p_{\min}-p_{\mathrm{tail}})]\bigr)$, $q$ below a bound set by the overlaps with $q<\alpha/3$ and $\alpha/q\in\mathbb N$, and $N=\Omega\bigl((p_{\min}-p_{\mathrm{tail}})^{-2}\log[(T/q+|\mathcal D|)/\eta]\bigr)$, every dominant eigenvalue lies within $\alpha/T$ of a selected maximum with probability $1-\eta$, so $T=\alpha/\epsilon$ suffices. When the dominant eigenvalues are separated by a gap, Theorems~3.2--3.3 of the same work allow shorter $T_{max}$.

In this work we sampled the renormalized density of Eq.\,\ref{eq:aT_qcels}, with $K=1$ and $T$, $\gamma$, $\alpha$, $q$ and $N$ as reported in Tab.\,\ref{tab:settingsA}. We can then write
\begin{align}
  T_{max}^{\text{QMEGS}} &= \max_n|t_n|\le\gamma T,\\
  T_{runtime}^{\text{QMEGS}} &= 2\sum_{n=1}^{N}|t_n|,\qquad \mathbb E\bigl[T_{runtime}^{\text{QMEGS}}\bigr]=2c_\gamma NT,\\
  N_{meas}^{\text{QMEGS}} &= 2N,
\end{align}
with $c_\gamma$ as for MM-QCELS.

\subsection{ODMD:  Observable Dynamic Mode Decomposition}

Here we summarize observable dynamic mode decomposition (ODMD) of Shen et al.\,\cite{shen2025estimating}. Dynamic mode decomposition (DMD)\,\cite{mezic2004comparison, schmid2010dynamic} approximates the evolution of a classical dynamical system by the linear map that best propagates a sequence of its snapshots by one time step. Since the quantum state is not accessible, ODMD uses as snapshots vectors of the autocorrelation function.

$C(t)$ is measured on the uniform grid $t_k=k\Delta t$, $k=1,\dots,N_t$, with $N_\text{shots}$ repetitions per circuit, and $\hat C_0=C(0)=1$ ($N_t=K+d$ in the notation of Ref.\,\cite{shen2025estimating}). The data are arranged into two Hankel matrices with $d$ rows,
\begin{equation}
  \mathbf X_{ij}=\hat C_{i+j},\qquad \mathbf X'_{ij}=\hat C_{i+j+1},\qquad i=0,\dots,d-1,\quad j=0,\dots,N_t-d ,
\end{equation}
which differ by one time step. For noiseless data $C(k\Delta t)=\sum_mp_m\mu_m^k$ with $\mu_m=e^{-iE_m\Delta t}$; if only $d$ levels were populated, $\mathbf X'=A\mathbf X$ would hold exactly for a $d\times d$ companion matrix $A$ whose eigenvalues are the $\mu_m$. ODMD determines $A$ by least squares,
\begin{equation}
  A=\mathbf X'\,\mathbf X_{\tilde\delta}^{+},\qquad \hat E_0=-\frac{1}{\Delta t}\max_\ell\arg\mu_\ell ,
  \label{eq:odmd}
\end{equation}
where $\mu_\ell$ are the eigenvalues of $A$ and $\mathbf X_{\tilde\delta}^{+}$ is the pseudoinverse of $\mathbf X$ after discarding its singular values below $\tilde\delta$ times the largest one.

The time step must satisfy $\Delta t<2\pi/[(E_{\max}-E_0)+(E_1-E_0)]$\,\cite[Eq.~(15)]{shen2025estimating}, with $E_{\max}$ the largest populated energy. In the noiseless case the error then decreases as $|\hat E_0-E_0|=O(N_te^{-\beta d})$, with $\beta>0$ set by the gap relative to the populated spectral range\,\cite[Eqs.~(27)--(28)]{shen2025estimating}; Ref.\,\cite{shen2025estimating} takes $d\simeq(N_t+1)/3$, one third of the $N_t+1$ samples. Noise robustness comes from the truncation threshold $\tilde\delta$, which should sit above the noise level $\epsilon_{\mathrm{noise}}$ of the samples ($\epsilon_{\mathrm{noise}}\le N_\text{shots}^{-1/2}$ by Eq.\,\ref{eq:shot_noise}); $\tilde\delta/\epsilon_{\mathrm{noise}}\approx10$ is reported as typically optimal for $\epsilon_{\mathrm{noise}}\lesssim10^{-2}$.

Ref.\,\cite{shen2025estimating} measures only $\Re\{C(t)\}$ ($\nu=1$). The matrix $A$ is then real and its eigenvalues come in conjugate pairs $e^{\pm iE_m\Delta t}$, so the spectrum must be shifted such that $|E_{\max}|\le|E_0|$ for the maximal phase to identify $E_0$. In this work we used the complex signal ($\nu=2$), for which this condition is not needed, with $\Delta t$, $N_t$, $N_\text{shots}$ and $\tilde\delta$ as reported in Tab.\,\ref{tab:settingsA}. Since $C(0)$ is known, only $k\ge1$ is measured and we can write
\begin{align}
  T_{max}^{\text{ODMD}} &= N_t\,\Delta t,\\
  T_{runtime}^{\text{ODMD}} &= \nu N_\text{shots}\,\Delta t\sum_{k=1}^{N_t}k=\nu N_\text{shots}\,\Delta t\,\frac{N_t(N_t+1)}{2},\\
  N_{meas}^{\text{ODMD}} &= \nu N_\text{shots}\,N_t .
\end{align}
At fixed $N_\text{shots}$, $T_{runtime}\propto T_{max}^2$.

\subsection{UVQPE: Ultracompact Variational Quantum Phase Estimation}

Here we summarize the unitary formulation of variational quantum phase estimation (UVQPE) of Klymko et al.\,\cite{klymko2022real}, a real-time Krylov method closely related to filter diagonalization\,\cite{parrish_quantum_2019, cortes_quantum_2022}. The ground state is approximated in the space spanned by the $N_t$ time-evolved states $\ket{\phi_j}=U(j\Delta t)\ket\Phi$, $j=0,\dots,N_t-1$ ($N_t=N_T+1$ in the notation of Ref.\,\cite{klymko2022real}). Instead of projecting $H$, which requires measuring its matrix elements, the unitary formulation projects the one-step propagator $U(\Delta t)$, giving the generalized eigenvalue problem (GEP)
\begin{equation}
  \mathbf U\mathbf c=\mu\,\mathbf S\mathbf c,\qquad
  \mathbf S_{jk}=\braket{\phi_j|\phi_k}=C\bigl((k-j)\Delta t\bigr),\qquad
  \mathbf U_{jk}=\braket{\phi_j|U(\Delta t)|\phi_k}=C\bigl((k-j+1)\Delta t\bigr),
  \label{eq:uvqpe}
\end{equation}
whose eigenvalues $\mu_\ell$ approximate $e^{-iE_m\Delta t}$. Both matrices are Toeplitz and, since $C(-t)=C(t)^\ast$, are built from $C(k\Delta t)$, $k=0,\dots,N_t$.

The shot noise makes $\mathbf S$ ill-conditioned. UVQPE therefore discards the singular values of $\mathbf S$ below a threshold $s_{\mathrm{SV}}$, solves the GEP in the retained subspace, and returns $\hat E_0=-\max_\ell\arg\mu_\ell/\Delta t$. This requires the populated spectral range times $\Delta t$ to be below $2\pi$, so that $E_m$ is not confused with its images $E_m\pm2\pi/\Delta t$. Klymko et al.\,\cite{klymko2022real} require the threshold $s_{\mathrm{SV}}$ to exceed the magnitude of the sampling noise, which scales as $N_\text{shots}^{-1/2}$.

The connection with $\mathcal F[C(t)]$ appears after rotating the Krylov basis with $W_{jl}=e^{ij\omega_l\Delta t}$\cite{cortes_quantum_2022}:
\begin{equation}
  \ket{\tilde\phi_l}=\sum_jW_{jl}\ket{\phi_j}=\sum_kc_k\,D_{N_t}\bigl((E_k-\omega_l)\Delta t\bigr)\ket{\psi_k},\qquad D_M(x)=\sum_{j=0}^{M-1}e^{-ijx}=\frac{1-e^{-iMx}}{1-e^{-ix}},
  \label{eq:dirichlet}
\end{equation}
with $c_k=\braket{\psi_k|\Phi}$. Each rotated state filters the spectrum around $\omega_l$ with the Dirichlet kernel, whose square is the Fej\'er kernel of Eq.\,\ref{eq:fejer}, $|D_M(x)|^2=M^2K_M(x)$. For $\omega_l=2\pi l/(N_t\Delta t)$ on the DFT grid, $W/\sqrt{N_t}$ is unitary, so $W^\dagger\mathbf SW$ and $W^\dagger\mathbf UW$ define the same GEP (with an absolute threshold rescaled by $N_t$)\,\cite{cortes_quantum_2022, parrish_quantum_2019}. The slowly decaying sidelobes of $D_M$ ($\propto1/(M|x|)$ relative to the peak) leak out-of-band weight into every basis state, which adds to the ill-conditioning of $\mathbf S$. QPD replaces the boxcar with optimally concentrated filters, as discussed in the main text. If the phase-cancellation conditions of Ref.\,\cite{klymko2022real} hold, the singular values of $\mathbf S$ grow linearly with $N_t$, so that weakly populated states eventually rise above $s_{\mathrm{SV}}$.

In this work we used the same grid as for ODMD, with $\Delta t$, $N_t$, $N_\text{shots}$ and $s_{\mathrm{SV}}$ as reported in Tab.\,\ref{tab:settingsA}. Since UVQPE requires the complex signal ($\nu=2$), we can write
\begin{align}
  T_{max}^{\text{UVQPE}} &= N_t\,\Delta t,\\
  T_{runtime}^{\text{UVQPE}} &= 2N_\text{shots}\,\Delta t\sum_{k=1}^{N_t}k=N_\text{shots}\,\Delta t\,N_t(N_t+1),\\
  N_{meas}^{\text{UVQPE}} &= 2N_\text{shots}\,N_t ,
\end{align}

\begin{table*}[h!]
\centering\small
\caption{Theoretical acquisition, estimator used in this work and cost of the compared methods.}
\label{tab:methods}
\renewcommand{\arraystretch}{1.25}
\begin{tabular}{@{}l p{0.209\linewidth} p{0.197\linewidth} l p{0.197\linewidth} c@{}}
\hline\hline
Method & Acquisition & Estimator & ${T_{\max}}$ & ${T_{\mathrm{runtime}}}$ \\
\hline
QPE & $N_{\mathrm{QPE}}$ shots, $n_a=\log_2N_T$ ancillas & Fej\'er histogram: mode, min, or (generalised) circular mean & $N_T\Delta t/2$ & $N_{\mathrm{QPE}}(N_T-1)\Delta t$ \\
MM-QCELS & $N_j$ i.i.d.\ $t\sim a_{T_j}$, $j=0,\dots,l$, $N_\text{shots}^n=1$, $\nu=2$ & Minimizer of Eq.\,\ref{eq:qcels_loss}, refined over levels & $\le\gamma T_l$ & $\simeq2c_\gamma\sum_jN_jT_j$ \\
QMEGS & $N$ i.i.d.\ $t\sim a_T$, $N_\text{shots}^n=1$, $\nu=2$ & $\arg\max_jG(\theta_j)$, Eq.\,\ref{eq:qmegs_G} & $\le\gamma T$ & $\simeq2c_\gamma NT$ \\
ODMD & $t_k=k\Delta t$, $k\le N_t$, $N_\text{shots}$, $\nu\in\{1,2\}$ & Hankel DMD, Eq.\,\ref{eq:odmd}, relative SVD cut $\tilde\delta$ & $N_t\Delta t$ & $\nu N_\text{shots}\Delta t\frac{N_t(N_t+1)}{2}$ \\
UVQPE & $t_k=k\Delta t$, $k\le N_t$, $N_\text{shots}$, $\nu=2$ & Toeplitz GEP, Eq.\,\ref{eq:uvqpe}, SVD cut $s_{\mathrm{SV}}$ & $N_t\Delta t$ & $N_\text{shots}\Delta t\,N_t(N_t+1)$ \\
QPD & $t_k=\pi k/W_s$, $k\le N_s$, $M_s=F\sqrt{N_s\log N_s}$ & Solution of Eq.\,\ref{eq:geneig_exact}, projection on principal subspace of \textbf{S} & $\pi N_s/W_s$ & $\nu M_s\frac{\pi}{W_s}\frac{N_s(N_s+1)}{2}$ \\
\hline\hline
\end{tabular}
\end{table*}

\subsection{Computational details for the phase estimation benchmark}
\label{app:comp-detail-benchmark}

The benchmark set consists of 28 different molecular systems with active spaces spanning 4 to 16 spin orbitals (qubits) (see Tab.\,\ref{tab:molecules}). In all our experiments the initial state has been chosen as the Hartree-Fock state, i.e. $|\Phi\rangle = |HF\rangle$, for systems that are not closed-shell singlets (i.e. odd hydrogen chains) canonical orbitals used to compute the initial mean field state and electron integrals are obtained from a ROHF calculation(ref.). Active spaces ($n_e, n_o$) are obtained choosing $n_o$ orbitals around the Fermi level as implemented in PySCF\cite{sun2018pyscf}. The autocorrelation function is computed exactly from the spectrum and overlaps obtained building the full active-space FCI matrix and the Hadamard test outcomes (i.e. shot noise) is simulated sampling the from a binomial distribution centered at the exact value of $C(t)$ (see Eq.\,\ref{binomial_distribution}). 

Implementation of the methods corresponding to Ref.\,\cite{klymko2022real,ding_even_2023, ding2024quantum, shen2025estimating} relies on the original code developed by the authors and has been modified only for the purpose of integration in the benchmark pipeline.

\begin{table}[h]
\centering\footnotesize
\caption{Molecular benchmark set. All values are read from the result files. Distances in \AA; $N=2n_o$ is the number of qubits; $2S+1$ is the spin multiplicity; $p_0$ is the HF ground-state overlap; $\lambda$ is the rescaling factor of Eq.~\ref{eq:rescaling}; $E_0$ is the CASCI ground-state energy. Polyenes labelled ``$r_{\mathrm{C=C}}$'' are planar all-trans chains with $r_{\mathrm{C-C}}=1.46$ and $r_{\mathrm{CH}}=1.09$; the \texttt{cc1p7A} variants stretch every double bond to 1.70.}
\label{tab:molecules}
\renewcommand{\arraystretch}{1.12}
\setlength{\tabcolsep}{4pt}
\begin{tabular}{@{}l l l c c c c c r@{}}
\toprule
Label & Geometry & Basis & $(n_e,n_o)$ & $N$ & $2S{+}1$ & $p_0$ & $\lambda$ (Ha) & $E_0$ (Ha) \\
\midrule
\multicolumn{9}{@{}l}{\emph{Main set} ($p_0\ge0.9$, 19 systems)} \\
\texttt{h2\_chain\_1A\_2e2o} & H$_2$, $r=1.00$ & cc-pVDZ & (2,2) & 4 & 1 & 0.9904 & 0.977 & $-1.108338$ \\
\texttt{h3\_chain\_1A\_3e3o} & linear H$_3$, $r=1.00$ & cc-pVDZ & (3,3) & 6 & 2 & 0.9870 & 1.511 & $-1.596683$ \\
\texttt{h4\_chain\_1A\_4e4o} & linear H$_4$, $r=1.00$ & cc-pVDZ & (4,4) & 8 & 1 & 0.9748 & 2.638 & $-2.186232$ \\
\texttt{h5\_chain\_1A\_5e5o} & linear H$_5$, $r=1.00$ & cc-pVDZ & (5,5) & 10 & 2 & 0.9645 & 3.389 & $-2.691071$ \\
\texttt{h6\_chain\_1A\_6e6o} & linear H$_6$, $r=1.00$ & cc-pVDZ & (6,6) & 12 & 1 & 0.9520 & 4.604 & $-3.272663$ \\
\texttt{h7\_chain\_1A\_7e7o} & linear H$_7$, $r=1.00$ & cc-pVDZ & (7,7) & 14 & 2 & 0.9355 & 5.479 & $-3.787992$ \\
\texttt{h8\_chain\_1A\_8e8o} & linear H$_8$, $r=1.00$ & cc-pVDZ & (8,8) & 16 & 1 & 0.9260 & 6.799 & $-4.362525$ \\
\texttt{nh3\_8e5o} & NH$_3$, $r_{\mathrm{NH}}=1.01$ & 6-31G & (8,5) & 10 & 1 & 0.9989 & 2.711 & $-56.163128$ \\
\texttt{h2co\_6e5o} & H$_2$CO$^{\dagger}$ & cc-pVDZ & (6,5) & 10 & 1 & 0.9874 & 4.557 & $-112.899551$ \\
\texttt{ethene\_2e2o} & C$_2$H$_4$$^{\dagger}$ & cc-pVDZ & (2,2) & 4 & 1 & 1.0000 & 1.292 & $-76.114582$ \\
\texttt{butadiene\_4e4o} & C$_4$H$_6$$^{\dagger}$ & cc-pVDZ & (4,4) & 8 & 1 & 0.9883 & 2.344 & $-151.490913$ \\
\texttt{hexatriene\_6e6o} & C$_6$H$_8$$^{\dagger}$ & cc-pVDZ & (6,6) & 12 & 1 & 0.9852 & 4.054 & $-226.861976$ \\
\texttt{octatetraene\_8e8o} & C$_8$H$_{10}$$^{\dagger}$ & cc-pVDZ & (8,8) & 16 & 1 & 0.9832 & 6.017 & $-302.232339$ \\
\texttt{butadiene\_ideal\_4e4o} & C$_4$H$_6$, $r_{\mathrm{C=C}}=1.34$ & cc-pVDZ & (4,4) & 8 & 1 & 0.9763 & 1.672 & $-154.945505$ \\
\texttt{hexatriene\_ideal\_6e6o} & C$_6$H$_8$, $r_{\mathrm{C=C}}=1.34$ & cc-pVDZ & (6,6) & 12 & 1 & 0.9405 & 2.613 & $-231.858119$ \\
\texttt{octatetraene\_ideal\_8e8o} & C$_8$H$_{10}$, $r_{\mathrm{C=C}}=1.34$ & cc-pVDZ & (8,8) & 16 & 1 & 0.9478 & 4.002 & $-308.747571$ \\
\texttt{formamide\_6e6o} & HCONH$_2$, $r_{\mathrm{CO}}=1.219$ & cc-pVDZ & (6,6) & 12 & 1 & 0.9791 & 5.173 & $-168.964592$ \\
\texttt{n2\_1p09A\_6e6o} & N$_2$, $r=1.09$ & cc-pVDZ & (6,6) & 12 & 1 & 0.9484 & 4.747 & $-109.021161$ \\
\texttt{porphyrin\_8e8o} & porphyrin (planar)$^{\ddagger}$ & STO-3G & (8,8) & 16 & 1 & 0.9489 & 3.719 & $-969.695851$ \\
\midrule
\multicolumn{9}{@{}l}{\emph{Multireference subset} ($p_0<0.9$, 9 systems)} \\
\texttt{h2\_chain\_2A\_2e2o} & H$_2$, $r=2.00$ & cc-pVDZ & (2,2) & 4 & 1 & 0.8050 & 0.441 & $-0.996699$ \\
\texttt{h4\_chain\_2A\_4e4o} & linear H$_4$, $r=2.00$ & cc-pVDZ & (4,4) & 8 & 1 & 0.6236 & 1.281 & $-1.997021$ \\
\texttt{h6\_chain\_2A\_6e6o} & linear H$_6$, $r=2.00$ & cc-pVDZ & (6,6) & 12 & 1 & 0.4754 & 2.454 & $-2.997842$ \\
\texttt{h8\_chain\_2A\_8e8o} & linear H$_8$, $r=2.00$ & cc-pVDZ & (8,8) & 16 & 1 & 0.3589 & 3.835 & $-3.998754$ \\
\texttt{butadiene\_cc1p7A\_4e4o} & C$_4$H$_6$, $r_{\mathrm{C=C}}=1.70$ & cc-pVDZ & (4,4) & 8 & 1 & 0.7293 & 1.211 & $-154.833689$ \\
\texttt{hexatriene\_cc1p7A\_6e6o} & C$_6$H$_8$, $r_{\mathrm{C=C}}=1.70$ & cc-pVDZ & (6,6) & 12 & 1 & 0.6084 & 2.282 & $-231.656428$ \\
\texttt{octatetraene\_cc1p7A\_8e8o} & C$_8$H$_{10}$, $r_{\mathrm{C=C}}=1.70$ & cc-pVDZ & (8,8) & 16 & 1 & 0.4156 & 3.543 & $-308.526547$ \\
\texttt{n2\_1p40A\_6e6o} & N$_2$, $r=1.40$ & cc-pVDZ & (6,6) & 12 & 1 & 0.8561 & 3.532 & $-108.916561$ \\
\texttt{formamide\_co2p2A\_6e6o} & HCONH$_2$, $r_{\mathrm{CO}}=2.20$ & cc-pVDZ & (6,6) & 12 & 1 & 0.8589 & 7.051 & $-168.723488$ \\
\bottomrule
\end{tabular}
\par\smallskip\raggedright\footnotesize
$^{\dagger}$PubChem 2D depiction coordinates: all heavy-atom nearest-neighbour distances are 1.00~\AA\ and the shortest C--H distance is 0.62~\AA \\
$^{\ddagger}$PubChem CID 5481287, 36 atoms, C$_{20}$H$_{12}$N$_4$, with 11 C--H and one N--H.
\end{table}

It is also important to notice that all the algorithms compared in this work require the frequency of $C(t)$ to lie within an interval $[a, b]$ which is, in general, algorithm dependent. In order to comply with the prescription given by the authors of each algorithm we carried our numerical experiments with scaled and reshifted Hamiltonian (and energies) $\tilde{H}=\frac{H-e_0\mathrm{I}}{\lambda},\, \tilde{E}_k=\frac{E_k-e_0}{\lambda}$. For MM-QCELS, QMEGS, ODMD and UVQPE the shift is the exact ground-state energy, $e_0=E_0$, and the scale is $\lambda=1.15\max_k|E_k-E_0|=1.15\,(E_{\max}-E_0)$, where the maximum runs over all eigenstates in the target spin sector. Because each method recovers $\widetilde E_k$ from the phase of $\sum_kp_k\,\mathrm{e}^{-i\widetilde E_k\tilde t}$, which is defined only modulo $2\pi$, the spectrum must fit in a window narrower than $2\pi$. QPE resolves phases on $[-\pi,\pi)$, and ODMD and UVQPE sample at unit steps, $\Delta\tilde t=1$. QPE uses the same $\lambda$ but the Hartree--Fock shift, $e_0=E_{\mathrm{HF}}$: with $e_0=E_0$ the ground state would fall exactly on a QPE bin, giving zero discretisation error. QPD is run without rescaling ($\lambda=1$) (the algorithmic consequences have been discussed in Sec.\,\ref{sec:benchmark_main} of the main text and in Sec.\,\ref{app:level-shift}) as also done in the original paper\,\cite{stroschein_ground_2025}.

All errors are evaluated after mapping back to physical units, $\hat E_0=\lambda\hat{\widetilde E}_0+e_0$.
Two features of this normalisation should be kept in mind. First, $\lambda$ is computed from the exact spectrum, which would not be available when the algorithm will be applied at production sccale. In practice, one would use a computable upper bound $\lambda_{\mathrm{bd}}\ge\lambda$, for instance the 1-norm of the qubit Hamiltonian, and costs would then grow by the factor $\lambda_{\mathrm{bd}}/\lambda$.

As additional information, we summarize the acquisition scheme and cost accounting in Tab.\,\ref{tab:settingsA}.

\begin{table*}[h!]
\centering\small
\caption{Acquisition and cost accounting for the methods studied in this work. All single-ancilla methods measure both real and imaginary part of $C(t)$. See \,\ref{app:methods_phase_estimation} for a summary of all the methods considered.}
\label{tab:settingsA}
\renewcommand{\arraystretch}{1.3}
\setlength{\tabcolsep}{4pt}
\begin{tabular}{@{}l p{0.322\linewidth} p{0.119\linewidth} p{0.092\linewidth} p{0.101\linewidth} p{0.155\linewidth}@{}}
\hline\hline
 & \multicolumn{3}{c}{Acquisition} & \multicolumn{2}{c}{Recorded cost} \\
\cline{2-4}\cline{5-6}
Method & Time grid & Time points & Shots & ${T_{\max}}^c$ & ${T_{\mathrm{runtime}}}$ \\
\hline
QPE & readout register of $N_T=2T_b$ bins ($n_a=7$--$12$), $\tau=1$; Fej\'er kernel ($\alpha=1$) & -- & $N_{\mathrm{QPE}}=60$ & $T_b$ & $60\,(2T_b-1)$ \\
MM-QCELS & i.i.d.\ $t\sim a_{T_j}$, Eq.~\ref{eq:aT_qcels}; $T_j=2^jT_0$, $T_0=T_b/4$, $T_l=T_b$; $\gamma=1$ & $N_0=1000$, $N_{j\ge1}=500$ & 1 & $\max_n|t_n|$ & $\sum_n|t_n|$ \\
QMEGS & i.i.d.\ $t\sim N(0,T^2)$ truncated to $|t|\le\sigma T$ and renormalised, i.e.\ Eq.~\ref{eq:aT_qcels} with $\gamma\to\sigma$; $T=T_b$; $\sigma=\max(\gamma,\sqrt{\ln(1/\bar p)})$ & $N^{a}$ & 1 & $\max_n|t_n|$ & $\sum_n|t_n|$ \\
ODMD & ${\tilde t}_k=k$, $k=0,\dots,T_b$ ($\Delta{\tilde t}=1$); complex signal & $T_b+1$ & 500 & $T_b$ & $500\sum_kk$ \\
UVQPE & as ODMD & $T_b+1$ & 500 & $T_b$ & $500\sum_kk$ \\
QPD & $t_k=\pi k/W_s$, $W_s=2.125$~Ha & $N_s+1$ & $\sqrt{N_s \text{ln}(N_s)}^{b}$ & $\pi N_s/W_s$ & $\tfrac{\pi M_s}{2W_s}N_s(N_s{+}1)$ \\
\hline\hline
\end{tabular}
\par\smallskip\footnotesize
$^{a}$\,$N=\min\{2\times10^5,\ \max(1,\lfloor4\bar p^{-2}\ln(10^3(T+K))\rfloor)\}$ with $\bar p=\min\{\max(\hat p_0,10^{-2}),0.99\}$.\\
$^{b}$\, $N_s=$27,41,68,101,169,338.\\
$^{c}$\, $T_b=$ 64, 128, 256, 512, 1024, 2048. \,\, Valid for all the methods but for QPD.
\end{table*}

\subsection{Additional plots for the phase estimation benchmark}
\label{app:additional_benchmark}

\begin{figure*}[h!]
    \centering
    \includegraphics[width=\textwidth]{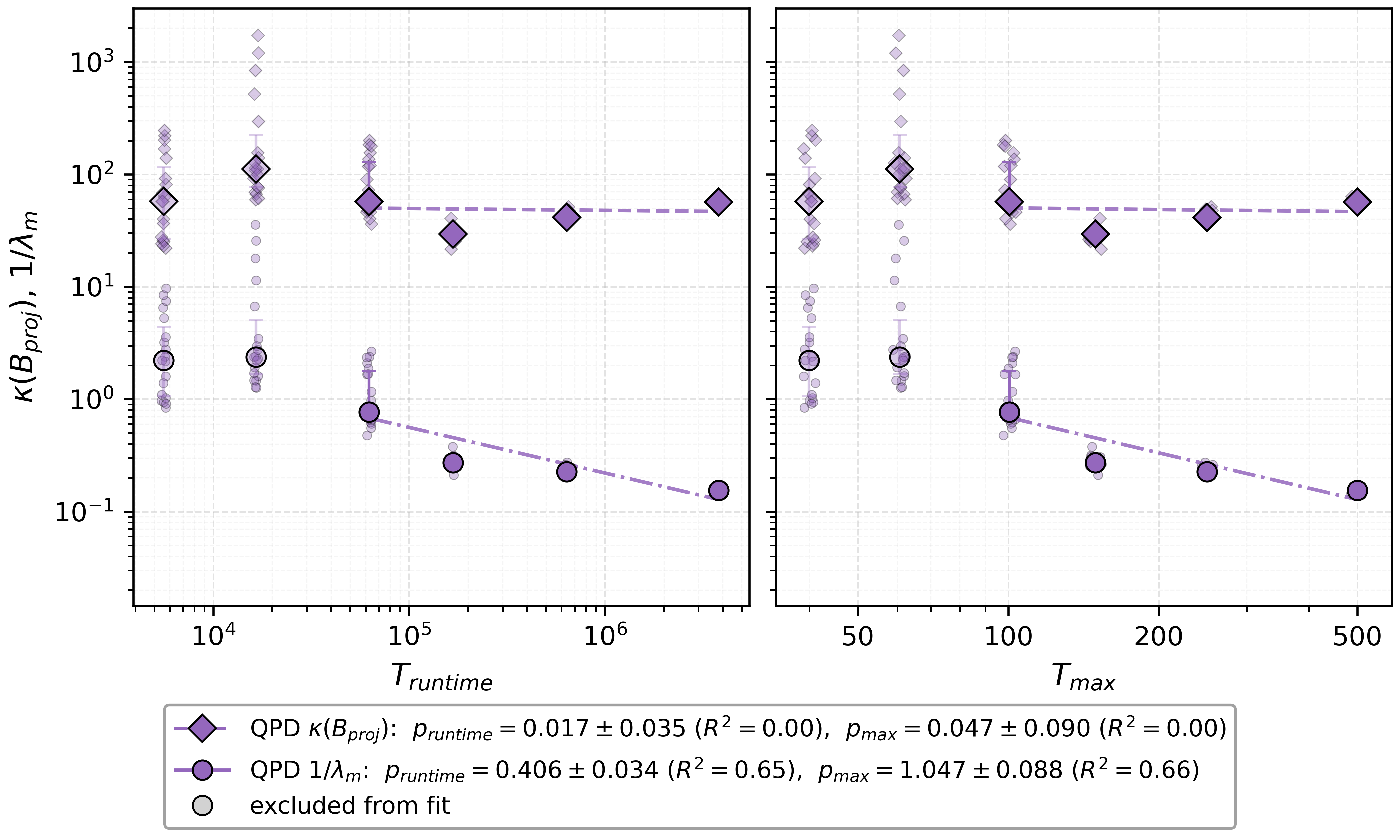}
    \caption{Conditioning of the GEP of Quantum Prolate Diagonalization (Eq.\,\ref{eq:h_pq_integral}-\ref{eq:s_pq_integral}) following the discretization procedure of Ref.\,\cite{stroschein_ground_2025}. Diamonds report the condition number of the overlap matrix $\textbf{S}$ projected on the subspace of the $m$-leading eigenvalues. Circles report the factor $1/\lambda_m$ where $\lambda_m$ is the smallest eigenvalue of the projected overlap matrix. Faint markers correspond to the individual molecules values, each averaged over 10 repetitions. Bold markers correspond to the median values and error bars are represented by the interquartile ranges. Dashed lines report power-law fit obtained with a least square procedure. Empty markers correspond to points excluded from the fit as belonging to a pre-asymptotic regime.}
    \label{fig:placeholder}
\end{figure*}

\begin{figure*}[h!]
    \centering
    \includegraphics[width=\textwidth]{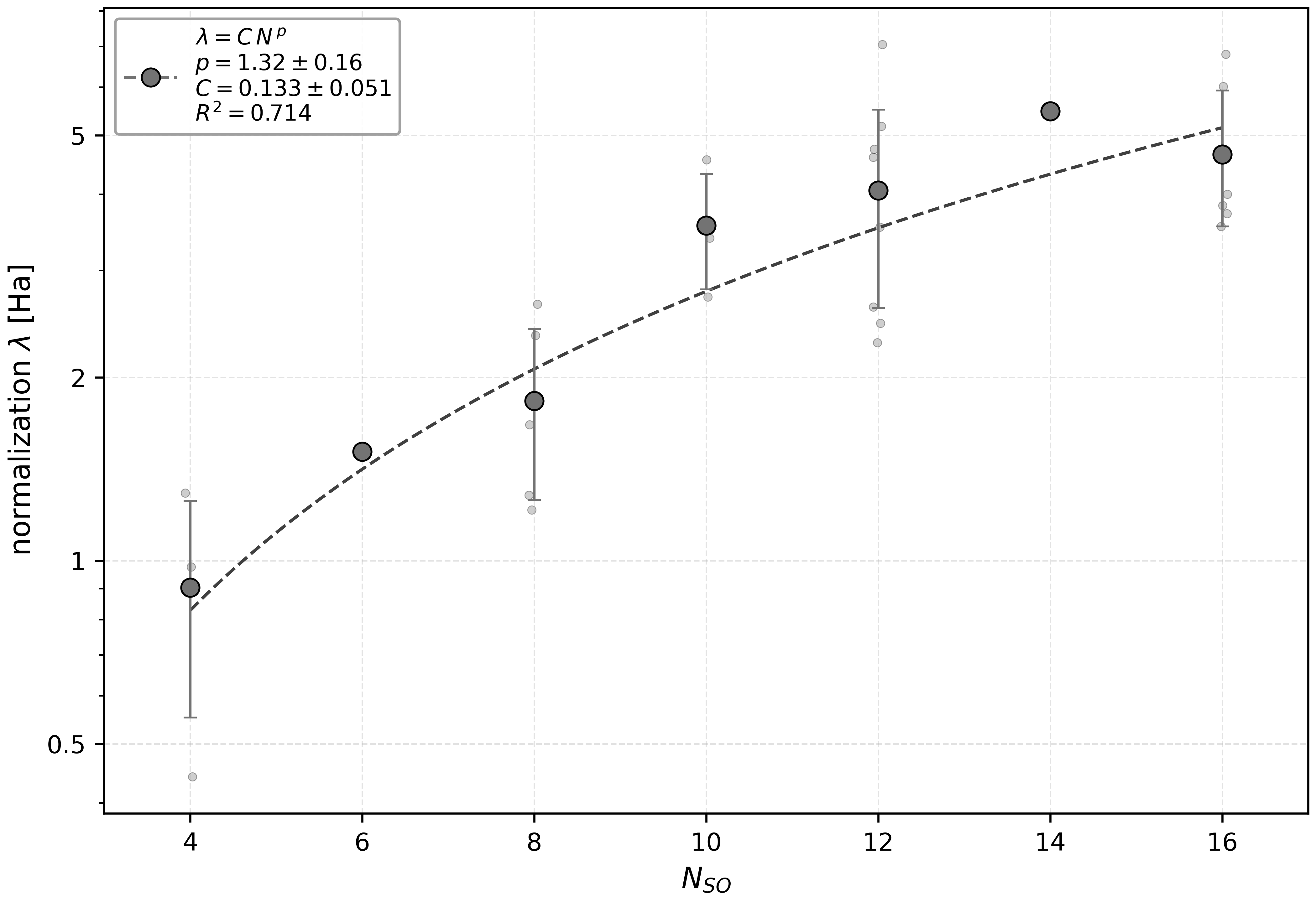}
    \caption{Normalization factor $\lambda$ of the rescaled Hamiltonian $\tilde{H} = \frac{H}{\lambda}$ as a function of the active space size for the dataset corresponding to the plots of the main text. $\lambda$ has been set in all our calculations as $\lambda = 1.15\,\Delta_{max}$ where $\Delta_{max} = \max_k |E_k - E_0|$ is the spectral range of the spin sector spanned by the $|HF\rangle$ state as obtained from the classical CASCI calculations done to compute the autocorrelation function (see \ref{app:comp-detail-benchmark}). Bold markers correspond to median values over molecules with the same number of spin orbitals, error bars are shown with interquartile ranges. Dashed lines correspond to a power-law fit $\lambda = C\,N^{p}$.}
    \label{fig:normalization_factor_scaling}
\end{figure*}

\newpage
\subsection{Phase estimation benchmark in a low overlap regime}

\begin{figure*}[h!]
    \centering
    \includegraphics[width=\textwidth]{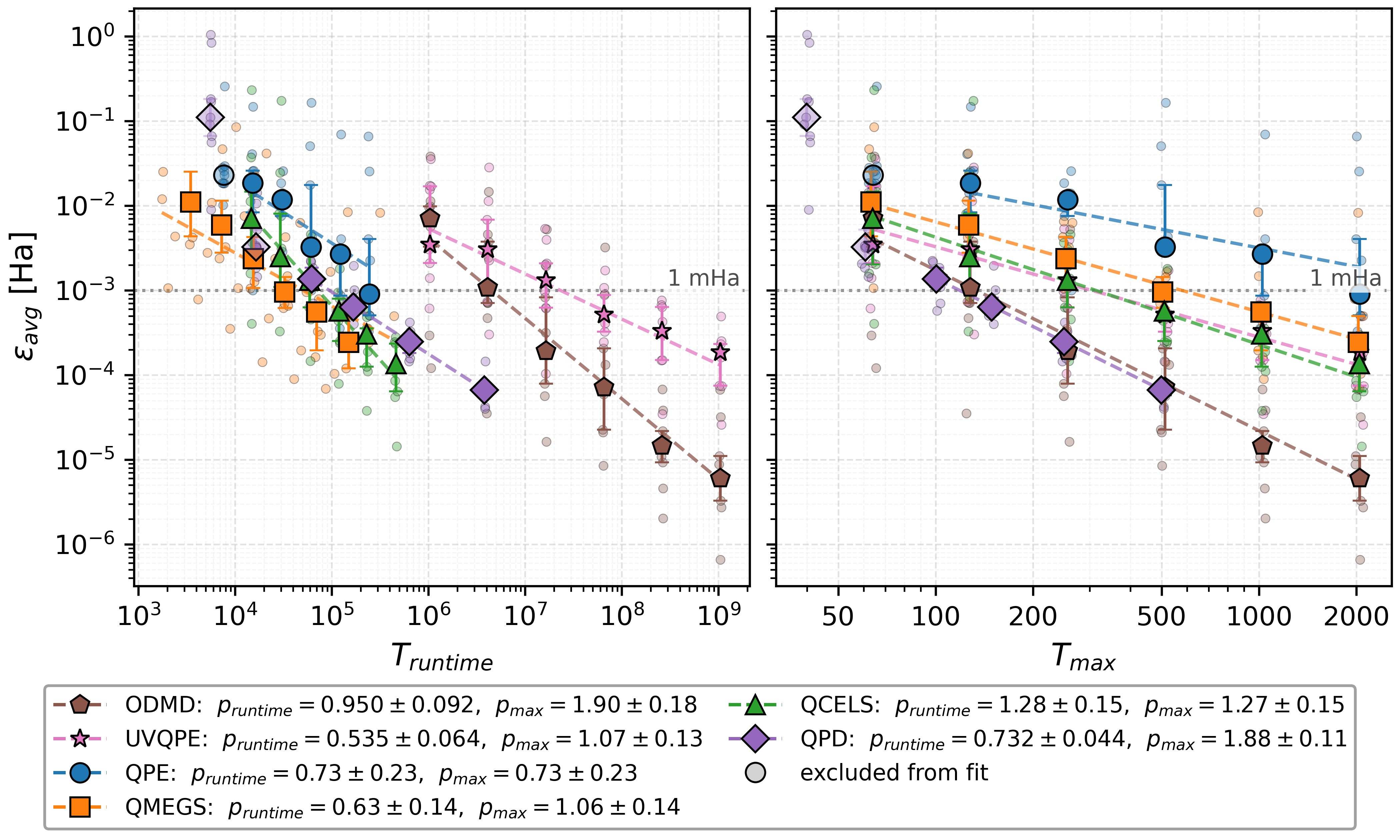}
    \caption{Benchmarking the computational cost of molecular ground-state phase estimation in the low-overlap regime. The legend identifies the different phase estimation algorithms compared in this study (see \ref{app:methods_phase_estimation}). (left) Average estimated ground-state energy as a function of $T_{\mathrm{runtime}}$ (see Eq.\,\ref{eq:Trun}). (right) Average estimated ground-state energy as a function of $T_{\mathrm{max}}$. Faint dots represent the average estimates obtained from 10 independent realizations of each algorithm for a given molecule in our benchmark. Larger markers indicate the median of the averaged errors, and error bars represent the interquartile range. Dashed lines show power-law fits of the form $\epsilon_{\mathrm{avg}}(T)=AT^{-p}$. Empty markers correspond to data point not considered for the fit. Further details on the computational procedure are provided in \ref{app:comp-detail-benchmark}.}
    \label{fig:low_overlap_computational_bench}
\end{figure*}

\begin{figure*}[h!]
    \centering
    \includegraphics[width=\textwidth]{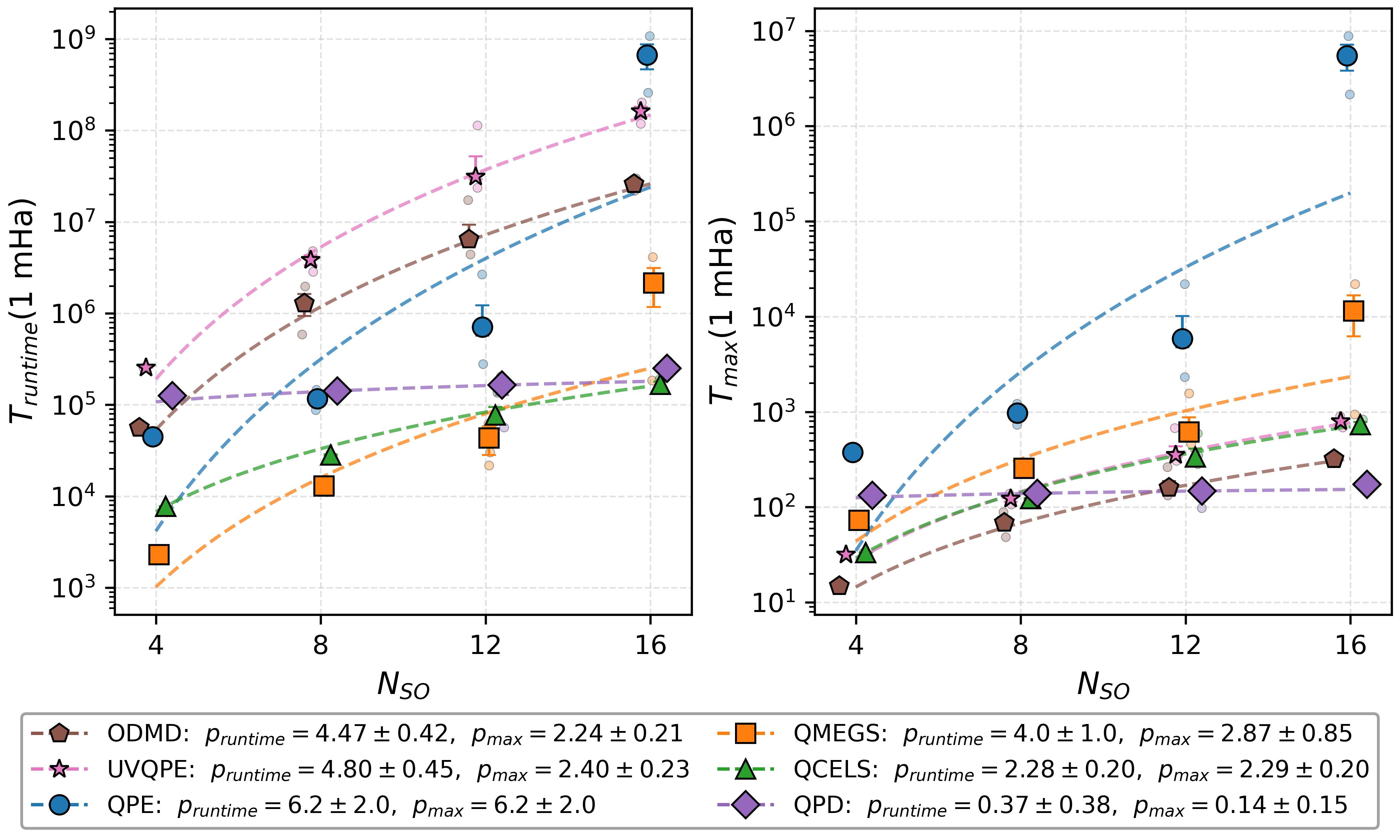}
    \caption{Computational cost scaling with system size in the low-overlap regime. The legend identifies the different phase estimation algorithms compared in this study (see \ref{app:methods_phase_estimation}). (left) $T_{runtime}$ needed to achieve an error estimate of $\epsilon = 1\,\text{mHa}$ of error as a function of the number of spin-orbitals $N_{SO}$. (right) $T_{max}$ needed to achieve an error estimate of $\epsilon = 1\text\,{mHa}$ of error as a function of the number of spin-orbitals $N_{SO}$. Faint dots represent the average estimates obtained from 10 independent realizations of each algorithm for a given molecule in our benchmark. Larger markers indicate the median of the averaged errors, and error bars represent the interquartile range. Dashed lines show power-law fits of the form $T(N_{SO})=AN_{SO}^{p}$. Further details on the computational procedure are provided in \ref{app:comp-detail-benchmark}.}
    \label{fig:scaling_vs_system_size_low_overlap}
\end{figure*}

\begin{figure*}[h!]
    \centering
    \includegraphics[width=\textwidth]{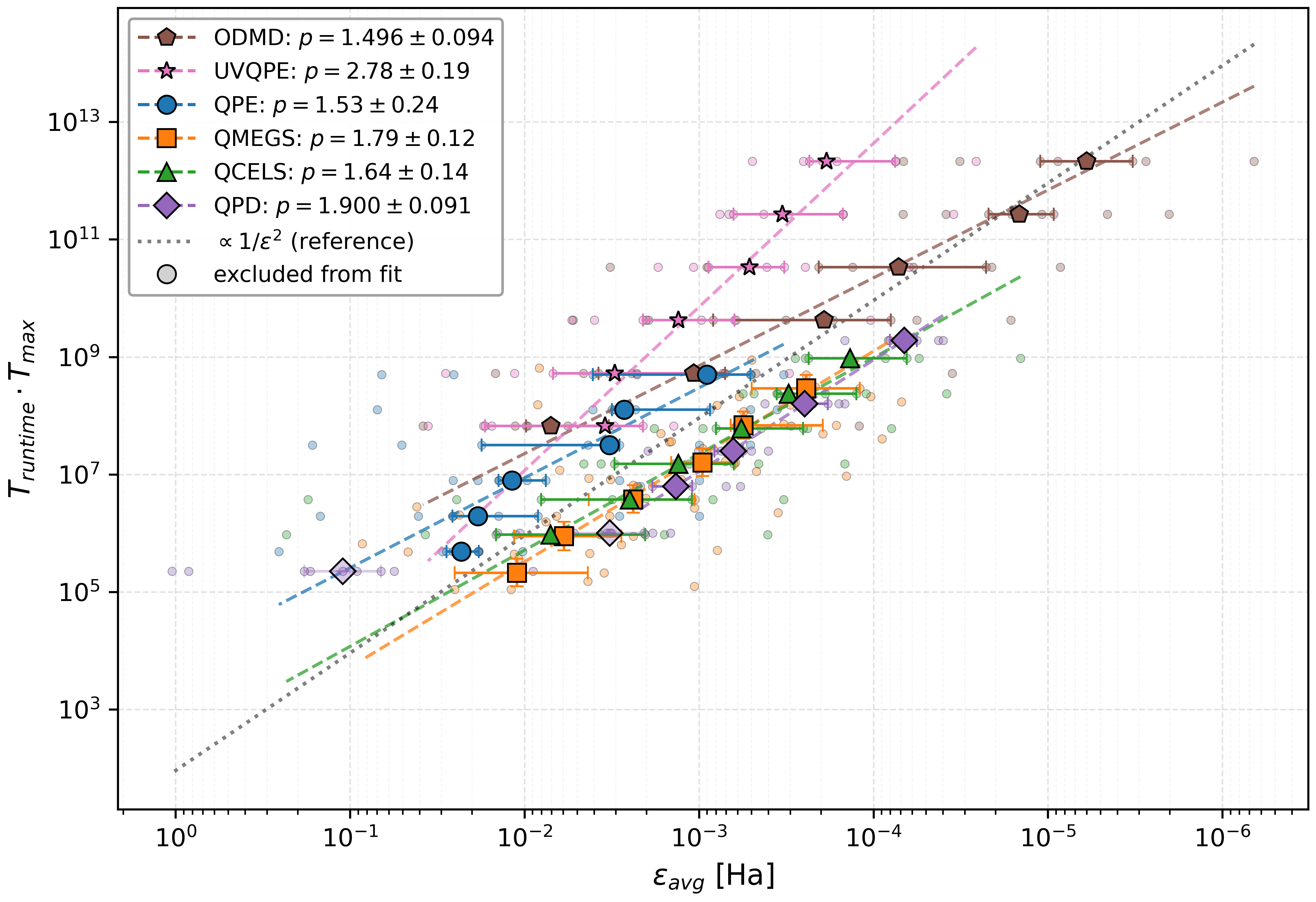}
    \caption{Combined cost $T_{runtime}\,T_{max}$ needed to reach an average ground-state energy error $\epsilon_{avg}$ in the low-overlap regime. Faint dots represent the individual molecules at each computational budget with error estimates averaged from 10 independent realizations of each algorithm. Larger markers indicate the median of the averaged errors, and error bars represent the interquartile range. Dashed lines represent power-law fits $T_{runtime}\,T_{max}(\epsilon_{avg}) = C\,\epsilon_{avg}^{-p}$. Empty markers correspond to data points not considered for the fit. Each fitting is performed independtly for each molecule and the legend gives the mean $\pm$ its standard error. Dotted line represent the reference scaling corresponding to the bound derived in Ref.\,\cite{kimura_fisher_2026}.}
    \label{fig:computational_tradeoff_plot_low_overlap}
\end{figure*}

\end{document}